\documentclass[12pt,a4paper,oneside]{article}
\usepackage[T2A]{fontenc}
\usepackage[utf8]{inputenc}
\usepackage[russian]{babel}
\usepackage{amsfonts}
\usepackage{amstext}
\usepackage{amsmath}
\usepackage{amssymb}
\usepackage[numbers,square,sort]{natbib}
\usepackage{lscape}
\usepackage{float}
\usepackage{alltt}
\usepackage[pdftex]{graphicx}
\usepackage{array}
\usepackage{booktabs}
\usepackage{rotating}
\usepackage[center]{subfigure}
\usepackage{captcont}
\usepackage{framed}
\usepackage{color}
\usepackage{bm}
\usepackage{comment}
\bibpunct{(}{)}{,}{a}{,}{,}
\usepackage{geometry}
\usepackage{tabularx}
\usepackage{hyperref}
\hypersetup{
    colorlinks=true,
    linkcolor=blue,
    filecolor=magenta,      
    urlcolor=cyan,
    pdftitle={Overleaf Example},
    pdfpagemode=FullScreen,
    }

\usepackage{hyperref}
\hypersetup{colorlinks,
citecolor=blue
}

\makeatletter
 {\par\unskip\endMakeFramed%
 \at@end@of@kframe}
\makeatother

\IfFileExists{upquote.sty}{\usepackage{upquote}}{}
\newcommand{\vc}[1]{\ensuremath{\ifcat #1a\mathbf{#1} \else\boldsymbol{#1} \fi}}

\begin{document}


\title{Тестирование на единичный корень в панельных данных: обзор}
\author{
Антон Скроботов$^{a}$ \\
{\small {$^{a}$ РАНХиГС, СПБГУ, Институт Гайдара}}\\
}
\date{5 августа 2024}
\maketitle

\begin{center}
    \LARGE{Panel Data Unit Root testing: Overview}
    
    

\end{center}


\begin{abstract}
В данном обзоре рассматриваются методы тестирования на единичный корень в панельных данных. Обсуждаются современные подходы к тестированию в пространственно-коррелированных панелях, предваряя анализ разбором независимых панелей. Кроме этого, приводятся методы тестирования при нелинейности в данных (например, в случае наличия структурных сдвигов), а также методы тестирования в коротких панелях, когда времнной диапазон небольшой и конечен. В заключении приводятся ссылки на существущие пакеты, которые позволяют реализовать некоторые из описанных методов.

\medskip

\indent \emph{Ключевые слова}: тестирование на панельный единичный корень, тестирование на стационарность панели, структурные сдвиги, детрендирование, общие факторы, пространственная корреляция.

\medskip

This review discusses methods of testing for a panel unit root. Modern approaches to testing in cross-sectionally correlated panels are discussed, preceding the analysis with an analysis of independent panels. In addition, methods for testing in the case of non-linearity in the data (for example, in the case of structural breaks) are presented, as well as methods for testing in short panels, when the time dimension is small and finite. In conclusion, links to existing packages that allow implementing some of the described methods are provided.
\medskip


\medskip


\end{abstract}

\section{Введение}

Тестирование наличия единичных корней в данных имеет большое значение для эмпирического анализа. Практически ни одно макроэкономическое исследование не обходится без тестирования того, является ли конкретный временный ряд стационарным относительно тренда (trend stationary, TS) или является стационарным в первых разностях (difference stationary, DS). В первом случае, если ряд является стационарным относительно тренда, то моделировать ряд необходимо в уровнях. В противном случае нужно перейти к первым разностям временного ряда, если моделируется именно этот конкретный ряд по отдельности, или переходить к анализу коинтеграции нескольких временных рядов, каждый из которых является нестационарным. Наличие коинтеграции позволяет дать экономическое обоснование долгосрочных зависимостей и краткосрочной корректировке к долгосрочным состояниям равновесия.

Однако если рассматривать не отдельные временные ряды для конкретного субъекта (например, одной конкретной страны или одного конкретного региона), а группу (панель) временных рядов по каждому субъекту, то могут возникнуть некоторые проблемы. Существует феномен, который заключается в том, что если оценить регрессию одного нестационарного ряда на другие нестационарные ряды, то оценки OLS-коэффициентов будут часто значимы, даже когда фактически зависимости между рядами нет. Этот феномен называется ложной регрессией. Аналогичный вывод можно получить, рассматривая нестационарные временные ряды, у которых ошибки будут зависимыми. Существует необоснованное заблуждение, что для панельных данных ложная регрессия не наблюдается, особенно при малом количестве наблюдений по времени. Однако эффект ложной регрессии для панельных данных намного более существенен, чем для отдельных временных рядов. Хотя оценка в панельной регрессии для нестационарных (независимых в долгосрочном смысле) временных рядов и будет состоятельной (из-за того, что мы оцениваем модель в виде пула и усредняем все оценки по временным рядам), соответствующая ей $t$-статистика будет расходиться к бесконечности, что означает почти стопроцентное отвержение нулевой гипотезы на больших выборках. Таким образом, выводы на основе этой $t$-статистики будут необоснованными, и необходимо проверять наличие коинтеграции между всеми временными рядами в панели, предполагая, что долгосрочная зависимость между различными показателями одинаковая.
Однако до проверки наличия коинтеграции между панелями временных рядов необходимо проверить наличие единичного корня в панелях, и проверка коинтеграции происходит, если гипотеза о наличии панельного единичного корня не была  отвергнута. Кроме этого, одна из причин перехода к панелям -- низкая мощность классических тестов на наличие единичного корня, и объединение нескольких временных рядов в панели может позволить найти свидетельство стационарности определенного числа временных рядов в этой панели. Соответственно в данном обзоре мы акцентируем внимание на проблемах тестирования наличия панельного единичного корня в данных.

Структура данной работы состоит из следующих разделов. В Разделе 2 обсуждаются как классические тесты на панельный единичный корень, так и недавно разработанные тесты. Гипотеза панельного единичного корня заключается в том, что все временные ряды в панели имеют единичный корень. Тестироваться эта гипотеза может, однако, против разных альтернатив: однородной и неоднородной. Однородная альтернатива заключается в том, что все временные ряды в панели являются стационарными. Для тестирования гипотезы о панельном единичном корне против этой альтернативы используются $t$-статистики на основе модели пула. С другой стороны, при неоднородной альтернативе предполагается, что существует как доля стационарных временных рядов в панели, так и доля нестационарных временных рядов в панели, и эти доли не равны нулю (то есть они должны быть значимы). Для тестирования гипотезы о панельном единичном корне против этой неоднородной альтернативы вычисляются тестовые статистики для каждого ряда в панели, а затем суммируются. При нулевой гипотезе оба этих теста, как против однородной альтернативы, так и против неоднородной альтернативы, имеют асимптотически нормальное распределение, что упрощает тестирование по сравнению с одномерными временными рядами. Оба теста также являются состоятельными против своей альтенативы, но также они являются состоятельными против другой альтернативы. Другими словами, тест против однородной альтернативы (тест по модели пула) будет также отвергать нулевую гипотезу, если какие-то временные ряды являются стационарными, а какие-то нестационарными. Кроме этого, тест, построенный против однородной альтернативы, даже если действительные данные являются неоднородными (в смысле инерционности), является более мощным (в смысле асимптотической локальной мощности), чем тесты, построенные против правильной неоднородной альтернативы. 

В Разделе 2.2 рассматривается проблема наличия детерминированной компоненты в данных. Самый простой вид детерминированной компоненты -- это фиксированные эффекты. Можно усложнить модель на случай индивидуальных трендов, структурных сдвигов и т.п. В отличие от анализа одномерных временных рядов, при тестировании наличия единичного корня в панелях простое детрендирование на первом шаге не будет очищать от влияния детерминированной компоненты. Тестовые статистики в этом случае не будут иметь асимптотически нормальное распределение, а будут асимптотически смещенными. Для решения этой проблемы предлагается несколько подходов. Первый заключается в корректировке тестовой статистики на асимптотическое смещение. Второй метод заключается не в обычном OLS-детрендировании, который нарушает мартингальное свойство ряда, а так называемым рекурсивном детрендировании, которое позволяет получать асимптотические выводы, аналогичные случаю отсутствия детерминированной компоненты. Еще один из подходов заключается в предварительном преобразовании ряда таким образом, чтобы не было смещения итоговой тестовой статистики. Все эти проблемы относятся только к тестам, построенным для тестирования против однородных альтернатив. При тестировании против неоднородных альтернатив каждую статистику для конкретного временного ряда нужно просто скорректировать на математическое ожидание и дисперсию, а затем сложить эти скорректированные статистики. Также можно отметить, что для предотвращения смещения можно использовать обобщенный метод моментов.

В Разделе 2.3 обсуждается асимптотическая локальная мощность различных тестов, а также приводятся оптимальные тесты. В данном разделе также акцентируется внимание на эффект от наличия индивидуальных трендов. В то время как при тестировании наличия единичного корня в одномерном временном ряде окрестности, в которых мощность была нетривиальной, имели один и тот же порядок вне зависимости от того, был ли тренд в данных или нет, в случае тестирования панельного единичного корня изменяется порядок окрестности, то есть при добавлении индивидуальных трендов окрестность единичного корня должна быть больше, чтобы мощность была нетривиальной, чем при учете только фиксированных эффектов.

В Разделе 2.4 обсуждается влияние слабой зависимости ошибок на тесты. В отличие от тестов на единичные корни во временных рядах, в панелях зависимость в ошибках не устраняется путем добавления запаздывающих разностей, и необходимо дополнительно корректировать статистику на смещение. Заметим, что это не требуется, если выполняется рекурсивное детрендирование. В Разделе 2.5 описываются другие тесты, по той или иной причине не вошедшие в предыдущие разделы. 

В Разделе 3 рассматривается обратная гипотезе о наличии панельного единичного корня, гипотеза о стационарности всех временных рядов в панели. Здесь, в отличие от классических тестов на стационарность типа KPSS, суммируются либо все KPSS-статистики для каждого временного ряда, либо только числители, и сумма в этом случае делится на долгосрочную дисперсию. Предельное распределение полученной статистики также, как и в случае тестирования наличия панельного единичного корня, является стандартным нормальным. Слабую зависимость в ошибках можно корректировать как параметрически, так и непараметрически.

В Разделе 4 обсуждается проблема кросс-секционной корреляции между ошибками в панели. Наличие такой корреляции может сильно исказить статистические выводы относительно наличия панельного единичного корня. Пространственная корреляция имеет место согласно некоторым макроэкономическим теориям, которые утверждают, что существуют некоторые общие факторы (например, технологические шоки), которые влияют не на одну, а на некоторое множество переменных. Наличие общих факторов, которые дополнительно могут быть нестационарными, предполагает наличие пространственной коинтеграции между субъектами, которая также часто может иметь место. Также источником кросс-секционной корреляции может быть пространственная (spatial) корреляция, которая основана на пространственных взаимосвязях и пространственной неоднородности. Здесь термин пространственный относится к географии субъектов, входящих в панель, к их местоположению, географическому расстоянию между ними, а также расстоянию в экономическом и социально-сетевом смысле.

Существует несколько способов борьбы с кросс-секционной коррелированностью (см. Разделы 4.1-4.6). Первый способ заключается в очистке от общего фактора при предположении, что данный фактор имеет место быть. Этот фактор может быть как стационарным, так и нестационарным, и последнее несколько затрудняет исследование. Очистка от стационарного фактора может производится как при помощи метода главных компонент, так и при помощи аппроксимации факторов средними значениями лагов. Если факторы может быть нестационарными, можно дополнительно тестировать их на наличие единичного корня и коинтеграции. Асимптотическая локальная мощность для первого случая не изменяется по сравнению с тестами при отсутствии пространственной корреляции. Кросс-секционную корреляцию можно также учесть путем модификации тестовых статистик, принимая тот факт, что они сами являются коррелированными. При предположении об отсутствии общих факторов можно использовать GLS-преобразование для тестирования гипотезы о панельном единичном корне. Еще один способ -- метод инструментальных переменных. Однако наиболее робастными методами являются методы ресемплинга и бутстрапа. Блочный бутстрап среди них оказывается наиболее робастным среди самого широкого класса предположений (в случае динамических нестационарных факторов), в то время как решетчатый бутстрап оказывается не всегда асимптотически обоснованным.

В Разделе 4.7 обсуждается проблема определения доли стационарных и нестационарных временных рядов в панели. Как уже было отмечено, отвержение нулевой гипотезы о наличии единичного корня во всех временных рядах в панели не является свидетельством того, что все временные ряды являются стационарными, а только свидетельством того, что существует статистически значимая пропорция стационарных временных рядов. То есть мы не можем сказать при отвержении нулевой гипотезы, являются ли все ряды стационарными, либо их какая-то часть. Тестировать наличие единичного корня в каждом конкретном временном ряде также несколько проблематично, поскольку при нулевой гипотезе мы будем отвергать ее с вероятностью, равной уровню значимости. То есть, даже если все ряды являются нестационарными, при тестировании на 5-процентном уровне значимости приблизительно для 5 процентов рядов гипотеза единичного корня будет отвергаться. Было предложено множество подходов как для оценвиания доли стационарных временных рядов, так и для разбиения этих временных рядов на классы. Часть методов основано на упорядочивании $p$-значений отдельных тестов и выборе некоторого уровнях значимости, основанного на этом упорядочивании. Другой способ основан на последовательном удалении наиболее ``стационарного'' временного ряда до тех пор, пока гипотеза о панельном единичном корне не будет не отвергаться. Похожая процедура использует методы бутстрапа. Можно поступить и наоборот, и оценить долю нестационарных временных рядов, анализируя поведение дисперсии. Можно показать, что кросс-секционная дисперсия будет расти с темпом, равном доле нестационарных временных рядов. Эту долю можно оценить обычными статистическими методами.

В Разделе 4.8 описываются остальные  методы тестирования гпиотезы о панельном единичном корне, робастные к кросс-секционной коррелированности. В Разделе 4.9 приводится сравнение различных методов.

В Разделе 5 описываются тесты на стационарность при кросс-секционной коррелированности. Существует как стандартная очистка от факторов панельной версии теста KPSS, так и другие подходы.

В Разделе 6 обуждаются основные подходы к тестированию наличия единичных корней в панельных данных при наличии нелинейности. В Разделе 6.1 описываются методы, связанные с тестированием наличия единичного корня в панелях при наличии сдвигов в детерминированной функции, в Разделе 6.2 обсуждается проблема, когда альтернатива является нелинейной: то есть тестируется гипотеза о наличии единичного корня против альтернативы, что некоторые из временных рядов в панели являюстя нелинейными, но стационарными, в Разделе 6.3 рассматривается проблема тестирования изменения инерционности временных рядов в панели (альтернативная гипотеза -- гипотеза о том, что некоторые временные ряды меняют порядок интегрированности в определенный момент времени).  В Разделе 6.4 описываются методы, связанные с тестированием стационарности панели. 

В Разделе 7 рассматривается важная проблема, связанная с тем, что тесты, изученные в работах, приведенных в предыдущих главах, предполагают, что	число наблюдений по времени является достаточно большим и намного большим, чем число объектов в панели. Соотвественно, асимптотическая теория, лежащая в основе рассмотренных тестов, основана либо на последовательной асимптотике, сначала по числу временных периодов, а затем по числу объектов, либо на совместной асимптотике, когда к бесконечности стремятся и число временных периодов, и число объектов. Однако во многих эмпирических приложениях нет достаточно большого временного периода при заведомо меньшем количестве объектов. Для решения этой проблемы существует определенный пласт литературы, в котором предполагается, что число временных периодов фиксировано, а асимптотика работает только по числу объектов -- это так называемые короткие панели. В Разделе 7.1 рассматриваются тесты на единичный корень в коротких панелях, в Разделе 7.2 -- их обобщения на случай структурных сдвигов, а в Разделе 7.3 рассматриваются тесты на стационарность в коротких панелях.

В последнем разделе приводятся ссылки на пакеты и программы, позволяющие реализовать часть методов, описанных в обзоре.

\section{Тесты на единичный корень для независимых панелей}

\subsection{LLC и IPS тесты}

Рассмотрим простейший случай, в котором временные ряды $\{y_{i0},\dots,y_{iT}\}$ для кросс секционных субъектов $i=1,2,\dots,N$ порождаются для каждого $i$ простой авторегрессией первого порядка
\begin{equation}\label{IndPan1}
y_{it}=\rho_iy_{i,t-1}+\varepsilon_{it},
\end{equation}
где начальное значение $y_{i0}$ является фиксированной константой\footnote{\citet{HHLS2010} и \citet{Westerlund2013} анализировали влияние начального значения на тесты, описываемые ниже, получая противополодные выводы относительно мощности по сравнению с временными рядами. \citet{Westerlund2013}, среди прочего, указывает на отсутствие необходимости разработки робастной тестовой стратегии для различных амплитуд начального значения, рекомендуя просто игнорировать его влияние. }, ошибки $\varepsilon_{it}$ являются независимыми и одинаково распределенными (i.i.d.) по всем $i$ и $t$ с $E(\varepsilon_{it})=0$, $E(\varepsilon^2_{it})=\sigma^2_i<\infty$ и $E(\varepsilon^4_{it})<\infty$. Эти процессы можно эквивалентно записать следующим образом, по аналогии с простой регрессией Дики-Фуллера:
\begin{equation}\label{IndPan2}
\Delta y_{it}=\phi_iy_{i,t-1}+\varepsilon_{it},
\end{equation}
где $\Delta y_{it}=y_{it}-y_{i,t-1}$, $\phi_i=\rho_i-1$. 

Нас интересует тестирование нулевой гипотезы
\begin{equation}\label{IndPan4}
H_0:\phi_1=\dots\phi_N=0,
\end{equation}  
то есть гипотезы о том, что все временные ряды имеют единичный корень (являются независимыми случайными блужданиями), против одной из следующих альтернатив, $H_{1a}$ и $H_{1b}$:
\begin{eqnarray}
H_{1a}&:&\phi_1=\dots=\phi_N\equiv\phi<0,\\
H_{1b}&:&\phi_1<0,\dots,\phi_{N_0}<0, \ N_0\leq N.
\end{eqnarray}
При первой альтернативе, $H_{1a}$, авторегрессионный параметр одинаковый для всех кросс-секционных субъектов. Эта альтернатива была рассмотрена в \citet{LLC2002} (далее LLC, см. также более раннюю версию работы,  \citet{LevinLin1993}) и была названа \textit{однородной альтернативой} (homogeneous alternative). При второй альтернативе предполагается, что все $N_0$ кросс-секционных объектов ($0<N_0\leq N$) являются стационарными с индивидуальными авторегрессионными коэффициентами. Эта альтернатива была рассмотрена в \citet{IPS2003} (далее IPS) и была названа \textit{неоднородной альтернативой} (heterogeneous alternative). В \citet{WesterlundLarsson2012} рассматривался третий вариант альтернативы, о наличии единичного корня в среднем, то есть при предположении, что пока некоторые субъекты могут быть нестационарными, вероятность, что такое случится, достаточно мала. Другими словами, предполагается, что авторегрессионный параметр является случайным с некоторым математическим ожиданием и дисперсией, а нулевая гипотеза заключается в том, что математическое ожидание равно единице, а дисперсия равна нулю. Авторы приводят преимущества такой спецификации, однако, неясно, соответствует ли такая постановка реальным данным.

\citet{ChangSong2009} рассматривают также гипотезу о том, что $\phi_i=0$ для некоторых $i$ против альтернативы о том, что $\phi_i<0$ для всех $i$. Для тестирования такой гипотезы \citet{ChangSong2009} предлагают брать максимум из всех индивидуальных левосторонних статистик (то есть ту статистику, которая показывает наименьшее отвержение нулевой гипотезы).

Для состоятельности тестовых статистик здесь предполагается, что доля стационарных временных рядов в панели сходится к фиксированной константе, то есть $N_0/N\rightarrow\kappa$ при $N\rightarrow\infty$. Отвержение нулевой гипотезы в пользу \textit{неоднородной} альтернативы не обязательно говорит о том, что наличие единичного корня отвергается для всех $i$, а только о том,  что гипотеза отвергается для доли $N_0<N$, и тест не дает каких-либо рекомендаций о величине $\kappa$ или о тех элементах панели, для которых гипотеза отвергается. С другой стороны, отвержение гипотезы единичного корня против \textit{однородной} альтернативы не обязательно означает, что все панельные субъекты являются стационарными, поскольку тест, построенный тким образом, чтобы иметь мощность против однородной альтернативы, также будет иметь мощность и при неоднородной альтернативе. 

\citet{KarlssonLothgren2000} проводят симуляции Монте-Карло в зависимости от доли стационарных временных рядов в панели. Авторы получили достаточно логичный результат, что мощность всех тестов увеличивается при увеличении пропорции стационарных временных рядов в панели. Также мощность увеличивается сильнее при росте $T$, чем при росте $N$. Следствием этого является то, что при больших $T$ и достаточно малых пропорциях стационарных временных рядов в панели мы будем часто отвергать нулевую гипотезу о наличии единичного корня, в то время как при малых $T$ даже при достаточно высокой пропорции стационарных временных рядов в панели из-за низкой мощности мы будем редко не отвергать нулевую гипотезу, после этого некорректно моделируя все временные ряды как нестационарные. Поэтому авторы рекомендуют не накладывать однородные ограничения на все временные ряды в панели, а анализировать эти временные ряды индивидуально.

Отметим, что существует также подход \citep{GOS2008}, в котором рассматривается трехмерная панель (например, время, страна и марка автомобиля или прогноз инфляции для различных индивидов с различными временными горизонтами; другими словами, вместе с числом временных периодов $T$ также имеются индексы $N$ и $M$, где $N$ может обозначать страны или отрасли, а $M$ может обозначать регионы или фирмы внутри страны или отрасли).

\subsubsection{LLC-тест}

Тест LLC основан на $t$-статистике для $\phi$ в регрессии пула
\[\Delta y_{it}=\phi y_{i,t-1}+\varepsilon_{it}\]
или, используя матричные обозначения,
\[\Delta \mathbf{y}_i=\phi \mathbf{y}_{i,-1}+\mathbf{\varepsilon}_{i},\]
где $\Delta \mathbf{y}_i=[\Delta y_{i1},\dots,\Delta y_{iT}]'$, $\mathbf{y}_{i,-1}=[y_{i0},y_{i1},\dots,y_{i,T-1}]'$ и $\mathbf{\varepsilon}_{i}=[\varepsilon_{i1},\dots,\varepsilon_{iT}]'$.

На первом шаге оценивается $\sigma_i^2$ для каждого панельного временного ряда:
\[\hat{\sigma}_i^2=\frac{\Delta \mathbf{y}'_i \mathbf{M}_i\Delta \mathbf{y}_i}{T-2},\]
где $\mathbf{M}_i=\mathbf{I}_T-\mathbf{X}_i(\mathbf{X}'_i\mathbf{X}_i)^{-1}\mathbf{X}'_i$ и $\mathbf{X}_i=(\mathbf{y}_{i,-1})$. Тогда $t$-статистика для проверки гипотезы \eqref{IndPan4} принимает вид:
\begin{equation}\label{IndPan5}
\tau_\phi=\frac{\sum_{i=1}^N{\Delta \mathbf{y}'_i\mathbf{y}_{i,-1}/\hat{\sigma}^2_i}}{\sqrt{\sum_{i=1}^N{\mathbf{y}'_{i,-1}\mathbf{y}_{i,-1}/\hat{\sigma}^2_i}}}.
\end{equation}  
Как отмечает \citet{Breitung2000}, в LLC предлагается дополнительно делить эту статистику на $\hat{\sigma}_{NT}$, общее (по всем $N$ и $T$) стандартное отклонение остатков, но эти остатки уже скорректированы на свои стандартные отклонения, поэтому это стандартное отклонение можно опустить.

\subsubsection{IPS-тест}

В отличие от LLC, рассматривая неоднородную альтернативу $H_{1b}$, состоящую из множества неравенств, IPS предлагают тест, основанный на среднем индивидуальных $t$-статистик,
\begin{equation}\label{IndPan6}
\bar{\tau}=\frac{1}{N}\sum_{i=1}^N{\tau_i},
\end{equation}  
где
\begin{equation}\label{IndPan7}
\tau_i=\frac{\Delta \mathbf{y}'_i\mathbf{y}_{i,-1}}{\hat{\sigma}_i\sqrt{\mathbf{y}'_{i,-1}\mathbf{y}_{i,-1}}}
\end{equation}
является $t$-статистикой Дики-Фуллера для $i$-ого кросс-секционного субъекта.

\citet{ChangSong2009} предлагают вместо усреднения тестовых статистик брать минимум из всех статистик, что позволяет получить более высокую мощность при очень малом количестве стационарных временных рядов в панели.

\subsubsection{LM-версии LLC и IPS}

Можно построить LM версии тестов LLC и IPS. Отличия этих тестов от стандартных LLC и IPS заключается только в способе оценивания дисперсии $\sigma_i^2$: в данном случае она оценивается при нулевой гипотезе. Однако тесты остаются асимптотически эквивалентными вне зависимости от способа оценивания дисперсии, поскольку обе оценки дисперсии являются состоятельными.

\subsubsection{Асимптотика тестов на единичный корень в панельных данных}

Рассмотрим получение предельного распределения для статистики LLC. Используя FCLT и CMT, можно показать, что при $T\rightarrow\infty$
\begin{equation}\label{IndPan8}
\tau_\phi=\frac{\sum_{i=1}^N{\Delta \mathbf{y}'_i\mathbf{y}_{i,-1}/\hat{\sigma}^2_i}}{\sqrt{\sum_{i=1}^N{\mathbf{y}'_{i,-1}\mathbf{y}_{i,-1}/\hat{\sigma}^2_i}}}\Rightarrow_{T\rightarrow\infty}\frac{\sum_{i=1}^N{\int_0^1{W_i(r)dW_i(r)}}}{\sqrt{\sum_{i=1}^N{\int_0^1{W_i(r)^2dr}}}}\equiv\zeta,
\end{equation}
где $W_i(r)$, $i=1,\dots,N$ -- независимые Винеровские процессы, а $\Rightarrow$ обозначает слабую сходимость. Применяя закон больших чисел, числитель \eqref{IndPan8} (деленный на $N$) сходится по вероятности к нулю, а знаменатель(деленный на $N$) сходится к $\frac{1}{2}$. Поэтому $t$-статистика на основе регрессии пула (по центральной предельной теореме) имеет стандартное нормальное распределение.

Рассмотрим получение предельного распределения для статистики IPS. Понятно, что при $T\rightarrow\infty$ каждая статистика $\tau_i$ в \eqref{IndPan7} сходится к обычному распределению Дики-Фуллера, 
\[\tau_i\Rightarrow\eta_i\equiv\frac{\int_0^1{W_i(r)dW_i(r)}}{\int_0^1{W^\mu_i(r)^2dr}}.\]
Поскольку статистика 
\[\bar{\tau}\Rightarrow\frac{1}{N}\sum_{i=1}^N{\eta_i}\]
при фиксированном $N$ и $T\rightarrow\infty$, то при $N\rightarrow\infty$ статистика $\sqrt{N}\bar{\tau}$  имеет стандартное нормальное предельное распределение, аналогично LLC. 

Результаты выше были основаны на том, что сначала применяется асимптотика при $T\rightarrow\infty$, а затем при $N\rightarrow\infty$. Можно рассмотреть альтернативный подход, когда одновременно $T,N\rightarrow\infty$ (IPS доказывают, что асимптотические результаты остаются неизменными по сравнению с последовательной асимптотикой) или когда $N$ является функцией от $T$, следуя \citep{PhillipsMoon1999}.\footnote{ Последний подход может принимать вид, например, $T(N)=cN$ для $c\neq0$. Этот подход в \citep{PhillipsMoon1999} называется асимптотикой диагональной траектории (diagonal path). Недостатком этого подхода является то, что он сильно специфичен и не дает соответствующей аппроксимации для заданных $T$ и $N$, а также зависит от конкретной функциональной формы зависимости $T=T(N)$. } \citet{PhillipsMoon1999} вводят понятия последовательной и совместной сходимости по вероятности и слабой сходимости и устанавливают соотношения между этими типами сходимости. Авторы заключают, что последовательная асимптотика не предполагает совместную, так что в некоторых ситуациях совместная асимптотика может не выполняться. Хотя  подходы, связанные с совместной асимптотикой, являются интересными с теоретической точки зрения и иногда дают полезные следствия, предельные результаты по существу такие же, как и в последовательной асимптотике, но требуют более сильных условий. С практической точки зрения последовательная асимптотика является достаточной для большинства случаев.


\subsection{Наличие детерминированной компоненты}

Более подробно остановимся на вопросе наличия детерминированных компонент в данных, а именно на спецификации детерминированной компоненты. Обычно нас интересуют два наиболее распространенных случая, 
\begin{eqnarray}
\Delta y_{it}&=&\mu_i+\phi_iy_{i,t-1}+\varepsilon_{it},\ \text{Случай 1}\label{IndPan12}\\
\Delta y_{it}&=&\mu_i+\beta_it+\phi_iy_{i,t-1}+\varepsilon_{it},\ \text{Случай 2}\label{IndPan13}
\end{eqnarray}
где уравнение \eqref{IndPan12} соответствует случаю индивидуально-специфических констант (фиксированных эффектов), а \eqref{IndPan13} соответствует случаю индивидуально-специфических трендов (``случайных трендов'', incidental trends в терминологии \citet{MoonPhillips1999b}). 

Отметим, что многие тесты требуют условия, что $N/T\rightarrow\infty$ для контроля размера, так как в противном случае ошибка, вызванная детрендированием, будет возрастать с ростом $N$; см. \citep{WesterlundBreitung2013}, Fact 3.

\subsubsection{Тест LLC}

Для простоты рассмотрим Случай 1 с фиксированными эффектами. Тест LLC основан на $t$-статистике для $\phi$ в регрессии с фиксированными эффектами
\[\Delta y_{it}=\alpha_i+\phi y_{i,t-1}+\varepsilon_{it}\]
или, используя матричные обозначения,
\[\Delta \mathbf{y}_i=\mathbf{1}\alpha_i+\phi \mathbf{y}_{i,-1}+\mathbf{\varepsilon}_{i},\]
где $\Delta \mathbf{y}_i=[\Delta y_{i1},\dots,\Delta y_{iT}]'$, $\mathbf{y}_{i,-1}=[y_{i0},y_{i1},\dots,y_{i,T-1}]'$, $\mathbf{\varepsilon}_{i}=[\varepsilon_{i1},\dots,\varepsilon_{iT}]'$ и $\mathbf{1}=[1,\dots,1]'$.

На первом шаге оценивается $\sigma_i^2$ для каждого временного ряда:
\[\hat{\sigma}_i^2=\frac{\Delta \mathbf{y}'_i \mathbf{M}_i\Delta \mathbf{y}_i}{T-2},\]
где $\mathbf{M}_i=\mathbf{I}_T-\mathbf{X}_i(\mathbf{X}'_i\mathbf{X}_i)^{-1}\mathbf{X}'_i$ и $\mathbf{X}_i=(\mathbf{1},\mathbf{y}_{i,-1})$. Тогда $t$-статистика для проверки гипотезы \eqref{IndPan4} принимает вид:
\begin{equation}\label{IndPan5}
\tau_\phi=\frac{\sum_{i=1}^N{\Delta \mathbf{y}'_i\mathbf{M}_1\mathbf{y}_{i,-1}/\hat{\sigma}^2_i}}{\sqrt{\sum_{i=1}^N{\mathbf{y}'_{i,-1}\mathbf{M}_1\mathbf{y}_{i,-1}/\hat{\sigma}^2_i}}},
\end{equation}  
где $\mathbf{M}_1=\mathbf{I}_T-\mathbf{1}(\mathbf{1}'\mathbf{1})^{-1}\mathbf{1}'$. 

Оценивание коэффициента $\phi$ эквивалентно оцениванию коэффициента $\phi$ в центрированной регрессии
\[\Delta\tilde{y}_{it}=\phi\tilde{y}_{i,t-1}+e_{it},\]
где $\tilde{y}_{it}=y_{it}-T^{-1}\sum_{j=0}^Ty_{i,j}$. При нулевой гипотезе мы получим
\[\lim_{T\rightarrow\infty}\frac{1}{T}\sum_{t=1}^Te_{it}\tilde{y}_{i,t-1}=-\sigma^2_i/2,\]
и
\[\lim_{T\rightarrow\infty}\frac{1}{T}\sum_{t=1}^T\tilde{y}_{i,t-1}\tilde{y}'_{i,t-1}=\sigma^2_i/6,\]
так что $\sqrt{N}T(\hat{\phi}-1)+3\sqrt{N}\Rightarrow N(0,51/5)$, то есть $T(\hat{\phi}-1)\stackrel{p}{\longrightarrow}-3$. Следовательно, оценка $\hat{\phi}$ является асимптотически смещенной, и $t$-статистика для проверки $\phi=0$ расходится к $-\infty$ при росте $T$ и $N$ (из-за коррелированности регрессора и ошибки). Это смещение называется смещением Никелла (Nickell bias), см. \citet{Nickell1981}.

Этот результат можно получить иначе следующим образом. Заметим, что при $T\rightarrow\infty$
\begin{equation}\label{IndPan8_det}
\tau_\phi=\frac{\sum_{i=1}^N{\Delta \mathbf{y}'_i\mathbf{M}_1\mathbf{y}_{i,-1}/\hat{\sigma}^2_i}}{\sqrt{\sum_{i=1}^N{\mathbf{y}'_{i,-1}\mathbf{M}_1\mathbf{y}_{i,-1}/\hat{\sigma}^2_i}}}\Rightarrow_{T\rightarrow\infty}\frac{\sum_{i=1}^N{\int_0^1{W^\mu_i(r)dW_i(r)}}}{\sqrt{\sum_{i=1}^N{\int_0^1{W^\mu_i(r)^2dr}}}}\equiv\zeta,
\end{equation}
где $W^\mu_i(r)=W_i(r)-\int_0^1{W_i(s)ds}$ -- центрированный Винеровский процесс, $W_i(r)$, $i=1,\dots,N$ -- независимые Винеровские процессы. Применяя закон больших чисел, числитель \eqref{IndPan8_det} (деленный на $N$) сходится по вероятности к $-\frac{1}{2}$, а знаменатель (деленный на $\sqrt{N}$) сходится к $\frac{1}{\sqrt{6}}$. Поэтому $t$-статистика на основе регрессии пула расходится.

Имея значения для математических ожиданий и дисперсий функционалов от Винеровских процессов, представленные в Таблице \eqref{tab:1} (см. \citet{LevinLin1993}), можно получить, что в Случае 1 
\[\tau_\phi\Rightarrow N(-\sqrt{1.875N},1.25),\]
а в Случае 2 
\[\sqrt{\frac{448}{277}}\tau_\phi\Rightarrow N(-\sqrt{3.75N},1).\]

 \begin{table}[h!]
  \caption{Значения для математических ожиданий и дисперсий функционалов от Винеровских процессов}%
  \label{tab:1}%
  \begin{center}%
 \small
\begin{tabularx}{0.8\textwidth}{ccccc} \toprule
& $E(\int{W^j_idW_i})$ & $Var(\int{W^j_idW_i})$ &  $E(\int{(W^j_i)^2})$ & $Var(\int{(W^j_i)^2})$\\
$\mu_i$ ($j=\mu$) & $-\cfrac{1}{2}$ & $\cfrac{1}{12}$ & $\cfrac{1}{6}$ & $\cfrac{1}{45}$\\
$(\mu_i,t)$ ($j=\tau$) & $-\cfrac{1}{2}$ & $\cfrac{1}{60}$ & $\cfrac{1}{15}$ & $\cfrac{11}{6300}$\\
 \bottomrule
\end{tabularx}
\end{center}
\end{table}

Данные выводы можно получить, применяя CLT и LLN согласно
\[\frac{1}{\sqrt{N}}\sum_{i=1}^N{\left(\int_0^1{W^j_i(r)dW_i(r)}-E\left(\int_0^1{W^j_i(r)dW_i(r)}\right)\right)}\Rightarrow_{N\rightarrow\infty}N\left(0,Var\left(\int_0^1{W^j_i(r)dW_i(r)}\right)\right)\]
и
\[\frac{1}{N}
\sum_{i=1}^N{\int_0^1{W^j_i(r)^2dr}}\stackrel{p}{\longrightarrow}_{N\rightarrow\infty}E\left(\int_0^1{W^j_i(r)^2dr}\right)
\]
и используя Таблицу \eqref{tab:1}. В общем случае тестовой статистикой для проверки гипотезы $H_0$ в \eqref{IndPan4}, которая имеет стандартное нормальное распределение, является скорректированная на смещение статистика
\begin{equation}\label{IndPan14}
Z_{LLC}=\frac{\sum_{i=1}^N{(\Delta \mathbf{y}'_i\mathbf{M}_1\mathbf{y}_{i,-1}/\hat{\sigma}^2_i-\mu_T^* T)}}{\sigma_T^*\sqrt{\sum_{i=1}^N{\mathbf{y}'_{i,-1}\mathbf{M}_1\mathbf{y}_{i,-1}/\hat{\sigma}^2_i}}}=\frac{\tau_\phi}{\sigma_T^*}-\frac{\mu_T^*TN}{\sigma_T^*\sqrt{\sum_{i=1}^N{\mathbf{y}'_{i,-1}\mathbf{M}_1\mathbf{y}_{i,-1}/\hat{\sigma}^2_i}}},
\end{equation}
где $\mu_T^*$ является выборочным аналогом $\mu_\infty^*=E(\int{W^j_idW_i^j})$, а $\sigma_T^{2*}$ является выборочным аналогом $\sigma_\infty^{2*}=Var(\int{W^j_idW_i^j})/E(\int{(W^j_i)^2})$. Значения $\mu_T^*$ и $\sigma_T^{2*}$ приведены в LLC, Table 2, для различных $T$. 

Для конкретного случая с фиксированными эффектами, применяя центральную предельную теорему к центрированному числителю в \eqref{IndPan8_det} при $N\rightarrow\infty$:
\[\frac{1}{\sqrt{N}}\sum_{i=1}^N{\left(\int_0^1{W^\mu_i(r)dW_i(r)}-E\left(\int_0^1{W^\mu_i(r)dW_i(r)}\right)\right)}\Rightarrow_{N\rightarrow\infty}N(0,\frac{1}{12}).\]
Тогда
\[\frac{\frac{1}{\sqrt{N}}\sum_{i=1}^N{\left(\int_0^1{W^\mu_i(r)dW_i(r)}-E\left(\int_0^1{W^\mu_i(r)dW_i(r)}\right)\right)}}{\sqrt{\frac{1}{N}\sum_{i=1}^N{\int_0^1{W^\mu_i(r)^2dr}}}}\Rightarrow_{N\rightarrow\infty}N(0,\frac{1}{2}).\]
Таким образом, статистика
\begin{equation}\label{IndPan10}
Z_{LLC}=\frac{\sum_{i=1}^N{(\Delta \mathbf{y}'_i\mathbf{M}_1\mathbf{y}_{i,-1}/\hat{\sigma}^2_i}+\frac{1}{2}T)}{\sqrt{\frac{1}{2}}\sqrt{\sum_{i=1}^N{\mathbf{y}'_{i,-1}\mathbf{M}_1\mathbf{y}_{i,-1}/\hat{\sigma}^2_i}}}=\frac{\tau_\phi}{\sqrt{\frac{1}{2}}}-\frac{-\frac{1}{2}TN}{\sqrt{\frac{1}{2}}\sqrt{\sum_{i=1}^N{\mathbf{y}'_{i,-1}\mathbf{M}_1\mathbf{y}_{i,-1}/\hat{\sigma}^2_i}}}
\end{equation}
будет иметь стандартное нормальное предельное распределение. Отметим, что при $\mu_i=0$ статистика $Z_{LLC}$ совпадает с $\tau_\phi$. Также следует заметить, что скорость сходимости для оценки $\hat{\phi}$ будет равна $T\sqrt{N}$ (то есть $T\sqrt{N}(\hat{\phi}-1)=O_p(1)$), то есть сходимость происходит быстрее при $T\rightarrow\infty$ (суперсостоятельность), чем при $N\rightarrow\infty$. Также из скорости сходимости следует более высокая мощность панельных тестов, поскольку она увеличивается не только с ростом $T$, но и с ростом $N$.

\subsubsection{Коррекция числителя оценки коэффициента}

Вспомним, что $T(\hat{\phi}-1)\stackrel{p}{\longrightarrow}-3$, и статистика \eqref{IndPan10} основана на коррекции оценки $\hat{\phi}$, $\hat{\phi}^{+}\equiv\hat{\phi}+3/T$. Можно, однако, скорректировать не саму оценку коэффициента $\hat{\phi}$, а только ее числитель:
\begin{equation}\label{phihat_cor}
\hat{\phi}^{\#}=\frac{\sum_{i=1}^N{\Delta \mathbf{y}'_i\mathbf{M}_1\mathbf{y}_{i,-1}/\hat{\sigma}^2_i}+\frac{NT}{2}\hat{\sigma}_i^2}{\sum_{i=1}^N{\mathbf{y}'_{i,-1}\mathbf{M}_1\mathbf{y}_{i,-1}/\hat{\sigma}^2_i}}
\end{equation}
(см. \citet{HahnKuersteiner2002} и \citet{MoonPerron2004}). Тогда (в случае фиксированных эффектов) $\sqrt{N}T(\hat{\phi}^{\#}-1)\Rightarrow N(0,3)$, и соответствующая $t$-статистика будет иметь стандартное нормальное распределение. Но в случае коррекции оценки коэффициента, как в LLC, $\sqrt{N}T(\hat{\phi}^{+}-1)\Rightarrow N(0,51/5)$, так что последняя оценка для $\phi$ будет менее эффективна в смысле дисперсии, но обе $t$-статистики будут асимптотически эквивалентны.

\subsubsection{Детрендирование, устраняющее смещение}

Альтернативный способ исключить смещение при оценивании был рассмотрен в \citep{BreitungMeyer1994}. Авторы в качестве оценки константы использовали начальное значение $y_{i0}$, так что регрессионная модель для Случая 1 (против однородной альтернативы $H_{1a}$) принимает вид
\begin{equation}\label{IndPan14_1}
\Delta y_{it}=\phi(y_{i,t-1}-y_{i0})+\varepsilon_{it}.
\end{equation}
При нулевой гипотезе величина $E[(y_{i,t-1}-y_{i0})\varepsilon_{it}]=0$, так что $t$-статистика для $\phi$ в регрессии \eqref{IndPan14_1} не будет смещенной и будет иметь асимптотическое стандартное нормальное распределение. Полученный тест, являющийся $t$-статистикой в регрессии для преобразованных рядов, называем $UB_{NT}$. 

В случае наличия линейного тренда, чтобы получить несмещенную тестовую статистику, имеющую стандартное нормальное распределение, \citep{Breitung2000} предложил рассмотреть следующую регрессию:
\begin{equation}\label{IndPan14_2}
\Delta y_{it}^*=\phi y_{i,t-1}^*+\varepsilon_{it}^*,
\end{equation}
где 
\begin{eqnarray*}
\Delta y_{it}^*&=&s_t\left[\Delta y_{it}-\frac{1}{T-t}(\Delta y_{it}+\dots+\Delta y_{iT})\right],\\
s_t^2 &=& (T-t)/(T-t+1),\\
y_{i,t-1}^*&=& y_{i,t-1}-y_{i0}-\frac{t-1}{T}(y_{iT}-y_{i0}).
\end{eqnarray*}
Это преобразование называется преобразованием Гельмерта (Helmert transformation). В данном преобразовании вычитание $y_{i0}$ удаляет константу, а $(y_{iT}-y_{i0})/T=(\Delta y_{i0}+\dots+\Delta y_{iT})/T$ является оценкой коэффициента при тренде. Смещения не возникает из-за ортогональности преобразованных регрессора и ошибки. 

\subsubsection{Тест IPS}

Для тестов IPS коррекция связана только с конкретными детерминированными компонентами для каждого $i$, то есть каждая статистика $\tau_i$ в \eqref{IndPan7} корректируется на некоторые средние и дисперсию, заданные в IPS, Table 3 (средние и дисперсии распределений Дики-Фуллера).

\subsubsection{Коррекция на основе GMM}

Другая возможность устранения смещения из-за детрендирования - использование альтернативных методов оценивания, таких как обобщенный метод моментов (Generalized Methods of Moments, GMM). \citet{Breitung1997} применяет вторые разности для получения теста на единичный корень без корректировки смещения, используя соответствующую GMM-оценку.

\subsection{Асимптотическая локальная мощность}

\citep{Breitung2000} рассмотрел локальные (однородные) альтернативы, если процесс $y_{it}$ порождается как
\begin{equation}\label{LocalAlt1}
y_{it}=\left(1-\frac{c}{T\sqrt{N}}\right)y_{i,t-1}+\varepsilon_{it},
\end{equation}
где $c>0$. Используя последовательную предельную теорию, автор показал, что при локальной альтернативе данного вида процесс $y_{it}$ ведет себя также, как и при нулевой гипотезе, то есть $T^{-1/2}y_{i,\lfloor rT\rfloor}\Rightarrow\omega_iW_i(r)$ при $0\leq c<\infty$, где $\omega_i^2$ -- долгосрочная дисперсия $\varepsilon_{it}$. Этот результат отличается от случая с обычным временным рядом, когда предельный процесс является процессом Орнштейна-Уленбека. Также при локальной альтернативе (и при отсутствии детерминированной компоненты) $t$-статистика для проверки $\phi=0$ имеет асимптотическое распределение $N(-c/\sqrt{2},1)$.

\citep{Breitung2000} отмечает, что если рассматривать обычную скорректированную статистику LLC, то локальная мощность будет состоять из двух различных компонент. Первая представляет асимптотический эффект смещения вследствие метода детрендирования, а вторая -- обычный параметр масштаба предельного распределения при последовательности локальных альтернатив. Если долгосрочная дисперсия оценивается состоятельно, обе компоненты взаимоуничтожаются, и тестовая статистика будет центрирована около нуля, так что она не будет иметь мощности. При фиксированной стационарной альтернативе оценка долгосрочной дисперсии, вычисленная на основе первых разностей, будет сходиться к нулю, так что тест будет состоятельным (что не так в одномерном случае, где долгосрочная дисперсия должна оцениваться для ряда в уровнях).

\citep{MPP2007} предложили более общий подход, в котором
\begin{equation}\label{LocalAlt2}
y_{it}=\left(1-\frac{c_i}{N^\eta T}\right)y_{i,t-1}+\varepsilon_{it},
\end{equation}
где $c_i$ являются $i.i.d.$ случайными величинами  с математическим ожиданием $\mu_c$ и дисперсией $\sigma_c^2$ на неотрицательном ограниченном интервале $[0,M_c]$. 

Сначала рассмотрим Случай 1 (фиксированные эффекты, отсутствие тренда). В этом случае
\begin{equation}\label{LocalAlt2.1}
Z_{LLC}\Rightarrow N\left(-\frac{3}{2}\sqrt{\frac{5}{51}}\mu_c,1\right),
\end{equation}
так что функция асимптотической локальной мощности будет $\Phi\left(\frac{3}{2}\sqrt{\frac{5}{51}}\mu_{c,1}-z_\xi\right)$, где $\Phi(\cdot)$ -- функция стандартного нормального распределения, $z_\xi$ -- критическое значение на уровне значимости $\xi$. Поэтому функция мощности зависит только от среднего $c_i$, но не от неоднородности локальных альтернатив около своего среднего.

\citep{MoonPerron2008} показали, что $t$-статистика, основанная на оценке $\hat{\phi}^{\#}$ в \eqref{phihat_cor}, в этом случае будет сходиться к нормальному распределению с математическим ожиданием $\mu_{c,2}/4\sqrt{3}$ и единичной дисперсией. Однако такая статистика будет иметь нетривиальную мощность только в окрестности $1/N^{1/4}T$ единицы, в отличие от статистики LLC, хотя оба коэффициента ($\hat{\phi}^{+}$ и $\hat{\phi}^{\#}$) будут асимптотически эквивалентными, а соответствующие $t$-статистики будут иметь стандартное нормальное распределение. Статистика, основанная на $\hat{\phi}^{\#}$, будет более эффективной в том смысле, что оценка коэффициента $\phi$ будет иметь меньшую дисперсию. Но статистика $Z_{LLC}$ будет иметь большую локальную мощность. Это происходит из-за того, что (скорректированная) оценка для $\phi$ в LLC, $\hat{\phi}^{+}$, является суммой (скорректированной) оценки $\hat{\phi}^{\#}$ и дополнительной случайной компоненты. Это приводит к эффективности при нулевой гипотезе оценки $\hat{\phi}^{\#}$. Но в окрестности $1/N^{1/2}T$ $t$-статистика, основанная на $\hat{\phi}^{\#}$, имеет то же самое распределение, что и при нулевой гипотезе (так что тест будет иметь тривиальную мощность), а дополнительная компонента сдвигается в среднем от этого распределения, так что мощность статистики LLC зависит от этой дополнительной компоненты, и $\hat{\phi}^{\#}$ не оказывает влияния асимптотически. Появляющаяся дополнительная компонента является панельной версией теста Саргана-Бхаргавы \citep{SarganBhargava1983}:
\begin{equation}\label{LocalAlt2.2}
V_0=\frac{\sqrt{N}}{\hat{\sigma}}\left(\frac{1}{NT^2}\mathbf{y}'_{i,-1}\mathbf{M}_1\mathbf{y}_{i,-1}-\frac{1}{6}\hat{\sigma}^2\right).
\end{equation}
Тест, отвергающий нулевую гипотезу при малых значениях  статистики $V_0$, имеет ту же самую асимптотическую локальную мощность, что и тест LLC.

\subsubsection{Наличие индивидуальных трендов}

В случае индивидуально-специфических трендов скорость сходимости локальной окрестности единицы должна быть ниже, чем $1/\sqrt{N}T$. Для получения нетривиальной локальной мощности она должна быть $1/N^{1/4}T$. Отличие связано с тем, что сложно обнаружить наличие единичных корней в панелях при наличии неоднородных трендов, называемых ``случайными трендами'' в терминологии \citep{MoonPhillips1999b}). Также в \citep{MPP2007} допускаются двухсторонние альтернативы, предполагая, что $c_i$ имеют ненулевые среднее и дисперсию. Функция локальной мощности точечно-оптимального теста зависит от четвертого момента параметра $c_i$, так что в панелях с более рассеянными авторегрессионными коэффициентами, как правило, нулевая гипотеза будет легче отвергаться.

\citet{MPP2007} сравнивают ряд тестов на основе их локальной мощности. Тест \citet{Breitung2000}, основанный на коррекции временных рядов, $UB_{NT}\Rightarrow N(\mu_{c,2}/6\sqrt{6},1)$ ($\mu_{c,i}$ обозначает $i$-ый момент случайной величины $c$), так что его асимптотическая мощность равна $\Phi(\mu_{c,2}/6\sqrt{6}-z_\xi)$. Для теста LLC, $Z_{LLC}\Rightarrow N(-\frac{15\sqrt{15}}{2}\frac{112}{193}\mu_{c,2}/420,1)$, то есть тесты LLC и \citet{MoonPhillips2004} имеет более низкую асимптотическую локальную мощность, чем тест Брайтунга, что также подтверждается симуляциями в \citet{Breitung2000}. Еще данные результаты говорят о том, что мощность будет зависеть от второго момента локализующего параметра, то есть при более неоднородной альтернативе мощность будет выше. Отметим, что в \citet{Breitung2000} ошибочно утверждаестя неправильная окрестность единицы, $1/\sqrt{N}T$, в которой, как показано в \citet{MPP2006}, тест Брайтунга имеет тривиальную мощность.

На малых выборках, однако, асимптотическая локальная мощность  все еще может плохо предсказывать действительную мощность. В \citet{WesterlundLarsson2015d} предлагают подход, который дает более аккуратную аппроксимацию фактической мощности на конечных выборках (при малых $N$), если авторегрессионный параметр задавать как
\[\rho_i=1+\frac{c_i}{N^{\kappa}T^\tau}\]
Авторы показывают, как моменты некоторых общеизвестных выборочных элементов, лежащих в основе тестовых статистик, записываются в виде разложения бесконечного порядка (до порядка $T^{-1}$) в моментах $c_i$ с коэффициентами, зависящими от скорости сокращения локальной альтернативы. Кроме этого, авторы задают относительную скорость роста $N$ и $T$, $\theta=\ln(T)/\ln(N)>0$, так что $T=N^\theta$. Другими словами, в отличие от предыдущих подходов, где задаются фиксированные значения $\tau=1$ и $\kappa=1/2$ или $\kappa=1/4$, авторы получают, что любая тройка $(\theta,\tau,\kappa)$, удовлетворяющая условию $\theta(1-\tau)+1/2-\kappa=0$ (или $\theta(1-\tau)+1/4-\kappa=0$ для случая с трендами), будет давать ту же самую асимптотическую локальную мощность.

Важно отметить, что мощность не будет теперь зависеть только от первого момента $c_i$, но будет зависеть от моментов более высокого порядка ($\mu_c^j=E(c_i^j)$). Например, если использовать левосторонний тест, и $\mu_c^2>0$, но $\mu_c^1>0$ или $\mu_c^1=0$, то можно некорректно заключить, что панель имеет единичный корень. То есть тест не может отличить с одной стороны наличие единичного корня и с другой стороны взрывного поведения ($\mu_c^1>0$) и/или случая, когда $c_i$ в среднем равны нулю ($\mu_c^1=0$), но имеют ненулевую дисперсию ($\mu_c^2>0$). Или, например, может возникнуть проблема, когда $\mu_c^1<0$, но более высокие моменты достаточно велики, чтобы устранить эффект отрицательного среднего, приводя к положительному сносу. Таким образом, в отличие от существующей практики использовать левосторонние тесты, авторы рекомендуют использовать двухсторонние.

\subsubsection{Оптимальные тесты}

Можно построить асимптотическую огибающую мощности, используя лемму Неймана-Пирсона, аналогично \citep{ERS1996}. (Почти) оптимальный тест (точечно-оптимальный тест), предложенный в \citet{MPP2007}, в случае отсутствия детерминированной компоненты будет иметь вид
\begin{equation}\label{LocalAlt3}
V_{NT}=\frac{1}{\hat{\sigma}^2}\left(\sum_{i=1}^N\sum_{t=1}^T(\Delta_{\bar{c}_i}y_{it})^2-(\Delta y_{it})^2\right)-\frac{1}{2}\mu_{c,2},
\end{equation}
где $\Delta_{\bar{c}_i}y_{it}=y_{it}-(1-\bar{c}_i/T\sqrt{N})y_{i,t-1}$ для $t=1,\dots,T$, $\bar{c}_i$, $i=1,\dots,N$ выбраны случайным образом вместо неизвестных значений $c_i$, $\mu_{c,k}=E(\bar{c}_i^k)$, а дисперсия $\sigma^2$ оценивается при нулевой гипотезе. При локальной альтернативе статистика $V_{NT}$ сходится к $N(-E(c_i\bar{c}_i),2\mu_{c,2})$. Огибающую мощности можно получить, полагая $\bar{c}_i=c_i$.

Сравнивая мощность оптимального теста с тестом LLC, авторы заключают, что LLC является оптимальным только в случае однородной альтернативы, $c_i=c$, так что $E(c_i)=\sqrt{E(c_i^2)}$. Если же случайным образом генерировать значения $\bar{c}_i$ для построения тестовой статистики, ее мощность будет ниже, чем если взять одинаковые значения $\bar{c}_i=\bar{c}$, но тогда тест будет идентичным тесту LLC. Кроме того, тест будет зависеть от значения $\bar{c}$, в отличие от теста \citep{ERS1996} для одного временного ряда из-за того, что в случае панели локальная альтернатива ближе к нулевой гипотезе (из-за деления на $N^\eta$), чем в случае единственного временного ряда.

\citet{MPP2007} также разработали оптимальные тесты для случая наличия фиксированных эффектов и индивидуально-специфических трендов.




\citet{MPP2014} предложили обобщение теста \citet{MPP2007} на случай серийно коррелированных ошибок. \citet{BDA2015a} на основе теории предельных экспериментов получают огибающую асимптотической локальной мощности для неоднородной панели с фиксированными эффектами. Авторы предлагают асимптотически оптимальный тест, являющийся ничем иным, как масштабированным числителем статистики $P_b$, предложенной в \citep{BaiNg2010}:
\[\hat{\Delta}=\frac{\sqrt{2}}{\sqrt{N}T\hat{\sigma}^2}\sum_{i=1}^N\sum_{t=3}^T{(y_{i,t-1}-y_{i1})\Delta y_{it}}.\]
Этот тест является равномерно наиболее мощным (UMP) тестом, как и тест отношения правдоподобий \citet{MPP2007} и тест \citet{BreitungMeyer1994}. Как отмечено в \citet{Westerlund2014}, $\hat{\Delta}=P_b+o_p(1)$. Отметим, что \citet{MPP2007} и \citet{BDA2015a} рассматривали локальную альтернативу вида
\[\rho_i=1+\frac{h}{\sqrt{N}T}H_i,\]
где $H_i$ является случайной со средним 1, а детерминированная величина $h$ описывает отклонение от единичного корня. \citet{BDAW2016} рассматривают другую постановку, в которой $H_i$ является случайной со средним 0 и дисперсией 1, однако, теперь локальная окрестность должна быть порядка $1/N^{1/4}T$, а не $1/N^{1/2}T$. Поскольку математическое ожидание $H_i$ неизвестно на практике, авторы рекомендуют использовать отвержение каждого из двух предложенных оптимальных тестов на основе двухсторонней гипотезы.

\subsubsection{Детрендирование}

\citet{Westerlund2015c} вместо обычного детрендирование рассматривал свойства рекурсивного детрендирования, допуская возможно нелинейную трендовую функцию (например, полиномиальный тренд). Кроме полиномиального тренда допускается также более сложная структура, такая как, например, гладкий сдвиг в уровнях на основе логистической функции или множественные сдвиги в трендах (см. \citet{Westerlund2014b}). Более ранние работы, такие как \citet{SKO2004} и \citet{Sul2009}, а также многие другие, акцентирующие внимание на различных подходах, изучали эффект рекурсивного детрендирования только на основе симуляций, в то время как в \citet{Westerlund2015c} более аккуратно анализируются асимптотические свойства рекурсивного детрендирования. Причина преимущества рекурсивного детрендирования, например, для обычных временных рядов, заключается в том, что обычное (по всей выборке) детрендирование нарушает мартингальное свойство данных, а рекурсивное его сохраняет, что приводит к менее смещенной оценке наибольшего авторегрессионного корня. Рекурсивное детрендирование можно проводить как для рядов в уровнях, так и в разностях (последнее приводит к большей мощности). Более конкретно, статистику $t\text{-}REC$ можно записать следующим образом:
\begin{equation}
t\text{-}REC=\frac{\sum_{i=1}^N\sum_{t=p+1}^TR_{i,t-1}r_{it}/\hat{\sigma}^2_{e,i}}{\sqrt{\sum_{i=1}^N\sum_{t=p+1}^TR_{i,t-1}^2/\hat{\sigma}^2_{e,i}}},
\end{equation}
где 
\[\hat{\sigma}^2_{e,i}=\frac{\mathbf{r}_i^{\prime}\mathbf{r}_i}{T},\]
\[r_{it}=y_{it}-\sum_{k=2}^t{y_{ik}a_{kt}},\]
а
\[a_{kt}=d^{\prime}_k(\sum_{n=1}^t{d_{n}d^{\prime}_{n}})^{-1}d_{t}\]
\[d_t=G\Delta D_t,\]
где $G$ является $p\times (p+1)$ матрицей, состоящей из нулей и единиц, чтобы не учитывать дифференцированную константу, которая становится вектором из нулевых элементов. Здесь $p+1$ является размерностью детерминированной компоненты. Переменная $R_{i,t-1}$ определяется как $R_{i,t-1}=\sum_{n=p+1}^t{r_{in}}$. 

Статистика в модели пула на основе рекурсивно детрендированных рядов будет иметь асимптотическое нормальное распределение без дополнительной коррекции числителя на факторы, связанные со средним и дисперсией, как делается в других тестах. Однако, сравнивая локальную мощность с тестами \citet{Breitung2000} и тестом $t^{+}$ в \citet{MoonPerron2008} в случае наличия трендов, \citet{Westerlund2015c} заключает, что асимптотическая локальная мощность теста, основанного на рекурсивном детрендировании, несколько ниже (хотя на конечных выборках выше, чем $t^{+}$). Более точно, $t^{+}\Rightarrow-0.053\mu_{c,2}+N(0,1)$, $UB_{NT}\Rightarrow0.068\mu_{c,2}+N(0,1)$ (используются правосторонние критические значения) и $t\text{-}REC\Rightarrow-0.052\mu_{c,2}+N(0,1)$ и оптимальный тест $V_{NT}\Rightarrow-0.075\mu_1+N(0,1)$, так что $0.075>0.068>0.053>0.052$. Напомним, что в случае отсутствия тренда  $t^{+}\Rightarrow-0.47\mu_{c,1}+N(0,1)$, $t\text{-}REC\Rightarrow-0.5\mu_{c,1}+N(0,1)$, а оптимальный тест $V_{NT}\Rightarrow-0.5\mu_{c,1}+N(0,1)$.

\citet{Westerlund2015c} задается вопросом, изменяется ли локальная окрестность, в которой тест имеет нетривиальную мощность, при добавлении тренда более высокого порядка (например, квадратичный). Как уже было отмечено ранее, при наличии индивидуальных трендов тест имеет мощность только в окрестности $N^{-1/4}T^{-1}$ единицы, в отличие от случая с только фиксированными эффектами, где такой окрестностью будет $N^{-1/2}T^{-1}$. Оказывается, что тренд более высокого порядка не меняет данную окрестность, и она остается такой же, как и в случае линейного тренда, то есть $N^{-1/2}T^{-1}$. Однако мощность все равно уменьшается (локальная мощность зависит от степени тренда), то есть ``предельные издержки'' линейного тренда намного выше, чем трендов более высокого порядка.

\citep{Westerlund2014} анализирует GLS-детрендирование\footnote{Единственная предшествующая работа, которая исследовала эффект от GLS-детрендирование на основе симуляций, была работа \citep{Lopez2009}.}, однако только для случая фиксированных эффектов. Вестерлунд рассматривает два варианта GLS-детрендирования: в первом случае детрендирование производится после взятия первых разностей, а во втором случае наоборот. В случае отдельного временного ряда порядок не играет роли. Автор заключает, что первый способ приводит к не только смещенной оценке, но и расходящейся, а при коррекции на среднее и дисперсию мощность сильно падает (рассматривались статистики \citet{MoonPerron2008}, $t^{\#}$ и $t^{+}$). В терминах локальной мощности рассматриваемые статистики также хуже $t^{\#}$ и $t^{+}$ и даже не лучше, чем обычные GLS-статистики для каждого временного ряда. С другой стороны, если брать разность после детрендирования, то тест будет несмещенным и более мощным, чем $t^{\#}$ и $t^{+}$ при OLS-детрендировании. В качестве объяснения данного феномена см. Remark 6 в \citet{Westerlund2014}.

Обозначим соответствующий GLS-оператор как $\bar{D}$, и этот оператор применяется к $y_{it}$ как $\bar{D}y_{it}=X_{it}-wy_{i1}-w(1-\bar{\rho})\sum_{k=2}^T{\Delta_{\bar{\rho}}y_{ik}}$, где $\Delta_{\bar{\rho}}y_{it}=y_{it}-\bar{\rho}y_{i,t-1}$ для $t\geq2$, а параметр
\[\bar{\rho}=1+\frac{\bar{c}}{N^{\bar{\kappa}}T}\]
теперь зависит не только от $\bar{c}\neq0$, но и от $\bar{\kappa}\geq0$\footnote{Например, в \citep{Lopez2009} выбиралось $\bar{\kappa}=\kappa$, а в \citep{Choi2001} -- $\bar{kappa}=0$.} Пусть (Q)GLS-оценка максимального авторегрессионного корня равна
\[\hat{\phi}_{QGLS}=\frac{\sum_{i=1}^N\sum_{t=2}^T{\bar{D}y_{i,t-1}\Delta(\bar{D}y_{it})}}{\sum_{i=1}^N\sum_{t=2}^T{(\bar{D}y_{i,t-1})^2}},\]
а соответствующая ей $t$-статистика равна
\[t_{QGLS}=\delta_{QGLS}\frac{\hat{\phi}_{QGLS}}{\hat{\sigma}_e/\sqrt{\sum_{i=1}^N\sum_{t=2}^T{(\bar{D}y_{i,t-1})^2}}},\]
где $\delta_{QGLS}=\sqrt{3/4}$. Локальная мощность теста будет равна $\Phi\left(-\frac{\sqrt{3}\mu_{c,1}}{2\sqrt{2}}+z_\xi\right)$. Асимптотические результаты не зависят от $\bar{\kappa}$ и $\bar{c}$ 
и остаются теми же самыми, даже когда $\bar{c}_i$, $i=1,\dots,N$, не равны. Сравнивая с асимптотической огибающей мощности, которая задается как (см. \citep{MPP2007})
\[\Phi\left(\frac{\sqrt{\mu_{c,2}}}{\sqrt{2}}+z_\xi\right)\geq\Phi\left(-\frac{\mu_{c,1}}{\sqrt{2}}+z_\xi\right)\]
с равенством при $c_1=\dots=c_N$, тест $t_{QGLS}$ имеет непренебрежимую локальную мощность внутри той же самой сокращающейся окрестности, что и огибающая, но будет ниже, чем огибающая, поскольку $\sqrt{3}/2\sqrt{2}\approx0.612<1/\sqrt{2}\approx0.707$. Однако $t_{QGLS}$ не лучше, чем статистики $P_a$ и $P_b$, предложенные в \citet{BaiNg2010}, являющиеся оптимальными, как и статистика $UB_{NT}$. 

Для практической реализации Вестерлунд рекомендует использовать  $\bar{\kappa}=0$ и $\bar{c}=-1$.


Подводя итог данного раздела, оптимальными тестами в случае наличия только фиксированных эффектов являются $V_{NT}$, $UB_{NT}$, $P_a$, $P_b$ и $t\text{-}REC$, а тест LLC имеет несколько более низкую мощность. С другой стороны, при наличии трендов  оптимальными тестами являются $V_{NT}$, $P_a$, $P_b$, затем идет $UB_{NT}$, а после статистика $t^{+}$.

\subsection{Наличие слабой зависимости ошибок}

Как и в случае временных рядов, логично было бы предположить, что ошибки $\varepsilon_{it}$ в \eqref{IndPan1} могут быть слабо зависимыми. Тогда, аналогично расширенному тесту Дики-Фуллера, можно аппроксимировать краткосрочную динамику добавлением запаздывающих разностей:
\begin{equation}\label{IndPan14_3}
\Delta y_{it}=d_{it}+\phi_{i1} y_{i,t-1}+\sum_{j=1}^{p_i}\psi_{i,p_i}\Delta y_{i,t-j}+\varepsilon_{it},
\end{equation}
где $d_{it}$ -- некоторая детерминированная компонента.

В случае однородных альтернатив в LLC рекомендуется сначала очистить переменные от краткосрочной динамики, получая остатки $e_{it}$ ($v_{it}$) от регрессии $\Delta y_{it}$ ($y_{it}$) на $\Delta y_{i,t-j}$, $j=1,\dots,p_i$, и $d_{it}$. Затем общий параметр $\phi$ можно оценить по регрессии пула
\begin{equation}\label{IndPan14_4}
(e_{it}/\hat{\sigma}_i)=\phi(v_{i,t-1}/\hat{\sigma}_i)+\nu_{it},
\end{equation}
где $\hat{\sigma}_i^2$ -- оцененная дисперсия $e_{it}$. Однако регрессия на первом шаге не удаляет всю зависимость в ошибках, поскольку
\[\lim_{T\rightarrow\infty}E\left[\frac{1}{T-p}\sum_{t=p+1}^T{e_{it}v_{i,t-1}/\sigma^2_i}\right]=\frac{\omega_i}{\sigma_i}\mu_\infty^*,\]
где $\omega_i^2$ -- долгосрочная дисперсия $e_{it}$. В LLC предлагается оценить $\omega_i^2$ непараметрически, используя ряды в первых разностях:
\[\hat{\omega}_i^2=\frac{1}{T}\left[\sum_{t=1}^T{\widehat{\Delta y}_{it}^2}+2\sum_{l=1}^K\left(\frac{K+1-l}{K+1}\right)\left(\sum_{t=l+1}^T{\widehat{\Delta y}_{it}\widehat{\Delta y}_{i,t-l}}\right)\right],\]
где $\widehat{\Delta y}_{it}=\Delta y_{it}-T^{-1}\sum_{t=2}^T\Delta y_{it}$ -- центрированный ряд разности, $K$ -- параметр усечения. Тогда статистика  LLC в \eqref{IndPan14} принимает вид
\begin{equation}\label{IndPan14_5}
Z_{LLC}=\frac{\sum_{i=1}^N{(\Delta \mathbf{y}'_i\mathbf{M}_1\mathbf{y}_{i,-1}/\hat{\sigma}^2_i-\mu_T^* T\hat{\sigma}_i/\hat{\omega}_i)}}{\sigma_T^*\sqrt{\sum_{i=1}^N{\mathbf{y}'_{i,-1}\mathbf{M}_1\mathbf{y}_{i,-1}/\hat{\sigma}^2_i}}}=\frac{\tau_\phi}{\sigma_T^*}-\frac{\mu_T^*T}{\sigma_T^*\sqrt{\sum_{i=1}^N{\mathbf{y}'_{i,-1}\mathbf{M}_1\mathbf{y}_{i,-1}/\hat{\sigma}^2_i}}}\times\sum_{i=1}^N{\frac{\hat{\sigma}_i}{\hat{\omega}_i}}.
\end{equation}
Снова отметим, что в контексте временных рядов данная оценка, основанная на первых разностях, не была бы состоятельной, поскольку при стационарной альтернативе сходится к нулю по вероятности. В панелях, однако, данная оценка улучшает мощность теста, поскольку корректирующая компонента пропадает, и статистика стремится к $-\infty$.

Чтобы избежать непараметрического оценивания долгосрочной дисперсии можно использовать подход \citet{BreitungDas2005}. На первом шаге предлагается оценить регрессию $\Delta y_{it}$ на детерминированную компоненту и лаги $\Delta y_{i,t-1},\dots,\Delta y_{i,t-p_i}$. При нулевой гипотезе очищенный от краткосрочной динамики ряд $\hat{\psi}_iy_{it}$ является случайным блужданием с некоррелированными приращениями. Этот подход также можно использовать для модификации несмещенной статистики $UB_{NT}$, так что асимптотическая стандартная нормальность сохраняется.

\citet{Westerlund2009} указывает недостаток подхода LLC, который связан с тем, что скорость, при которой оценка долгосрочной дисперсии $\hat{\omega}_i$ сходится к нулю при альтернативе, очень низка, если параметр ширины окна не слишком велик. Таким образом, как показано автором на симуляциях, большинство методов для выбора ширины окна не являются адекватными. Все это приводит к сдвигу распределения вправо на конечных выборках, что приводит к потере мощности. В \citet{WesterlundBlomquist2013b} предлагают подход, похожий на \citet{BreitungDas2005}, который основан на следующей регрессии:
\[\Delta y_{it}=d_t+\phi_iy^{*}_{i,t-1}+\sum_{j=1}^{p_i}\psi_{i,p_i}\Delta y_{i,t-j}+\varepsilon_{it},\]
где $y^{*}_{i,t-1}=(\hat{\sigma}_i/\hat{\omega}_i)y_{t,i-1}$, и долгосрочная дисперсия $\omega_i$ оценивается параметрически на основе авторегрессионного представления. В этом случае, при использовании скорректированного запаздывания объясняющей переменной, $y^{*}_{i,t-1}$, статистика LLC корректируется аналогично случаю отсутствия краткосрочной динамики.

Отметим, что статистика, основанная на рекурсивном детрендировании, $t\text{-}REC$, предложенная \citet{Westerlund2015c}, не требует коррекции статистики на смещение, а требует простой очистки переменных от серийной корреляции после выполнения рекурсивного детрендирования.

GLS-тест $t_{QGLS}$, предложенный в \citep{Westerlund2014}, хотя и не является смещенным из-за наличия детерминированной компоненты, будет смещенным из-за серийной корреляции. Пусть
\[\hat{\sigma}^2_\varepsilon=N^{-1}\sum_{i=1}^N\hat{\sigma}^2_{\varepsilon i}, \ \hat{\omega}^2_\varepsilon=N^{-1}\sum_{i=1}^N\hat{\omega}^2_{\varepsilon i}, \ \hat{\lambda}_\varepsilon=N^{-1}\sum_{i=1}^N\hat{\lambda}_{\varepsilon i}, \ \hat{\phi}^4_\varepsilon=N^{-1}\sum_{i=1}^N\hat{\omega}^4_{\varepsilon i}, \]
где $\hat{\sigma}^2_{\varepsilon i}$, $\hat{\omega}^2_{\varepsilon i}$ и $\hat{\lambda}_{\varepsilon i}=(\hat{\omega}^2_{\varepsilon i}-\hat{\sigma}^2_{\varepsilon i})/2$ -- оценки дисперсии, долгосрочной дисперсии и односторонней долгосрочной дисперсии процесса $\varepsilon_{it}$. Тогда скорректированная статистика будет иметь вид
\begin{equation}
\hat{\phi}_{QGLS}^*=\hat{\phi}_{QGLS}+\frac{NT\hat{\lambda}_{\varepsilon}}{{\sum_{i=1}^N\sum_{t=2}^T{(\bar{D}y_{i,t-1})^2}}},
\end{equation}
и соответствующая $t$-статистика будет задаваться как
\begin{equation}
t_{QGLS}^*=\delta^{*}_{QGLS}\frac{\hat{\phi}_{QGLS}}{\hat{\omega}_{\varepsilon}/\sqrt{\sum_{i=1}^N\sum_{t=2}^T{(\bar{D}y_{i,t-1})^2}}},
\end{equation}
где $\delta^{*}_{QGLS}=\sqrt{3\hat{\omega}^4_\varepsilon/4\hat{\phi}^4_\varepsilon}$.

Статистика IPS при наличии серийной корреляции строится точно также, как и в случае ее отсутствия, основываясь на статистиках расширенного теста Дики-Фуллера (а не обычного теста Дики-Фуллера), используя количество запаздывающих разностей, равное $p_i$, для каждого $i$-го временного ряда. Для построения статистики IPS можно использовать  средние и дисперсии, соответствующие каждому значению $p_i$.

\subsection{Другие тесты}

\subsubsection{Тесты, основанные на комбинации $p$-значений}\label{p-values}

Отметим некоторые недостатки наиболее популярных тестов LLC и IPS. Во-первых, данные тесты требуют, чтобы число временных рядов в панели было бесконечным, но в то же время чтобы число групп должно было быть достаточно мало относительно временного интервала (формально $N/T\rightarrow0$). В противном случае тесты не будут иметь корректный размер и в случае очень малых $N$, и в случае очень больших $N$. Во-вторых, для каждой группы требуется тот же самый тип детерминированной компоненты. В-третьих, предполагается одинаковый временной интервал для каждой группы (для IPS допускается ослабление этого предположения, но возникает проблемы вычисления моментов для тестовой статистики). Кроме этого, хотя IPS утверждают, что их тест превосходит тест LLC, эти два типа тестов не совсем корректно сравнивать между собой, поскольку тест IPS является просто способом комбинирования свидетельства наличия единичного корня $N$ тестов на единичный корень, примененных к $N$ группам. Также с точки зрения асимптотической локальной мощности тест LLC более мощный, чем тест IPS (см. \citet[Fact 2]{WesterlundBreitung2013}).

\citet{Choi2001} и \citet{MaddalaWu1999} независимо предлагают подход для решения данных проблем. \citet{MaddalaWu1999} предлагают тест против неоднородной альтернативы $H_{1b}$, основанный на $p$-значениях статистик, индивидуальных для каждого временного ряда, используя подход \citet{Fisher1932}. Обозначим некоторый тест на единичный корень для $i$-го временного ряда как $G_i$, тогда $p_i=F(G_i)$ -- $p$-значение для этого теста, где $F(\cdot)$ -- его функция распределения. Тогда предложенная авторами (правосторонняя) тестовая статистика будет иметь вид
\begin{equation}\label{IndPan15}
P=-2\sum_{i=1}^N{\ln(p_i).}
\end{equation}
Альтернативная возможность скомбинировать тесты была предложена \citet{Choi2001}, использующая обратный нормальный (левосторонний) тест, определенный как
\begin{equation}\label{IndPan16}
Z=\frac{1}{\sqrt{N}}\sum_{i=1}^N{\Phi^{-1}(p_i)},
\end{equation}
где $\Phi(\cdot)$ является функцией стандартного нормального распределения.

При нулевой гипотезе и при фиксированном $N$ статистика $P$ имеет распределение $\chi^2$ с $2N$ степенями свободы, а $Z$ имеет стандартное нормальное распределение. При $N\rightarrow\infty$ тест $P$ расходится к бесконечности и его необходимо модифицировать следующим образом:
\begin{equation}\label{IndPan15}
P^*=\frac{\frac{1}{\sqrt{N}}\sum_{i=1}^N{(-2\ln(p_i)-2)}}{2},
\end{equation}
поскольку $E(-2\ln(p_i))=2$ и $Var(-2\ln(p_i))=4$. Однако при $N\rightarrow\infty$ тест $Z$ все еще имеет стандартное нормальное распределение, то есть его можно использовать как при малых, так и при больших $N$.

\citet{Hanck2008} дает объяснение контринтуитивного феномена, что у тестов, основанных на комбинации $p$-значений, увеличиваются искажения размера при росте $N$. Пусть мы получаем $N$ $p$-значений для последующей их комбинации. Однако для конечных $T_i$ тесты не будут иметь размер, равный номинальному. Это приводит к тому, что $p$-значения не будут равномерно распределены на единичном интервале. Фактически $p$-значения будут равномерно распределены на интервале $[a,b]$, где $a\geq0$ и $b\leq 1$. Автор рекомендует использовать выборочное распределение статистик, например, при помощи критических значений для конечных выборок.

\subsubsection{Альтернативные подходы}

\citet{SLKN2004} предлагают модификации теста IPS, используя подход \citep{Leybourne1995}, беря максимум из двух статистик для каждого конкретного ряда, первая из которых строится для самого ряда, а вторая -- для обращенного во времени этого же ряда. Далее, центрируя и нормируя сумму из таких индивидуальных статистик, можно получить стандартное нормальное распределение. Можно также аналогичным образом использовать подход \citet{PGFF1994}, где вместо обычной статистики Дики-Фуллера берется взвешенная симметричная (WS) тестовая статистика. Еще один подход, уже упомянутый ранее, это использовать $LM$ статистику или, аналогично \citep{Leybourne1995}, беря минимум из двух $LM$ статистик.

\citet{OhSo2004} предлагают в качестве статистики для $i$-го субъекта брать сумму знаков величины под суммой в числителе в обычной статистике Дики-Фуллера, игнорируя знаименатель. То етсь такая статистика будет иметь вид
\[S_T^i=\sum_{t1}^Tsign(\Delta y_{it})sign(y_{i,t-1}-\hat{\mu}_{i,t-1}),\]
где $\hat{\mu}_{i,t-1}$ -- рекурсивная оценка среднего. Суммируя эти статистики по $N$, как $PS=\sum_{i=1}^NS_T^i$, \citet{OhSo2004} получают, что $(PS+N)/2$ имеет точное биномиальное распределение с параметрами $N$ (число испытаний) и 0.5 (вероятность успеха), а статистика $PS/\sqrt{N}$ имеет асимптотически стандартное нормальное распределение. При наличии автокоррелированности $\Delta y_{it}$ просто заменяется на остатки от регрессии $\Delta y_{it}$ на свои  запаздывающие значения. Тест \citet{OhSo2004}, как было отмечено на симуляциях, является робастным к тяжелым хвостам распределения ошибок (при наличии больших выбросов). 

\citet{OLS2010} предлагают статистику на единичный корень для неоднородных панелей, основанную на функции вклада. Полученный тест является более мощным, чем IPS, особенно в случае больших $N$ и малых $T$.

В \citet{LLW2014} обсуждается проблема, что во многих исследованиях относительно тестирования наличия панельного единичного корня требуется предположение, что $N/T\rightarrow0$, что на практике означает, что $T$ должно быть намного больше, чем $N$. Однако данное условие может быть и намного сильнее, как, например, в \citet{DemetrescuHanck2012a}, где предполагалось $N^5/T$, что на практике означает, что $T$ должно быть больше, чем $N^5$ (в этом случае при $N=10$ значение $T$ должно быть больше, чем 100000, что невозможно гарантировать на практике). \citet{LLW2014} разрабатывают тест отношения правдоподобий (LR) в модели
\[y_{it}=\rho y_{i,t-1}+\gamma+\mu_i+\lambda_t+\varepsilon_{it},\]
где $\mu_i$ и $\lambda_t$ -- индивидуальные и временные эффекты. Однако $\mu_i$ и $\lambda_t$, в отличие от большинства исследований по проверке панельного единичного корня, предполагаются не фиксированными эффектами, а случайными. Это значительно уменьшает количество параметров, которые нужно оценить. Это может привести к уменьшению смещения для корректировки оценки коэффициента, тем самым уменьшая зависимость от $N/T$. Авторы предлагают скорректированную на смещение оценку $\hat{\phi}_{BC}$ для $\rho-1$ и получают следующий асимптотический результат:
\[\sqrt{N}T^{3/2}\hat{\phi}_{BC}+\sqrt{N}T^{-3/2}6r_{\mu}^{-2}\Rightarrow\frac{\sqrt{12}}{r_\mu}N(0,1),\]
где $r_{\mu}=\sigma_\mu^2/\sigma^2_\varepsilon$. Важно отметить тот факт, что скорректированная на смещение оценка $\hat{\phi}_{BC}$ асимптотически эквивалентна MLE и не требует для этого предположения, что $N\rightarrow\infty$. Поэтому для получения предельного распределения выше требуется только условие $N/T^5\rightarrow0$, что допускает и случай $T/N\rightarrow0$. Также скорость сходимости $\sqrt{N}T^{3/2}$ выше, чем скорость для обычной суперсостоятельной панельной оценки $\sqrt{N}T$. Причина последнего заключается в том, что $\mu_i$ при нулевой гипотезе порождает линейный тренд. Углы наклона $(\mu_1,\dots,\mu_N)$ имеют нулевое среднее и, следовательно, не влияют на среднее $y_{it}$. Однако дисперсия увеличивается, приводя к сильному сигналу от наличия единичного корня. Обозначая концентрированную  функцию правдоподобия как $l_c(\phi)$, тест отношения правдоподобий будет иметь следующий вид
\[LR=-2(l_c(0)-l_c(\hat{\phi}_{BC})).\]
Тогда, если $N/T^3\rightarrow0$, $LR$ будет иметь асимптотическое распределение $\chi^2(1)$. Если $N/T^3\rightarrow c\in(0,\infty)$, то это же распределение будет иметь статистика $(\sqrt{LR}-\sqrt{3c}r_\mu^{-3/2})^2$. Если же $N/T^3\rightarrow \infty$, но $N/T^5\rightarrow 0$, то распределение $\chi^2(1)$ будет иметь статистика $(Tr_\mu)^3(LR-3NT^{-3}r_\mu^{-3})^2/(12N)$. Отметим, что данная статистика робастна к неправильной спецификации эффектов, при условии что $\sigma^2_\mu$ заменяется на $\bar{\sigma}^2_\mu=\lim_{N\rightarrow\infty}N^{-1}\sum_{i=1}^N{\mu_i}$.

\citet{FarkasMatyas2015} предлагают непараметрический и неасимптотический метод, основанный на событиях пересечения границы (boundary crossing events), то есть основанный на подсчете количества пересечений некоторых границ.


\section{Тестирование на стационарность}

В данном разделе мы опишем тесты, в которых в качестве нулевой гипотезы принимается гипотеза о стационарности всех временных рядов в панели. Для построения тестовой статистики, аналогично работе \citet{KPSS1992}, \citet{Hadri2000} использует представление ненаблюдаемых компонент вида
\begin{eqnarray}\label{Stat1}
y_{it}&=&d_{it}+u_{it}+\varepsilon_{it},\label{Stat1}\\
u_{it}&=&u_{i,t-1}+v_{it},\label{Stat2}
\end{eqnarray}
где $d_{it}$ -- детерминированная компонента, $\varepsilon_{it}$ и $v_{it}$ являются $i.i.d.$ и взаимно независимыми. Гипотезу о стационарности можно сформулировать как $\sigma^2_v=0$, если предположить однородность дисперсий, или $\sigma^2_{v,i}=0$ в противном случае. Альтернативная гипотеза заключается в том, что $\sigma^2_{v,i}>0$ для некоторой доли кросс-секционных объектов. Пусть $\hat{\varepsilon}_{it}$ -- остатки от регрессии $y_{it}$ на соответствующую детерминированную компоненту. Тогда LM (LBI, локально наилучшая инвариантная) статистика, предложенная \citet{Hadri2000}, принимает вид
\begin{equation}\label{Stat3}
LM=\frac{\frac{1}{N}\sum_{i=1}^N\eta_{iT}}{\hat{\sigma}^2_\varepsilon}=\frac{\frac{1}{N}\sum_{i=1}^N\frac{1}{T^2}\sum_{t=1}^TS^2_{it}}{\hat{\sigma}^2_\varepsilon},
\end{equation}
где $S_{it}$ является частичной суммой остатков,
\[S_{it}=\sum_{j=1}^t\hat{\varepsilon}_{jt},\]
а $\hat{\sigma}^2_\varepsilon$ -- состоятельная оценка $\sigma^2_\varepsilon$ при нулевой гипотезе, например,
\[\hat{\sigma}^2_\varepsilon=\frac{1}{N(T-1)}\sum_{i=1}^N\sum_{t=1}^T\hat{\varepsilon}^2_{it}.\]
Можно допустить гетероскедастичнось среди кросс-секционных субъектов, и тогда тестовая статистика будет иметь вид
\begin{equation}\label{Stat4}
LM=\frac{1}{N}\sum_{i=1}^N\frac{\frac{1}{T^2}\sum_{t=1}^TS^2_{it}}{\hat{\sigma}^2_{\varepsilon,i}},
\end{equation}
где $\hat{\sigma}^2_{\varepsilon,i}$ -- оценка дисперсии для конкретного $i$. Эту оценку можно также заменить на состоятельную оценку долгосрочной дисперсии, которая вычисляется стандартным способом.

При $T\rightarrow\infty$
\[LM\Rightarrow\frac{1}{N}\sum_{i=1}^N\int_0^1(W_i(r)-rW_i(1))^2dr\]
(в случае отсутствия тренда; если мы допускаем наличие тренда, то Броуновский мост заменяется на Броуновский мост второго порядка).
Поэтому статистика
\begin{equation}\label{Stat5}
Z_{LM}=\frac{\sqrt{N}(LM-\mu)}{\sigma}\Rightarrow N(0,1),
\end{equation}
при $N\Rightarrow\infty$, где $\mu=E(\int_0^1(W_i(r)-rW_i(1))^2dr$ и $\sigma=Var(\int_0^1(W_i(r)-rW_i(1))^2dr$. В случае отсутствия тренда $\mu=1/6$ и $\sigma^2=1/45$. При наличии тренда $\mu=1/15$ и $\sigma^2=11/6300$.

\citet{YinWu2001} дополнительно анализируют свойства панельного теста типа Лейбурна-МакКейба \citep{LM1994} и теста типа Фишера, основанного на комбинации $p$-значений индивидуальных тестов (имеющему асимптотическое распределение $\chi^2$), аналогично \citet{Choi2001} и \citet{MaddalaWu1999}). Тесты Фишера имеют лучший размер, чем тесты, основанные на групповом среднем.

\citet{ShinSnell2006}  рассматривают связь между последовательной и совместной предельными теориями для получения асимптотических распределений в контексте тестов на стационарность. Совместный предел требует условие, что $N/T\rightarrow0$, в то время как последовательный -- нет. Интуиция этого факта заключается в следующем. Обычно для каждой панельной единицы $i$ можно увидеть, что $E(\eta_{iT})=E(\eta_i)+T^{-1/2}B_i$, где $\eta_{iT}$ определяется в \eqref{Stat3}, $\eta_i$ -- тестовая статистика, являющаяся $i.i.d.$ среди кросс-секционных единиц, а $T^{-1/2}B_i$ -- компонента смещения, которая уменьшается со скоростью $\sqrt{T}$. Статистика группового среднего, таким образом,  будет содержать сумму из $N$ смещений, деленных на $\sqrt{NT}$. Если допустить $T\rightarrow\infty$ \textit{сначала}, как в последовательной асимптотике, то каждое панельное смещение будет сокращаться, и в последовательном пределе не будет никакого смещения. В отличие от этого, если предположить, что $N$ и $T$ растут \textit{одновременно}, то смещение не обязательно сократится. Есть $N$ смещений, деленных на $\sqrt{NT}$ в статистике группового среднего, и достаточным условием того, что смещения сократятся в последовательной асимптотике, будет условие $N/T\rightarrow0$.

В отличие от \citet{Hadri2000} и \citet{YinWu2001}, \citet{ShinSnell2006} предлагают параметрическую коррекцию слабой зависимости ошибок $v_{it}$ с дополнительной коррекцией дисперсии в \eqref{Stat5}, тем самым получая статистику, асимтотически эквивалентную $Z_{LM}$.

\section{Тесты на единичный корень для пространственно-\ коррелированных панелей} 

Предположение об отсутствии пространственной корреляции в ошибках является достаточно сильным и, вероятно, не выполняется во многих приложениях. Многие макроэкономические теории утверждают, что существуют одни и те же ненаблюдаемые общие факторы (такие как шоки технологии, привычки и фискальная политика). Соответственно, логично ожидать, что эти общие факторы влияют на многие  макроэкономические переменные, такие как процентные ставки, инфляция, выпуск и другие. Как отмечают \citet{PSY2013}, например,  при тестировании на наличие единичного корня в панели реальных выпусков можно было бы ожидать ненаблюдаемый общий шок выпуска (который происходит из-за технологии), который также проявляется в занятости, потреблении и инвестициях.  При тестировании на наличие единичного корня в межстрановых данных по инфляции можно было бы ожидать ненаблюдаемые общие факторы, которые коррелированы с темпами инфляции среди стран, которые также влияют на краткосрочные и долгосрочные процентные ставки среди различных рынков и экономик.

\citet{OConnell1998} и \citet{MaddalaWu1999} показали, что в пространственно коррелированных панелях наблюдаются сильные искажения размера для панельных тестов на наличие единичного корня. \citet{StraussYigit2003} демонстрируют, что степень искажений размера вследствие одновременной корреляции увеличивается при увеличении величины коэффициента этой кросс-корреляции и ее вариабельностью. Скорректированные на размер критические значения, соответственно, становятся более отрицательными. Авторы показывают, что усреднение согласно IPS (вычитание кросс-секционного среднего для устранения общего временного или агрегированного эффекта) не устраняет проблему искажений размера, вызванного вариацией кросс-корреляций, оно только частично уменьшает корреляцию. Кроме того, мощность не увеличивается при росте $\sqrt{N}$, и панели с большим значением $N$ показывают более серьезные искажения размера.

\citep{BMO2004,BMO2005} анализируют наличие коинтеграции между кросс-секционными объектами (cross-unit cointegration), которая увеличивает размер теста, так что отвержение гипотезы о наличии панельного единичного корня не является следствием более высокой мощности по сравнению с одномерными тестами, а лишь следствием наличия мешающих параметров.

Это привело к разработке множеств тестов, учитывающих пространственную корреляцию в панелях. Такие тесты принято называть тестами на панельный единичный корень второго поколения. Хотя корреляционная структура ошибок в общем случае неизвестна, и ее оценивание в общем случае недоступно вследствие ограничений на степени свободы,  упрощенное задание некоторой формы зависимости является обычной практикой в теоретических работах.

\citet{BBP2007} обсуждают еще один источник кросс-секционной корреляции -- пространственная (spatial) корреляция, популярная в региональных исследованиях и в экономике города, которая основана на пространственных взаимосвязях и пространственной неоднородности. Здесь термин пространственный относится к географии субъектов, входящих в панель, к их местоположению, географическому расстоянию между ними, а также расстоянию в экономическом и социально-сетевом смысле.

\subsection{Тесты, основанные на дефакторизации}

Одним из самых удобных способов упрощения структуры зависимости является включение общей временной дамми переменной (common time effects, CTE) в панельную регрессию. Обоснованием этого является то, что неокторое со-движение в многомерных временных рядах может происходить из-за общего фактора (common factor). Например, в страновых панелях временная дамми представляет собой общий международный эффект (например, глобальный шок или общий фактор делового цикла) или валюту, используемую в качестве меры стоимости (numeraire currency), в исследованиях паритета покупательской способности.

В модели, исследуемой \citet{PhillipsSul2003}, ошибка регрессии $u_{it}$ имеет вид
\begin{equation}\label{CSC1}
u_{it}=\lambda_if_t+\varepsilon_{it}, \ f_t\sim i.i.d. N(0,1) \text{ по всем $t$},
\end{equation}
где $f_t$ является общим временных эффектом, дисперсия которого нормализована равной единице для идентификации, а коэффициенты $\lambda_i$, предполагающиеся неслучайными, можно рассматривать как параметры  ``идиосинкразической доли'', которые измеряют вклад общего временного эффекта на временной ряд $i$. Ошибки $\varepsilon_{it}$ предполагаются взаимно независимыми и не зависящими от $\theta_s$ для всех $s$. В такой постановке источником пространственной корреляции является общий стохастический временной ряд $f_t$, и степень зависимости измеряется коэффициентами $\lambda_i$. В частности, ковариация между $u_{it}$ и $u_{jt}$ ($i\neq j$) задается как $E(u_{it}u_{jt}=\lambda_i\lambda_j)$. Если $\lambda_i=0$ для всех $i$, то пространственной корреляции нет в данных, а если $\lambda_i=\lambda_j=\lambda_0$ для всех $i$ и $j$, то пространственная корреляция одинакова. Таким образом, пространственная корреляция контролируется компонентами $\bm{\lambda}=(\lambda_1,\dots,\lambda_N)$. Полагая $\bm{u}_t=(u_{1t},\dots,u_{Nt})'$, мы можем определить условную ковариационную матрицу
\begin{equation}\label{CSC2}
\bm{V}_u=E(\bm{u}_t\bm{u}'_t | \sigma_1^2,\dots,\sigma_N^2)=\bm{\Sigma}+\bm{\lambda}\bm{\lambda}', \ \bm{\Sigma}=diag(\sigma_1^2,\dots,\sigma_N^2).
\end{equation}
О модели \eqref{CSC1} можно думать как о простой факторной модели, в которой $f_t$ является общим фактором, а $\lambda_i$ является факторной нагрузкой для ряда $i$.  \citet{PhillipsSul2003} предлагают исключить общий фактор $f_t$ и построить тесты на единичный корень на основе очищенных переменных.

\citet{BaiNg2004} рассматривали более общую модель, нежели \eqref{CSC1}. \citet{BaiNg2004} предлагают так называемый подход PANIC (Panel Analysis of Nonstationary in the Idiosyncratic and Common components, панельный анализ нестационарности в общей и идиосинкразической компонентах). Их модель имеет вид
\begin{eqnarray}
Y_{it}&=&\alpha_i+\beta_i t+\lambda'_iF_t+\varepsilon_{it},\label{CSC3}\\
(I-L)F_t&=&C(L)u_t,\label{CSC3.1}\\
(1-\rho_iL)\varepsilon_{it}&=&D_i(L)e_{it},\label{CSC3.2}
\end{eqnarray}
где $C(L)$ и $D_i(L)$ -- матричные полиномы. Идиосинкразическая ошибка $\varepsilon_{it}$ является I(1), если $\rho_i=1$, и является стационарной, если $|\rho_i|<1$. Факторы также могут быть как стационарными, так и интегрированными первого порядка (ранг матрицы $C(1)$ равен количеству общих трендов). Аналогичный процесс порождения данных (DGP) использовался в \citet{PhillipsSul2003} и \citet{Choi2006}. В \citet{MoonPerron2004} и \citet{MPP2007} использовался DGP вида
\begin{eqnarray}
Y_{it}&=&\alpha_i+\beta_i t+\varepsilon_{it},\\
\varepsilon_{it}&=&\rho_i \varepsilon_{i,t-1}+u_{it},\\
u_{it}&=&\lambda'_if_t+e_{it},
\end{eqnarray}
где $f_t$ и $e_{it}$ являются стационарными линейными процессами, и $e_{it}$ пространственно независима. Данный DGP отличается от рассмотренного в \citet{BaiNg2004} в том смысле, что он специфицирует динамику наблюдаемых переменных, в то время как процесс в \citet{BaiNg2004} специфицирует динамику ненаблюдаемых компонент. Как показано в \citet{BaiNg2010}, при $\varepsilon_{i0}=0$ и $\rho_i=\rho$ для всех $i$ последний процесс можно выразить через первый, так что фактор и идиосинкразическая компонента имеют один и тот же порядок интегрированности. Первая модель является более общей, чем вторая при нулевой гипотезе, что $\rho_i=1$ для всех $i$. Модель, рассмотренная в \citet{Pesaran2007}, идентична модели в \citet{MoonPerron2004}, за исключением того, что детерминированная компонента явно включена в регрессию:
\begin{eqnarray}
Y_{it}&=&\alpha_i+\beta_i t+\rho_iY_{i,t-1}+ u_{it},\\
u_{it}&=&\lambda'_if_t+e_{it},
\end{eqnarray}

Вернемся к модели \eqref{CSC3}-\eqref{CSC3.2}. Оценить факторы можно методом главных компонент. Если ошибки $\varepsilon_{it}$ являются I(0), то оценивая методом главных компонент, мы получаем состоятельные оценки $F_t$ и $\lambda_i$, когда все факторы являются I(0) или некоторые из них I(1). Но в случае, когда $\varepsilon_{it}$ являются I(1), регрессия $Y_{it}$ на $F_t$ будет ложной, даже если бы факторы являлись наблюдаемыми. Для получения состоятельных оценок авторы предлагают сначала привести данные к стационарному виду: в случае $\beta_i=0$  в \eqref{CSC3} (случай отсутствия трендов) определим $y_{it}=\Delta Y_{it}$, $f_{t}=\Delta F_{t}$ и $z_{it}=\Delta \varepsilon_{it}$. Тогда мы можем оценить модель
\begin{equation}\label{CSC4}
y_{it}=\lambda'_if_t+z_{it}
\end{equation}
методом главных компонент. Затем можно выполнить обратное преобразование $\hat{\varepsilon}_{it}=\sum_{s=2}^t\hat{z}_{is}$ и $\hat{F}_{t}=\sum_{s=2}^t\hat{f}_{s}$, получая состоятельные оценки факторов. Отметим, что хотя $z_{it}$ может быть передифференцированным, если исходные данные $\varepsilon_{it}$ были стационарными, оценки факторов все равно будут состоятельными, хотя и неэффективными. После оценивая мы можем использовать ряды $\hat{\varepsilon}_{it}$ для тестирования панельного единичного корня, поскольку эти ряды являются некоррелированными. \citet{BaiNg2004} также предлагают тестировать общие факторы на наличие единичного корня на основе процедуры \citet{StockWatson1988}. Единичный корень может наблюдаться или в идиосинкразической ошибке, или в общих факторах, или в обоих. Если только общий фактор содержит единичный корень, то мы не сможем выиграть в мощности относительно одномерных тестов, поскольку используется только информация относительно временного ряда $\{F_t\}$.

Кроме метода главных компонент \citet{Kapetanios2007} рассматривает другие варианты дефакторизации, такие как динамический метод главных компонент и метод, основанный на оценивании подпространства факторов. Кроме этого \citet{Kapetanios2007}  доказывает асимптотическую эквивалентность тестовых статистик, основанных на дефакторизованных данных и статистик при условии отсутствия кросс-секционной зависимости (в смысле сходимости по вероятности) для любого метода дефакторизации, удовлетворяющего некоторым стандартным условиям.

При наличии трендов ($\beta_i\neq0$ в \eqref{CSC3}) взятие первой разности приводит к
\[\Delta Y_{it}=\beta_i+\lambda'_i\Delta F_t+\Delta\varepsilon_{it}.\]
Усредняя по $t$, получим модель
\[\Delta Y_{it}-\overline{\Delta Y_{i}}=\lambda'_i(\Delta F_t-\overline{\Delta F})+(\Delta\varepsilon_{it}-\overline{\Delta\varepsilon_{i}}),\]
которую можно оценить методом главных компонент. 

Отметим, что количество факторов выбирается согласно информационным критериям, предложенным \citet{BaiNg2002}, в которых штрафная функция зависит и от $N$, и от $T$. Другими словами, количество факторов $r$ выбирается согласно
\[\hat{r}=\arg\min_{k=0,\dots,k_{\max}}IC_1(k),\]
где
\[IC_1(k)=\log\hat{\sigma}^2(k)+k\log\left(\frac{NT}{N+T}\right)\frac{N+T}{NT},\]
$\hat{\sigma}^2(k)=N^{-1}T^{-1}\sum_{i=1}^N\sum_{t=1}^T\hat{z}^2_{it}$. Максимальное количество факторов $k_{\max}$ предлагается выбирать равным 6.

\citet{BaiNg2002} акцентировали внимание на тесте \citet{Choi2001}, описанном в уравнении \eqref{IndPan16}:
\begin{equation}\label{CSC4.1}
P_{\hat{e}}=\frac{-2\sum_{i=1}^N{\ln(p_i)}-2N}{\sqrt{4N}},
\end{equation}
где $p_i$, $i=1,\dots,N$ являются $p$-значениями ADF-теста без детерминированной компоненты для $i$-ого субъекта, используя очищенные от факторов остатки. Напомним, что данный тест является асимптотически стандартным нормальным.


\citet{BaiNg2010} предлагают дополнительные тесты, основанные на очистке от факторов согласно подходу \citet{BaiNg2004}. Первый тест, $P_b$, является аналогом статистики $t^*_b$, предложенной в \citet{MoonPerron2004}, за исключением того, что она основана на другом множестве остатков и на другом методе дефакторизации. Данный тест основан на регрессии пула
\begin{equation}\label{CSC4.2}
\hat{e}_{it} = \rho\hat{e}_{i,t-1}+\varepsilon_{it}.
\end{equation}
Пусть
\[\hat{\sigma}^2_\varepsilon=N^{-1}\sum_{i=1}^N\hat{\sigma}^2_{\varepsilon i}, \ \hat{\omega}^2_\varepsilon=N^{-1}\sum_{i=1}^N\hat{\omega}^2_{\varepsilon i}, \ \hat{\lambda}_\varepsilon=N^{-1}\sum_{i=1}^N\hat{\lambda}_{\varepsilon i}, \ \hat{\phi}^4_\varepsilon=N^{-1}\sum_{i=1}^N\hat{\omega}^4_{\varepsilon i}, \]
где $\hat{\sigma}^2_{\varepsilon i}$, $\hat{\omega}^2_{\varepsilon i}$ и $\hat{\lambda}_{\varepsilon i}=(\hat{\omega}^2_{\varepsilon i}-\hat{\sigma}^2_{\varepsilon i})/2$ -- оценки дисперсии, долгосрочной дисперсии и односторонней долгосрочной дисперсии процесса $\varepsilon_{it}$. Эти оценки можно получить согласно \citet{AndrewsMonahan1992}, используя квадратичное спектральное ядро и предбеливание. Тогда для случая фиксированных эффектов
\begin{eqnarray}\label{CSC4.3}
P_a&=&\frac{\sqrt{N}T(\hat{\rho}^{+}-1)}{\sqrt{2\hat{\phi}^4_\varepsilon/\hat{\omega}^2_\varepsilon}},\\
P_b&=&\sqrt{N}T(\hat{\rho}^{+}-1)\sqrt{\frac{1}{NT^2}\left(\sum_{i=1}^N\sum_{t=2}^T\hat{e}^2_{i,t-1}\right)\frac{\hat{\omega}^2_\varepsilon}{\hat{\phi}^4_\varepsilon}},
\end{eqnarray}
где 
\[\hat{\rho}^{+}=\frac{\sum_{i=1}^N\sum_{t=2}^T\hat{e}_{i,t-1}\hat{e}_{i,t}-NT\hat{\lambda}_\varepsilon}{\sum_{i=1}^N\sum_{t=2}^T\hat{e}^2_{i,t-1}},\]
а для случая индивидуально-специфических трендов
\begin{eqnarray}\label{CSC4.4}
P_a&=&\frac{\sqrt{N}T(\hat{\rho}^{+}-1)}{\sqrt{(36/5)\hat{\phi}^4_\varepsilon\hat{\sigma}^4_\varepsilon/\hat{\omega}^8_\varepsilon}},\\
P_b&=&\sqrt{N}T(\hat{\rho}^{+}-1)\sqrt{\frac{1}{NT^2}\left(\sum_{i=1}^N\sum_{t=2}^T\hat{e}^2_{i,t-1}\right)\frac{5}{6}\frac{\hat{\omega}^6_\varepsilon}{\hat{\phi}^4_\varepsilon \hat{\sigma}^4_\varepsilon}},
\end{eqnarray}
где $\hat{\omega}^4_\varepsilon=(\hat{\omega}^2_\varepsilon)^4$, $\hat{\omega}^6_\varepsilon=(\hat{\omega}^2_\varepsilon)^3$, $\hat{\omega}^8_\varepsilon=(\hat{\omega}^2_\varepsilon)^4$ и
\[\hat{\rho}^{+}=\frac{\sum_{i=1}^N\sum_{t=2}^T\hat{e}_{i,t-1}\hat{e}_{i,t}}{\sum_{i=1}^N\sum_{t=2}^T\hat{e}^2_{i,t-1}}+\frac{3}{T}\frac{\hat{\sigma}^2_\varepsilon}{\hat{\omega}^2_\varepsilon}.\]
Важно отметить, что оценивание факторов на основе модели в первых разностях удаляет фиксированные эффекты, так что тестовые статистики для случаев отсутствия и наличия фиксированных эффектов совпадают.

Кроме этого, в \citet{BaiNg2010} был обобщен тест Саргана-Бхаргавы \citet{SarganBhargava1983}, $SB$, и модифицированный тест Саргана-Бхаргавы \citet{Stock1999}, $MSB$. Особенность теста $MSB$ заключается в том, что он не требует оценивания параметра $\rho$, что позволяет сравнить, происходит ли отличие в мощности из-за оценивания этого параметра. Выражение панельного теста $MSB$ ($PSMB$) в случае фиксированных эффектов (или отсутствия детерминированной компоненты) имеет вид:
\begin{equation}\label{CSC4.5}
PMSB=\frac{\sqrt{N}\left(\cfrac{1}{NT^2}\sum_{i=1}^N\sum_{t=2}^T\hat{e}_{i,t}^2-\hat{\omega}^2_\varepsilon/2\right)}{\sqrt{\hat{\phi}^4_\varepsilon/3}},
\end{equation}
а при наличии индивидуально-специфических трендов имеет вид
\begin{equation}\label{CSC4.5}
PMSB=\frac{\sqrt{N}\left(\cfrac{1}{NT^2}\sum_{i=1}^N\sum_{t=2}^T\hat{e}_{i,t}^2-\hat{\omega}^2_\varepsilon/6\right)}{\sqrt{\hat{\phi}^4_\varepsilon/45}},
\end{equation}


На основе симуляций на конечных выборках авторы заключают, что когда строится оценка $\hat{\rho}$, использующая данные, детрендированные методом наименьших квадратов, вне зависимости от того, как удаляются факторы, то это приводит к отсутствию мощности у тестов $P_b$.


\citet{WesterlundLarsson2009} анализируют возможное смещение предельного распределения статистики IPS ($Z_{\hat{e}}$), вызванное дефакторизацией. Результат \citep{BaiNg2004} не гарантирует, что статистика $Z_{\hat{e}}$ будет иметь асимптотически стандартное нормальное распределение. Это вызвано тем, что порядок смещения, вызванного заменой остатков $e_{it}$ на очищенные от факторов остатки $\hat{e}_{it}$, не является настолько большим, чтобы смещение исчезало асимптотически при росте $N$.

На основе разложения более высокого порядка \citet{WesterlundLarsson2009} предлагают корректировку смещения статистики $Z_{\hat{e}}$ для конечных выборок,
которая делает размер теста более близким к номинальному без одновременной потери в мощности.


\citet{Westerlund2015} анализирует асимптотическую локальную мощность тестов \citet{BaiNg2004,BaiNg2010}. 
Результаты заключаются в следующем. Во-первых, тесты \citep{BaiNg2004,BaiNg2010} проявляют то же самое сокращение скорости приближения к локальной альтернативе в случае наличия индивидуальных трендов, что и тесты первого поколения (см. \citet{MoonPhillips1999b}\footnote{В \citet{MoonPhillips1999b} показывается, что оценка максимального правдоподобия авторегрессионного параметра (при локальности к единице) с трендами, специфичными для каждого субъекта, не является состоятельной. Авторы называют этот феномен, возникающий из-за наличия бесконечного числа мешающих параметров, ``проблемой случайных трендов'', поскольку она аналогична известной проблеме случайных параметров в динамических панелях с фиксированным $T$.}). Как замечает \citet{Westerlund2015}, это очень интересно, поскольку подход PANIC не требует явного детрендирования посредством метода наименьших квадратов, что, как считалось, является причиной проблемы низкой мощности. Это подтверждает результаты \citet{MPP2007}, где показывается, что проблема не состоит в методе детрендирования, а состоит в наличии трендов, эффект от которых нужно устранить. Во-вторых, в то время как локальная мощность зависит от гетероскедастичности и серийной корреляции инноваций, на нее не оказывает влияние наличие общих факторов. Другими словами, тесты \citet{BaiNg2004,BaiNg2010} имеют то же самое асимптотическое распределение, что и их аналоги при отсутствии кросс-секционной коррелированности. В-третьих, многие тесты PANIC имеют одинаковое предельное распределение, и отличия наблюдаются только на конечных выборках. Наконец, модифицированные тесты Саргана-Бхаргавы, хотя и являются легко вычисляемыми, имеют более низкую асимптотическую локальную мощность.

\citet{BDA2015b} также анализируют асимптотическую локальную мощность в модели с общими наблюдаемыми факторами на основе теории предельных экспериментов. \citet{BecheriAkker2016} с использованием этой же теории анализируют тесты, основанные на подходе PANIC. \citet{BecheriAkker2016} предлагают новый асимптотически равномерно наиболее мощный (UMP) тест, поскольку существующие тесты не достигают огибающей мощности в случае неоднородных долгосрочных дисперсий идиосинкразических компонент. 


\citet{WesterlundBlomquist2013} анализируют проблему неопределенности относительно наличия тренда. В литературе по единичным корням во временных рядах общепринятой практикой является сначала предварительно тестировать наличие тренда, а на основе полученной информации строить тест либо с, либо без тренда. В качестве предварительного теста часто используют визуальный осмотр графика. Другой подход заключаться в использовании двух тестов, с и без тренда. Для панелей такая процедура становится гораздо более сложной, и часто  исследователи просто игнорируют возможное наличие тренда. С другой стороны, при неучете трендов мощность падает до нуля, если эти тренды фактически присутствуют в данных. Но если трендов нет, то мощность тестов с учетом этих трендов значительно ниже чем мощность тестов, не учитывающих тренды (и ухудшение гораздо более значительно, чем в одномерном случае). \citet{WesterlundBlomquist2013} предлагают алгоритм для тистирования гипотезы о панельном единичном корне, используя предварительное тестирование тренда в факторах.

Этот алгоритм работает хорошо на конечных выборках, но имеет несколько либеральный размер. Авторы предлагают также тест, основанный на подходе объединения отвержений, популяризованный в \citep{HLT2009b}: отвергать нулевую гипотезу (о наличии единичного корня), если хотя бы один из тестов ее отвергает. Используя консервативные критические значения для контроля размера (полученные на основе анализа поверхности отклика), такая процедура также хорошо работает на конечных выборках.

\subsection{Тесты, основанные на аппроксимации факторов}

\citet{Pesaran2007} исследует модель ошибок в виде \eqref{CSC1}, с одним фактором, но вместо того, чтобы строить тесты на единичный корень для очищенных от общего фактора данных, Песаран предлагает расширить стандартную ADF регрессию кросс-секционными средними лагированных уровней и первых разностей индивидуальных временных рядов. В отличие от подхода \citet{BaiNg2004}, фактор предполагается стационарным. Как отмечается в \citet{PSY2013}, если бы компонента $\lambda_if_t$ в регрессии ниже была бы I(1), то вне зависимости от того, равны ли значения $\phi_i$ нулю или нет все ряды были бы нестационарными. 
В простой модели
\begin{equation}\label{CSC5}
\Delta y_{it}=\alpha_i+\phi_i y_{i,t-1}+\lambda_if_t+\varepsilon_{it}
\end{equation}
для тестирования гипотезы $\phi_i=0$ для всех $i$ против неоднородной альтернативы (как в IPS) предлагается использовать следующую кросс-секционно расширенную (cross-sectionally augmented DF, CADF) регрессию:
\begin{equation}\label{CSC6}
\Delta y_{it}=\alpha_i+\phi_i y_{i,t-1}+\delta_i\bar{y}_{t-1}+\gamma_i\Delta\bar{y}_t+\varepsilon_{it}.
\end{equation}
Чтобы понять, как кросс-секционные средние лагированных уровней и первых разностей помогают устранить пространственную зависимость, усредним уравнение \eqref{CSC5} (для простоты при отсутствии фиксированных эффектов) по $N$:
\begin{equation}\label{CSC7}
\frac{1}{N}\sum_{i=1}^N\Delta y_{it}=\frac{\phi_i}{N}\sum_{i=1}^N y_{i,t-1}+\frac{f_t}{N}\sum_{i=1}^N\lambda_i+\frac{1}{N}\sum_{i=1}^N\varepsilon_{it}
= \frac{\phi_i}{N}\sum_{i=1}^N y_{i,t-1}+\frac{f_t}{N}\sum_{i=1}^N\lambda_i+o_p(1),
\end{equation}
где аппроксимация выполняется при больших $N$ (по закону больших чисел). Если $\sum_{i=1}^N\lambda_i\neq0$ (что достаточно вероятно на практике)\footnote{Если $\sum_{i=1}^N\lambda_i=0$, то $\Delta\bar{y}_t=\bar{\varepsilon}_t$ сходится к нулю при $N\rightarrow\infty$, а $\bar{y}_t$ является фиксированной константой для всех $t$.}, то фактор $f_t$ можно выразить через линейную комбинацию средних лагированных уровней и первых разностей. 

Распределение $t$-статистики для $i$-ого субъекта не будет зависеть от мешающих параметров, когда $N\rightarrow\infty$ независимо от того, является ли $T$ фиксированным или стремящимся к бесконечности совместно с $N$. Если $T$ является фиксированным, то чтобы $t$-статистика не зависела от мешающих параметров, необходимо применять тест к отклонениям от начального значения, $y_{it}-\bar{y}_0$. Также распределение (обозначаемое как $CADF_{if}$-распределение) является более скошенным влево, чем стандартное распределение Дики-Фуллера. Однако распределение стандартизованной статистики выглядит похожим на стандартное нормальное, хотя и имеет значимые ненулевые значения коэффициента асимметрии и (куртозиса-3).

Для построения тестов на панельный единичный корень Песаран предлагает использовать либо версию IPS,
\[CIPS=N^{-1}\sum_{i=1}^N{t_i},\]
где $t_i$ -- индивидуальная статистика для $i$-го ряда, либо тесты, предложенные \citet{MaddalaWu1999} и \citet{Choi2001}. Для исключения проблемы вычисления моментов Песаран предлагает использовать так называемые усеченные тесты, $CIPS^*$: то есть если тестовая статистика для $i$-го ряда $\leq K_1$, присваивать ей значение $K_1$, а если $\geq K_2$, присваивать ей значение $K_2$. Константы $K_1$ и $K_2$ должны быть таковы, чтобы вероятность тестовой статистики попасть в интервал $(K_1,K_2)$ была очень большая. Предлагается использовать значения $K_1$ равным, соответственно, 6.12, 6.19 и 6.42 для модели без детерминированной компоненты, модели с константой и модели с трендом, а значения $K_2$ равным 4.16, 2.61 и 1.70 для тех же самых моделей.\footnote{Как отмечается в \citep{Pesaran2007}, p. 278, усечение фактически никогда не оказывает влияние на статистику.} Однако распределение среднего из усеченных статистик все еще остается нестандартным из-за Винеровского процесса $W_f$, в отличие от IPS, где рассматривалась пространственно независимая панель. Распределение будет сходиться (почти наверное) к распределению, которое зависит от $K_1$, $K_2$ и $W_f$, и критические значения для которого можно получить посредством симуляций. Те же самые аргументы применимы и к тестам \citet{MaddalaWu1999} и \citet{Choi2001}. Степень отклонения от нормальности зависит от типа детерминированной компоненты, и плотность в случае отсутствия детерминированной компоненты является бимодальной и показывает наибольшее отклонение от нормальности. 

\citet{Pesaran2007} обсуждает наличие серийной корреляции, источником которой может быть зависимость в общем факторе $f_t$, автокорреляция в идиосинкразической ошибке $\varepsilon_{it}$ или автокорреляция в ошибках регрессии $u_{it}$. Все эти спецификации приводят в одной и той же регрессии вида
\begin{equation}\label{CSC8}
\Delta y_{it}=\alpha_i+\phi_i y_{i,t-1}+\delta_i\bar{y}_{t-1}+\sum_{j=0}^p\gamma_{ij}\Delta\bar{y}_{t-j}+\sum_{j=1}^p\beta_{ij}\Delta y_{i,t-j}+\varepsilon_{it}.
\end{equation}
Тогда все асимптотические результаты, полученные для случая с отсутствие автокорреляции, продолжают выполняться и в случае слабой зависимости ошибок. Как было показано в \citet{Westerlund2014e}, данный подход кажется необоснованным в том случае, когда авторегрессионная структура ошибок неоднородная. Однако на самом деле предельное распределение тестовой статистики $CADF_{if}$ остается тем же самым даже при неправильном моделировании авторегрессионной динамики как однородной.

Отметим, что тест Песарана обоснован в случае, если $N$ и $T$ имеют одинаковый порядок, в то время как \citet{BaiNg2004} и \citet{MoonPerron2004} требуют, чтобы $N/T\rightarrow0$. Также отметим, что для получения предельного распределения для  $CADF_{if}$ требуется условие, что $N\rightarrow\infty$, хотя $CADF_{if}$ - это тест для $i$-го временного ряда. Причина заключается в том, что из-за того, что общий фактор заменяется на кросс-секционное среднее, это создает ошибку аппроксимации, которая уменьшается с ростом $N$ (см. \citet{Westerlund2014e}).

Поскольку модель в \citet{Pesaran2007} допускает наличие только одного фактора (так что при наличии более одного фактора наблюдаются искажения размера теста), в \citet{PSY2013} было предпринято обобщение  на случай наличия большего количества факторов. Авторы предполагают, что существует $k$ дополнительных переменных, которые зависят от по крайней мере того же самого множества общих факторов, хотя и с различными факторными нагрузками. Например, в анализе сходимости выпуска (output convergence) разумно было бы утверждать, что выпуск, инвестиции, потребление, реальные цены на активы и цены на нефть оказывают одинаковый эффект на факторы в общем. Аналогично краткосрочные и долгосрочные процентные ставки и инфляция среди стран, вероятно, обладают рядом общих факторов.

Учитывая рассуждение выше, в случае наличия многофакторной структуры \citet{PSY2013} предлагают расширить регрессию \eqref{CSC6} или (в случае автокоррелированности) \eqref{CSC8} путем добавления дополнительных кросс-секционных средних лагированных уровней и первых разностей, связанных с экономически обоснованными другими переменными $\bm{x}_{it}$ ($k$ переменных):
\begin{equation}\label{CSC9}
\Delta \bm{y}_{i}=\alpha_i+\phi_i \bm{y}_{i,-1}+\bar{\bm{W}}_{ip}\bm{h}_{ip}+\bm{\varepsilon}_{i},
\end{equation}
где $\bar{\bm{W}}_{ip}=(\Delta \bm{y}_{i,-1},\Delta \bm{y}_{i,-2},\dots,\Delta \bm{y}_{i,-p};\Delta\bar{\bm{Z}},\Delta\bar{\bm{Z}}_{-1},\dots,\Delta\bar{\bm{Z}}_{-p};\bar{\bm{Z}}_{-1})$, $\Delta\bar{\bm{Z}}=(\Delta\bar{\bm{z}}_1,\dots,\Delta\bar{\bm{z}}_T)'$, $\bar{\bm{z}}_t=N^{-1}\sum_{i=1}^N\bm{z}_{it}$, $\bm{z}_{it}=(y_{it},\bm{x}'_{it})'$.

Также \citet{PSY2013} строят статистику Саргана-Вхаргавы, $CSB$. В случае отсутствия трендов эта статистика строится на основе регрессии
\begin{equation}\label{CSC10}
\Delta y_{it}=\sum_{j=0}^p\bm{c}'_{ij}\Delta\bar{\bm{z}}_{t-j}+\sum_{j=1}^p\beta_{ij}\Delta y_{i,t-j}+\varepsilon_{it},
\end{equation}
и сама статистика для $i$-го субъекта имеет вид
\begin{equation}\label{CSC11}
CSB_i=T^{-2}\sum_{t=1}^T\hat{u}_{it}^2/\hat{\sigma}^2_i,
\end{equation}
где $\hat{u}_{it}^2=\sum_{j=1}^t\hat{\varepsilon}_{ij}$ и $\hat{\sigma}^2_i=\sum_{t=1}^T\hat{\varepsilon}_{ij}/[T-p-(p+1)(k+1)]$, $\hat{\varepsilon}_{ij}$ -- OLS-остатки в регрессии \eqref{CSC10}. При наличии трендов в регрессию \eqref{CSC10} следует включить константу и скорректировать соответствующим образом количество степеней свободы в оценке дисперсии. Поскольку $CSB_i$ при нулевой гипотезе сходится к функционалу от Винеровского процесса и не зависит от мешающих параметров (а именно, от факторов и факторных нагрузок), $CBS$ тест можно построить как
\begin{equation}\label{CSC12}
CSB=N^{-1}\sum_{i=1}^N CBS_i.
\end{equation}
Критические значения, как и для теста $CIPS$, будут зависеть от количества факторов $k$. На основе стимуляций авторы заключают, что $CIPS$ и $CSB$ не показывают искажений размера для любых рассмотренных значений $T$ и $N$ независимо от того, является ли идиосинкразическая компонента коррелированной или нет, а также что $CSB$ имеет более высокую мощность, чем  $CIPS$, при меньших $T$.

Отметим, что в обсуждении выше количество факторов $k$ предполагается известным. Если оно неизвестно, то предлагается выбирать его либо на основе информационных критериев \citet{BaiNg2002}, либо равным $m_{\max}-1$.

\citet{WHS2015} анализировали асимптотическую локальную мощность тестов Песарана, полагая $\rho_i=\phi_i-1=\exp(T^{-1}c_i)$. Обычно локальная мощность должна зависеть только от среднего $c_i$, но не от дисперсии или других моментов. Для малых выборок мощность обычно уменьшается при росте дисперсии $c_i$, но при больших $N$ это эффект исчезает. В \citet{WHS2015} проводятся симуляции, на которых обнаруживается, что мощность фактически возрастает с ростом $\sigma^2_c$, и это остается верным даже для $N=100$. Этот эффект выше для больших $\lambda=\lim N^{-1}\sum_{i=1}^N\lambda_i$, что является достаточно неожиданным, учитывая тот факт, что асимптотическое распределение тестов не должно зависеть от мешающих параметров даже при локальной альтернативе.


Результаты \citet{WHS2015} следующие. Мощность $CADF_{i}$ и $CIPS$ увеличивается при росте $\beta_i=\lambda_i/\sigma_{\varepsilon,i}$ (О величине $\beta_i^2$ можно думать как об отношении дисперсии $\lambda_if_t$ относительно $\varepsilon_{i,t}$, или мере относительной важности общей компоненты), но не зависит от $\beta_i$ при $\sigma_c^2=0$. Однако в то время как мощность зависит от $\beta_i$, на размер это не оказывает влияние. Другими словами, если $\beta_i$ велико, то лучше использовать тест $CIPS$. Однако на практике значение $\beta_i$ неизвестно, но его можно легко оценить. То есть нужно оценить произведение $\lambda_if_t$ (поскольку по отдельности эти величины неидентифицируемы), например, полагая $\hat{f}_t=\Delta\bar{y}_t$ и оценивая $\hat{\lambda}_i$ методом наименьших квадратов в регрессии $\Delta y_{it}$ на константу и $\hat{f}_t$. Ооценкой для $\beta_i$ будет $\hat{\beta}_i=\hat{\lambda}_i\hat{\sigma}_f/\hat{\sigma}_{\varepsilon,i}$, где $\hat{\sigma}_f^2=T^{-1}\sum_{t=1}^T\hat{f}^2_t$. В качестве общей меры авторы рекомендую использовать среднее $\hat{\beta}_i$. Другим важным параметром является $\sigma_c^2$, который можно вывести, смотря вариативность оцененных значений $\rho_1,\dots,\rho_N$.

\citep{Westerlund2014b} предлагает в некотором смысле обобщение тестов Песарана. Пусть
\[y_{it}=\lambda'_if_t+\varepsilon_{it}.\]
Если бы $\lambda_i$ были бы известны, оценку $\hat{f}_t$ можно получить на основе метода наименьших квадратов. При неизвестных $\lambda_i$ предлагается использовать некоторый инструмент, $Z_i$, который некоррелирован с $\varepsilon_{it}$, но сильно коррелирован с $\lambda_i$, так что кросс-секционная OLS-оценка $Z_iy_{it}$ на $Z_i\lambda'_i$ была бы состоятельной для $f_t$, или эквивалентно,  чтобы $\hat{f}_t=N^{-1}\sum_{i=1}^N{Z_iy_{it}}$ была состоятельной оценкой для пространства, порожденного $f_t$. Эта оценка является инструментальной оценкой (пространства, порожденного) $f_t$. Тогда оценкой идиосинкразической ошибки будет
\[R_{it}=\sum_{s=1}^t{y_{is}-Z'_i\hat{f}_s}.\]
В качестве инструментов можно взять просто $Z_i=1$, что приводит к модели Песарана \citep{Pesaran2007}, рассмотренной выше. Однако вследствие предварительного рекурсивного детрендирования результирующая $t$-статистика для проверки гипотезы о панельном единичном корне будет асимптотически стандартной нормальной, что отличается от результата в \citep{Pesaran2007}. Однако  \citet{Westerlund2014b} рассматривает $t$-статистику для однородной альтернативы, и требует предположения, что $T$ намного больше, чем $N$.

Относительно выбора инструментов $Z_i$ для $\lambda_i$, Вестерлунд отмечает, что выбрать соответствующие достаточно легко, например, на основе экономического содержания 
или просто взять состоятельные оценки, полученные по методу главных компонент. Выбрать подходящие инструменты можно на основе информационных критериев, предложенных, например, \citet{BaiNg2002}.

\citet{WesterlundHosseinkouchack2016} решают проблему нестандартности распределения для теста \citet{Pesaran2007}. Авторы предлагают построить LM-тест, требующий оценивания модели только при нулевой гипотезе, для каждого субъекта. Этот тест будет также иметь нестандартное распределение при аппроксимации факторов, но если вычесть из каждой такой статистики статистику $CIPS_i$ в квадрате, то полученная статистика будет иметь распределение хи-квадрат, то есть зависимость от Винеровских процессов, связанных с факторами, удалится. Нормированная сумма таких скорректированных статистик будет иметь асимптотически стандартное нормальное распределение, как и в стандартных панельных тестах на единичный корень.

\subsection{Тесты, основанные на комбинации нескольких тестов}\label{CombinationTests}

В то время как подходы \citet{MaddalaWu1999} и \citet{Choi2001}, основанные на комбинации нескольких тестов, предполагают независимость отдельных временных рядов, в \citet{DHT2006} делается попытка устранить эту проблему. Используя идею \citet{Hartung1999}, \citet{DHT2006} акцентируют внимание на обратной стандартной нормальной тестовой статистике, $t_i=\Phi^{-1}(p_i)$, где $p_i$ обозначает $p$-значение теста Дики-Фуллера. Пусть статистики $t_i$ будут зависимыми, например как
\[Cov(t_i,t_j)=\delta_{ij} \text{ для $i\neq j$, $i,j=1,\dots,N$,}\]
с $-1/(N-1)<\rho<1$. Модифицированная обратная нормальная тестовая статистика будет иметь вид
\begin{equation}
t(\hat{\rho},\kappa)=\frac{\sum_{i=1}^Nt_i}{\sqrt{N+(N^2-N)\left(\hat{\rho}^*+\kappa\sqrt{\frac{2}{N+1}(1-\hat{\rho}^*)}\right)}},
\end{equation}
где $\hat{\rho}^*=\max\left(-\frac{1}{N-1},\hat{\rho}\right)$ с $\hat{\rho}=1-\frac{1}{N-1}\sum_{i=1}^N\left(t_i-\frac{1}{N}\sum_{i=1}^N{t_i}\right)^2$, и $\kappa$ является константой (\citet{DHT2006} предлагают выбирать $\kappa=0.2$. Данная статистика имеет асимптотически стандартное нормальное распределение, если выполняются следующие два условия:
\[\lim_{N\rightarrow\infty}\frac{1}{N(N-1)}\sum\sum_{i\neq j}{\rho_{ij}}=\tilde{\rho}, \ \tilde{\rho}\in(0,1),\]
и
\[\lim_{N\rightarrow\infty}\frac{1}{N(N-1)}\sum\sum_{i\neq j}{(\rho_{ij}-\tilde{\rho})^2}=0.\]
Второе условие говорит о том, что $\rho_{ij}$, ковариация между $t_i$ и $t_j$, не сильно изменяется по отношению к асимптотическому среднему. \citet{DHT2006} на основе симуляций показывают, что предложенная статистика работает хорошо, когда ковариационная матрица ошибок имеет постоянные внедиагональные элементы, а также превосходит тесты \citet{MaddalaWu1999} и \citet{Chang2002}.

\citet{ShengYang2013} применяют подход \citet{ZZWW2002} усеченных $p$-значений (truncated product method). Их тестовая статистика имеет вид
\begin{equation}
\prod_{i=1}^N{p_i^{\mathbb I(p_i\leq \tau)}},
\end{equation}
где $\mathbb I$ -- индикатор-функция. Когда некоторые временные ряды в панели явно нестационарны, так что $p$-значения будут близки к 1, обычный метод комбинации $p$-значений будет иметь меньшую мощность из-за этих больших $p$-значений. Путем усечения эти большие $p$-значения удаляются, что позволяет увеличить мощность. Значение $\tau$ предлагается выбирать равным 0.1. Критические значения предлагается получать либо на основе симуляций Монте-Карло, либо на основе бутстрапа. Тест \citet{ShengYang2013} иногда превосходит тесты \citet{Pesaran2007} и \citet{DHT2006}.

\subsection{Тесты, основанные на коррекции дисперсии в построении тестовых статистик}

Наличие общего фактора предполагает, что ошибки кросс-секционных субъектов коррелированы одинаковым способом и, следовательно, внедиагональные элементы ковариационной матрицы ошибок являются идентичными.  Как утверждают \citet{BreitungDas2005}, такая спецификация кажется слишком ограниченной во многих приложениях, использующих региональные данные, где обычно предполагается, что корреляция зависит от расстояния между регионами в выборке. 

\citet{OConnell1998} предлагает использовать тест Дики-Фуллера на основе $t$-статистики в GLS-регрессии. Другими словами, в модели
\[\Delta y_{it}=\phi y_{i,t-1}+\varepsilon_{it}\]
предполагается, что ковариационная матрица ошибок $E(\bm{\varepsilon}_t\bm{\varepsilon}'_t)=\bm{\Omega}$ положительно определена с конечными собственными значениями, где $\bm{\varepsilon}_t=[\varepsilon_{1t},\dots,\varepsilon_{Nt}]'$. Умножая эту регрессию слева на $\bm{\Omega}^{-1/2}$, мы получим тестовую статистику для проверки гипотезы $\phi=0$ (при однородной альтернативе) вида
\begin{equation}\label{CSC13}
t_{gls}=\frac{\sum_{t=1}^T\bm{y}'_{t-1}\bm{\Omega}^{-1}\Delta\bm{y}_t}{\sqrt{\sum_{t=1}^T\bm{y}'_{t-1}\bm{\Omega}^{-1}\bm{y}_{t-1}}},
\end{equation}
где неизвестная ковариационная матрица $\bm{\Omega}$ заменяется на ее оценку
\[\hat{\bm{\Omega}}=\frac{1}{T}\sum_{t=1}^T\hat{\bm{\varepsilon}}_t\hat{\bm{\varepsilon}}'_t.\]
Отметим, что (доступную) GLS-оценку можно построить только при $T\geq N$, в противном случае она будет вырожденной. \citet{Chang2004} показывает, что полученная статистика будет некорректной, поскольку предельное распределение будет зависеть от ковариационных параметров ошибки, так что критические значения можно получить посредством симуляций (бутстрапа). Преимуществом GLS-оценки будет то, что она не требует предположений о структуре ковариационной матрицы (например, о количестве общих факторов). Однако следствием такой обобщенности является плохое поведение на малых выборках. \citet{BreitungDas2008} предлагают методы оценивания ковариационной матрицы на основе предположения о факторной структуре ошибок и способах дефакторизации.  Альтернативный подход, основанный на коррекции стандартной ошибки, был предложен в \citet{Jonsson2005} и \citet{BreitungDas2005} (см. также \citet[Section 3.3.1]{HerwartzSiedenburg2008}), где дисперсию коэффициента можно состоятельно вычислить как
\begin{equation}\label{CSC14}
\widehat{var}(\hat{\phi})=\frac{\sum_{t=1}^T\bm{y}'_{t-1}\hat{\bm{\Omega}}\bm{y}_{t-1}}{\left(\sum_{t=1}^T\bm{y}'_{t-1}\bm{y}_{t-1}\right)^2}.
\end{equation}
Тогда робастную $t$-статистику, $t_{rob}$, можно записать как
\begin{equation}\label{CSC15}
t_{rob}=\frac{\hat{\phi}}{\sqrt{\widehat{var}(\hat{\phi})}}\frac{\sum_{t=1}^T\bm{y}'_{t-1}\Delta\bm{y}_t}{\sqrt{\sum_{t=1}^T\bm{y}'_{t-1}\hat{\bm{\Omega}}\bm{y}_{t-1}}},
\end{equation}
и эта статистика имеет стандартное нормальное распределение.

При наличии краткосрочной динамики можно на первом шаге очистить переменные от нее (строя регрессии ADF-типа), как в LLC. Однако такой подход не будет работать для GLS-регрессии, когда мы умножаем регрессию слева на  $\bm{\Omega}^{-1/2}$ (см. также \citet{Chang2004}). citet{BreitungDas2005} предлагают сначала очистить переменную $y_{tt}$ от всей краткосрочной динамики, строя авторегрессии в уровнях, а затем к полученному очищенному ряду, скажем, $\bm{y}_t^*$ , применять рассмотренные выше тесты. Тогда оба теста, $t_{gls}$ и $t_{rob}$ будут иметь асимптотически стандартное нормальное распределение. Для учета влияние детерминированной компоненты предлагается использовать подходы \citet{BreitungMeyer1994} и \citet{Breitung2000}. 

Симуляции, проведенные в \citet{BreitungDas2005}, показали, что GLS-тест можно применять только если $T$ гораздо больше, чем $N$. Когда $T$ больше $N$, бутстраповские тесты \citet{Chang2004} превосходят $t_{gls}$ и $t_{rob}$. Тест $t_{rob}$ имеет большую мощность при малых $N$ и $T$.

\citet{BreitungDas2008} рассматривают поведение тестов $t_{gls}$ и $t_{rob}$ при наличии общего фактора. Как мы отмечали ранее, источником нестационарности может быть как нестационарность общего фактора, так и нестационарность идиосинкразической компоненты. Если и общий фактор, и идиосинкразическая ошибка нестационарны, $t_{gls}$ имеет стандартное нормальное распределение асимптотически, а $t_{rob}$ является смещенной (гипотеза чаще отвергается на больших выборках, хотя искажения и не слишком высоки). Аналогичная ситуация происходит и в случае, когда общий фактор является стационарным, а идиосинкразическая ошибка нестационарной. Если только общий фактор нестационарный, а идиосинкразическая ошибка является стационарной, то все тесты являются некорректными, и их размер приближается к единице при росте выборки.

\subsection{Тесты, основанные на методах с инструментальными переменными}

Для учета пространственной корреляции был предложен ряд подходов, основанных на инструментальном оценивании. IV-оценка для каждого $i$-го субъекта, предложенная \citet{Chang2002} (для случая отсутствия детерминированной компоненты и слабой зависимости ошибок), строится как
\begin{equation}\label{IVChang}
\hat{\rho}_{i,IV}=\frac{\sum_{t=2}^TF(y_{i,t-1})y_{it}}{\sum_{t=2}^TF(y_{i,t})y_{it}},
\end{equation}
где $F(x)$ является интегрируемой функцией, удовлетворяющей $\int_{-\infty}^{\infty}xF(x)\neq0$. Другими словами, функция $F$ является производящей функцией  инструментов (instrument generating function, IGF), и инструмент $F(y_{i,t-1})$ должен коррелировать с регрессором $y_{i,t-1}$. Примерами регулярно интегрируемых IGF являются $\mathbb I(|x|\leq K)$, любая индикатор-функция на компактном интервале, определенная параметром усечения $K$, ее варианты, такие как $sgn(x)\mathbb I(|x|< K)$ и $x\mathbb I(|x|< K)$. Также можно привести пример функции типа $x\exp\{-|x|\}$. Заметим, что функция вида $F(x)=\mathbb I(|x|\leq K)$ приводит к усеченной OLS-оценке, которая использует только наблюдения в интервале $[-K,K]$.

Полученная на основе $\hat{\rho}_{i,IV}$ статистика имеет стандартное нормальное распределение, в отличие от асимптотики единичных корней, и это распределение является независимым от других кросс-секционных субъектов. При наличии лагов эти лаги используются сами в качестве инструментов. Для тестирования единичного корня в панели, аналогично IPS, индивидуальные статистики суммируются и делятся на $\sqrt{N}$. Асимптотически стандартное нормальное распределение нормализованной суммы индивидуальных статистик получено при условии только $T\rightarrow\infty$ и при некотором балансе величин выборок в несбалансированной панели (см. Assumption 4.1). При наличии детерминированной компоненты \citet{Chang2002} предлагает использовать предварительное рекурсивное детрендирование, так что в этом случае сохраняется асимптотически стандартное нормальное распределение. В качестве функции $F(\cdot)$ предлагается использовать $F(y_{i,t-1})=y_{i,t-1}\exp\{-c_i|y_{i,t-1}|\}$, где фактор $c_i$ обратно пропорционален выборочному стандартному отклонению $\Delta y_{it}$ и $T_i^{1/2}$:
\[c_i=KT_i^{-1/2}s^{-1}(\Delta y_{it}) \text{ с } s^2(\Delta y_{it})=T^{-1}_i\sum_{t=1}^{T_i}(\Delta y_{it})^2,\]
$K=3$ как приводящее к лучшим результатам на конечных выборках.

\citet{ImPesaran2003} подвергают сомнению хорошие результаты симуляций и асимптотической теории, полученных в \citet{Chang2002}. Хотя \citet{Chang2002} утверждает, что асимптотическая нормальность выполняется и для малых, и для больших $N$, в \citet{ImPesaran2003} указывается, что для этого требуется более сильное условие, что $N\ln T/sqrt{T}\rightarrow0$ при $N,T\rightarrow\infty$. Это происходит потому, что хотя индивидуальные $t$-статистики и являются асимптотически независимыми, дисперсия среднего этих $t$-статистик стремится к ненулевому пределу при $N,T\rightarrow\infty$, который зависит от мешающих параметров. В итоге тест, основанный на $\hat{\rho}_{i,IV}$, будет работать только при очень малых $N$ по сравнению с $T$. Кроме этого, Чанг использует в своих симуляциях достаточно слабую кросс-секционную коррелированность в ошибках, однако, если предположить факторную структуру ошибок, как было сделано в \citet{ImPesaran2003}, тест, основанный на $\hat{\rho}_{i,IV}$, будет иметь существенные искажения размера даже при очень малых $N$ по сравнению с $T$ (например, при $N=5$ и $T=100$). Все это говорит о недопустимости использования теста \citet{Chang2002} на практике (кроме тех случаев, когда он работает хорошо, но это обычно априорно неизвестно для конкретной панели).

С другой стороны, результат работы \citet{Demetrescu2009}, где автор рассматривает тест \citet{Chang2002} в случае авторегрессионного процесса ошибок бесконечного порядка, говорит о более низкой скорости сходимости тестируемого авторегрессионного параметра (нет гарантии, что он даже состоятелен со скоростью $\sqrt{T}$ в случае обычного временного ряда), что влечет недостаточную скорость сходимости оценки дисперсии (возможным решением является оценивание дисперсии при нулевой гипотезе или по ADF-регерссии, а не на основе IV-оценки). Также требуется более сильное предположение об инновациях: они должны быть $i.i.d.$ с дополнительными ограничениями в терминах их характеристических функций.

\citet{ShinKang2006}\footnote{В \citet{SPO2009} предлагается аналогичный тесту \citet{ShinKang2006} тест, основанный на знаках.} разрабатывают новую инструментальную оценку на основе LU разложения ковариационной матрицы $\mathbf{\Omega}^{-1}=\mathbf{\Gamma}\mathbf{\Gamma}^{\prime}$. SUR-оценку можно записать как
\begin{equation}
\hat{\rho}_{SUR}=\frac{\sum_{t=2}^T\mathbf{X}_t^{*\prime}\mathbf{y}_t^*}{\sum_{t=2}^T\mathbf{X}_t^{*\prime}\mathbf{X}_t^*},
\end{equation}
где $\mathbf{X}_t^*=\mathbf{\Gamma} \mathbf{X}_t$, $\mathbf{X}_t=diag(y_{1,t-1},\dots,y_{N,t-1})$, $y_t^*=\mathbf{\Gamma}y_t$, $\mathbf{y}_t=(y_{1t},\dots,y_{Nt})'$,
а соответствующую инструментальную оценку можно записать как
\begin{equation}
\hat{\rho}_{IV}=\frac{\sum_{t=2}^T\mathbf{H}_t^{\prime}\mathbf{y}_t^*}{\sum_{t=2}^T\mathbf{H}_t^{\prime}\mathbf{X}_t^*},
\end{equation}
где  $\mathbf{H}_t=diag(h_{1t},\dots,h_{Nt})$, $h_{it}=h_m(\sigma_{ii}^{-1/2}y_{i,t-1})$, 
\[h_m(x)=
\begin{cases}
1, &\text{если $x>m$}\\
\cfrac{x}{m}, &\text{если $|x|\leq m$}\\
-1, &\text{если $x<-m$}
\end{cases},\]
$\sigma_{ii}$ -- $(i,i)$-й элемент матрицы $\mathbf{\Omega}$ и $m\geq0$ является вещественным числом, таким как $0,1$ или 2. Инструмент $h_{it}$ называется инструментом Хубера (по аналогии с функцией Хубера для робастного оценивания). Функция Хубера уменьшает большие регрессоры до их знаков. Авторы строят статистику Вальда и доказывают, что она сходится к распределению хи-квадрат с $N$ степенями свободы. Скоростью сходимости $\hat{\rho}_{IV}$ является $T$, что отличается от скорости $T^{1/4}$, полученной в \citet{Chang2002} (что, в свою очередь, показывает большую эффективность $\hat{\rho}_{IV}$).
Можно ``хуберизовать'' не только регрессоры, но и зависимую переменную, $\mathbf{y}_t^*$, как было предложено в \citet{ShinPark2010}:
\begin{equation}
\tilde{\rho}_{IV}=\frac{\sum_{t=2}^T\mathbf{H}_t^{\prime}\mathbf{y}_t^{*H}}{\sum_{t=2}^T\mathbf{H}_t^{\prime}\mathbf{X}_t^*},
\end{equation}
где $y_{it}^{*H}=h_l(y_{it}^*)$

Для учета серийной корреляции \citet{ShinKang2006} предлагают очищать от краткосрочной динамики каждый временной ряд в панели (при нулевой гипотезе), а затем на основе полученных панельных остатков вычислять ковариационную матрицу. Тогда все асимптотические результаты будут выполняться. Для учета детерминированной компоненты предлагается рекурсивное детрендирование. Также \citet{ShinKang2006} рассмотрели IV-оценку, основанную на усреднении индивидуальных статистик, как в IPS, а также статистики, основанные на $p$-значениях \citet{Choi2001} для этих IV-статистик. Асимптотика в \citet{ShinKang2006}, однако, также чувствительна к предположению, что $T$ должно быть намного больше, чем $N$, поскольку преобразование основано на оценивании ковариационной матрицы ошибок.

\citet{DemetrescuHanck2012a} обощают подход \citet{ShinKang2006} на случай больших $N$ и факторной структуры ошибок\footnote{В \citet{WWYL2010} предлагается сначала очищать от факторов процесс согласно подходу \citet{BaiNg2004}, а затем строить тест, предложенный в \citet{Chang2002}, который описан выше.}. Асмптотическая нормальность нормированного среднего $t$-статистик достигается при предположении, что $N/T^{1/5}\rightarrow0$ при $N,T\rightarrow\infty$. То есть $T$ снова должно быть намного больше, чем $N$, что является свидетельством вычисления оценок $N(N-1)/2$ ковариаций, и в любом случае $N$ должно быть меньше $T$ для гарантии того, что оценка ковариационной матрицы будет положительно определенной. Для того, чтобы тест хорошо работал при соизмеримых $T$ и $N$, в \citet{DemetrescuHanck2012a} предлагается использовать усеченную оценку ковариационной матрицы вида
\[\mathbf{S}_T=\kappa_{1T}\mathbf{I}+\kappa_{2T}\hat{\mathbf{\Omega}},\]
где $\kappa_{1T}=m_T\cdot b_T^2/d_T^2$, $\kappa_{2T}=a_T^2/d_T^2$, 
$b_T^2=\min(\bar{b}_T^2,d_T^2)$, $a_T^2=d_T^2-b_T^2$, 
$d_T^2=tr[(\hat{\mathbf{\Omega}}-m_T\mathbf{I})(\hat{\mathbf{\Omega}}-m_T\mathbf{I})']/N$, $m_T=tr(\hat{\mathbf{\Omega}})/N$,
\[\bar{b}_T^2=\frac{1}{N}\left[\sum_{t=p+2}^T\left(\frac{\hat{\mathbf{\varepsilon}}'_t\hat{\mathbf{\varepsilon}}_t}{T}\right)^2-\frac{1}{T}tr\left(\hat{\mathbf{\Omega}}\right)\right],\]
где $\hat{\mathbf{\varepsilon}}_t$ -- остатки при нулевой гипотезе (очищенные от краткосрочной динамики). Единичная матрица $\mathbf{I}$ имеет полный ранг, что гарантирует обратимость матрицы $\mathbf{S}_T$ даже при $T<N$. В общем случае неправильно специфицированная, но обратимая матрица $\mathbf{S}_T$, приводит к смещению из-за компоненты $\kappa_{1T}\mathbf{I}$. Но веса $\kappa_{1T}$ и $\kappa_{2T}$ оптимальны в том смысле, что $\mathbf{S}_T$ асимптотически (при $N,T\rightarrow\infty$ совместно) имеет минимальную ожидаемую функцию потерь в классе линейных комбинаций $\mathbf{I}$ и $\hat{\mathbf{\Omega}}$. Скорость сходимости у $\mathbf{S}_T$ та же самая, что и у $\hat{\mathbf{\Omega}}$.

Как и тест IPS, IV-тест имеет нетривиальную мощность в окрестности $1/\sqrt{N}T$ нулевой гипотезы и локально более мощный, чем тест IPS. Авторы предлагают также использовать подход \citet{DHT2006}  для комбинирования нескольких коррелированных друг с другом тестовых статистик. Также допускается безусловная гетероскедастичность ошибок, указывая, что тест, основанный на $\hat{\rho}_{IV}$, устраняет сильную вариабельность лагированных уровней через знаковое ограничение.

В \citet{DemetrescuHanck2012b} предлагается альтернативный подход, в котором вместо процедуры ортогонализации стандартные ошибки для $t$-статистик строятся как робастные стандартные ошибки Уайта (подход \citet{Jonsson2005} приводит к худшим свойствам на конечных выборках), что позволяет сделать тесты более робастными к нестационарной волатильности.

\citet{ChangSong2009} обсуждают недостаток подхода \citet{Chang2002} при наличии кросс-секционной коинтеграции (наличие нестационарных общих факторов). В этом случае для функции $F(\cdot)$ добавляется дополнительное предположение, что $\int_{-\infty}^{\infty}{F_i(x)F_j(x)dx}=0$ для $i=j$, где $F_i(\cdot)$ -- инструментальная функция для $i$-го субъекта. Данное дополнительное предположение позволяет учесть коинтеграцию между субъектами. Тестовая статистика, предложенная авторами, основана на следующей регрессии:
\begin{equation}
y_{it}=\rho_i y_{i,t-1}+\sum_{j=1}^{p_i}{\phi_{i,j}\Delta y_{i,t-j}}+\sum_{j=0}^{q_i}{\beta'_{i,j}w_{i,t-j}}+\varepsilon_{it}
\end{equation}
для $i=1,\dots,N$, где $w_{it}$ интерпретируются как ковариаты (covariates), в качестве которых можно взять $\Delta y_{jt}$ для некоторых $j\neq i$. Это является обоснованным, поскольку такую модель можно представить в виде модели коррекции ошибок для $y_{it}$. Однако проблема выбора ковариат не рассматривается в работе \citet{ChangSong2009}, и включение всех лагированных разностей может стать проблематичным, особенно при больших $N$. На основе этой регрессии строится IV-статистика, аналогичная \eqref{IVChang}, в которой в качестве функции $F(\cdot)$ для $i$-го субъекта используется функция Эрмита нечетного порядка $k=2i-1$, $i=1,\dots,N$:
\[G_k(x)=(2^kk!\sqrt{\pi})^{-1/2}H_k(x)\exp{(-x^2/2)},\]
где $H_k$ является полиномом Эрмита порядка $k$, который задается как
\[H_k(x)=(-1)^k\exp{(x^2)}\frac{d^k}{dx^k}\exp{(-x^2)}.\]
Важно отметить, что использование функций Эрмита разного порядка требует упорядочивания кросс-секционных субъектов. Теоретически упорядочивание не имеет значения, но на конечных выборках результаты могут быть различными. В \citet{ChangSong2009} предлагается упорядочивать объекты согласно их оцененным долгосрочным дисперсиям, стандартизованным на краткосрочные ($i=1$ соответствует наименьшей долгосрочной дисперсии). После этого можно усреднить индивидуальные $t$-статистики аналогично IPS, независимость которых следует из ортогональности инструментальных функций.

\citet{Chang2012} предлагает еще один альтернативный подход, основанный на контуре равных сумм квадратов инструментальных переменных (contour given by the equi-squared-sum of the IV's). Этот контур задается как
\begin{equation}
S_T=\inf_{K\geq 1}\left\{\frac{1}{T\kappa^2(\sqrt{T})}\sum_{t=1}^K{F(y_{t-1})^2}\geq c\right\}
\end{equation}
для некоторой фиксированной константы $c$. Для каждого $T$ величина $S_T$ является временем остановки, для которого сумму квадратов инструментальных переменных достигает уровня $cT\kappa^2(\sqrt{T})$. Таким образом, если мы будем использовать вместо выборки размером $T$ выборку размером $S_T$ для вычисления статистики \citet{Chang2002}, задаваемой выражением \eqref{IVChang}, мы получим то же самое стандартное нормальное предельное распределение. При локальной альтернативе это распределение сдвигается с параметром масштаба, задаваемым как функция уровня контура $c$ и параметра локальности к единице.

Частным случаем является случай, когда $F(x)=sgn(x)$, что приводит к оценке Коши, предложенной в \citet{SoShin1999}. В этом случае $F(y_{t-1})^2\equiv1$ Для всех $t$, и $\kappa(\lambda)=1$. Только с соответствующим значением $c$ мы имеем $S_T=T$.

Данный тест, основанный на контуре, неприменим, если $S_T$ становится больше, чем $T$. Фактически это может произойти с ненулевой вероятностью для любого $c>0$  для всех подходящих  инструментальных функций, отличных от функции знака, если выборка конечная. Для решения этой проблемы предлагается расширить выборку при помощи бутстрапа, порождая последующие наблюдения, $T+1,\dots,S_T$, при нулевой гипотезе на основе бутстраповских остатков. Полученный тест на основе расширенной выборки имеет те же самые асимптотические свойства, что и теоретический тест, если бы мы имели выборку размером $S_T$. Однако мощность такого теста будет ниже, если бы $T$ было бы меньше, чем $S_T$, поскольку расширенная выборка строится при нулевой гипотезе. Обобщение на панельный случай с ненулевой детерминированной компонентой и краткосрочной динамикой ошибок аналогичен работе \citet{Chang2002}. Для практического применения в \citet{Chang2012} в качестве ограничения суммы квадратов инструментальных переменных предлагается использовать значение $cT\kappa^2(\sqrt{T})=cT^2$ с $c=3.5,2.5,1.5,0.5$ для $T=25,50,100,200$, соответственно.

\subsection{Тесты, основанные на методах ресемпелинга}

\citet{Chang2004} рассматривает применение бутстрапа для тестирования панельного единичного корня как против однородной, так и против неоднородной альтернатив.\footnote{Изначально в \citet{MaddalaWu1999} (см. также \citet{CerratoSarantis2007}) предлагается использовать бутстрап с фиксированным кросс-секционным индексом, но авторы не доказывают обоснованность применения бутстрапа.} Автокорреляционная динамика для каждого временного ряда в панели предполагается различной. Авторегрессионный параметр оценивался в системе либо посредством OLS, либо GLS (на основе симуляций авторы заключают, что GLS превосходит OLS). Для получения состоятельной оценки ковариационной матрицы для каждого временного ряда при нулевой гипотезе подбиралась авторегрессия (возможно разного порядка). В итоге строились $F$-статистики для OLS и GLS-оценок авторегрессионных коэффициентов. Поскольку $F$-статистики по своей природе являются двухсторонними, была предложена модификация, разработанная в \citep{Andrews1999}, которая заключается в ограничении сверху оценок авторегрессионных коэффициентов, чтобы не допустить возможность взрывных процессов. Предельные распределения полученных статистик при наличии пространственной корреляции и неоднородной серийной зависимости будут зависеть от мешающих параметров, что затрудняет использование фиксированных критических значений для случая отсутствия такой корреляции. Для получения критических значений автор предлагает использовать обычный остаточный бутстрап с восстановлением авторегрессий (re-colouring) по оцененным коэффициентам для каждого временного ряда в панели. Отметим, что в работе не предполагается факторная структура ошибок.

В \citet{ChoiChue2007} вместо бутстрапа предлагается использовать сабсемплинг для тестов LLC и IPS. Статистики строятся по подвыборкам определенной длины, $\{y_{is},\dots,y_{i,s+b-1}\}_{i=1}^N$, а на основе значений этих статистик строится распределение для конечных выборок. Если длина подвыборки $b\rightarrow\infty$ и $b/T\rightarrow0$ при $T\rightarrow\infty$, то эмпирическое распределение на основе подвыборок аппроксимирует предельное распределение равномерно при фиксированном $N$, так что можно использовать его для тестирования гипотез. В \citet{ChoiChue2007} также предлагают оптимальный выбор длины подвыборки $b$.

\citet{ChoiChue2007} применяют тесты LLC, IPS и обратный нормальный тест \citet{Choi2001} и показывают, что сабсемплинг является робастным даже при перекрестной коинтегрированности между разными объектами в панели. Тесты \citet{Pesaran2007} и \citet{MoonPerron2004}, с другой стороны, не учитывают возможную перекрестную коинтегрированность и имеют искажения размера.

В \citet{PSU2011} предлагается более общий подход, в котором процесс порождения данных аналогичен \citet{BaiNg2004}, но факторы не предполагаются не зависящими от идиосинкразической компоненты. Также предполагается более широкий класс зависимости между идиосинкразическими компонентами (динамическая пространственная зависимость). Отметим, что хотя в \citet{BaiNg2004} предлагается принцип тестирования обоих компонент, факторной и идиосинкразической, на наличие единичного корня, в \citet{PSU2011} тестируется наличие единичного корня в исходном временном ряде.

\citet{PSU2011} рассматривают упрощенные версии статистик LLC и IPS (без кореркции на среднее и дисперсию), а также не учитывают в построении статистик краткосрочную динамику посредством лагов. Последнее не имеет значения, поскольку в любом случае при неизвестном типе кросс-секционной зависимости ошибок невозможно получить тестовую статистику, которая не зависит от мешающих параметров, что дало бы асимптотическое уточнение (asymptotic refinement). Кроме этого авторы используют тесты, основанные на коэффициентах, а не на $t$-статистиках, поскольку наивная стьюдентизация может привести к некоторым проблемам в смысле точности тестов.

\citet{PSU2011} предлагают блочный бутстрап, который является бутстрапом со скользящим окном, предложенным в \citet{Kunsch1989}.

Хотя блочный бутстрап обоснован для очень широкого класса процессов порождения данных, решетчатый (sieve) бутсрап, используемый в \citet{Chang2004}, не будет работать при некоторых достаточно общих предположениях, как показано в \citet{SmeekesUrbain2014}. Причина заключается в том, что хотя и возможно записать систему в виде набора одномерных AR моделей бесконечного порядка, инновации этих уравнений не будут составлять векторный белый шум, что является необходимым для обоснованности решетчатого бутстрапа. На малых выборках решетчатый бутстрап практически неотличим от бутстрапа со скользящими блоками, и только на больших выборках отличия становятся заметными. Причина, по которой в предыдущих исследованиях решетчатый бутстрап работал хорошо -- это специфические комбинации параметров в DGP.

\citet{HwangShin2015} предлагают использовать стационарный бутстрап (см. \citet{PolitisRomano1994}) и доказывают обоснованность предложенного алгоритма, однако, не рассматривают случай кросс-секционной коинтеграции (факторных ошибок). Для учета детерминированной компоненты предлагается использовать рекурсивное детрендирование.

В \citet{HerwartzSiedenburg2008} предлагается использовать дикий бутстрап при предположении о наличии единственного общего фактора или пространственной зависимости в ошибках. Однако авторы предполагают автокорреляцию в ошибках только первого порядка.

\subsection{Определение доли стационарных временных рядов в панели}

Как уже было отмечено выше, отвержение нулевой гипотезы о наличии единичного корня во всех временных рядах в панели не является свидетельством того, что все временные ряды являются стационарными, а только свидетельством того, что существует статистически значимая пропорция стационарных временных рядов. То есть может быть так, что небольшая доля рядов является стационарной, а может быть и все ряды в панели являются стационарными (см. \citet[Fact 2]{WesterlundBreitung2013}). Соответственно, при отвержении нулевой гипотезы исследование должно сопровождаться оценкой доли стационарных временны рядов, и экономическая важность отвержения зависит от величины этой доли (см. \citet{Pesaran2012}). Ряд недавно разработанных тестов рассматривали проблему оценивания доли стационарных временных рядов в панелях. \citet{Pesaran2007b} в контексте тестирования сходимости выпуска и роста рассматривал долю временных рядов, для которых гипотеза единичного корня была отвергнута (или гипотеза стационарности была не отвергнута) на заданном уровне значимости. Хотя отдельные тесты на единичный корень и не являются пространственно независимыми, при нулевой гипотезе о нестационарности всех рядов эта доля будет сходиться к уровню значимости при $N,T\rightarrow\infty$. Хотя мы и имеем результат о частоте отвержения, равной уровню значимости, но если фактическая частота отвержения больше этого уровня значимости, то все равно мы не можем определить, для каких  временных рядов гипотеза о наличии единичного корня отвергалась верно, а для каких нет.

Было предложено ряд подходов для определения доли стационарных и нестационарных временных рядов в панели, а также методы кластеризации этих рядов на две группы. Подходы, основанные на множественном тестировании, широко используемые в статистике, были предложены в работах \citet{Hanck2009}, \citet{DeckerHanck2014} и \citet{MoonPerron2012} в контексте панельных данных.

\citet{Hanck2009} показывает, что при $N=20$ вероятность по крайней мере одного неверного отвержения нулевой гипотезы (Familywise Error Rate, FWER) равна
\[P_{k\geq1}=\sum_{j=1}^20\begin{bmatrix}
20\\
j
\end{bmatrix}0.05^j(1-0.05)^{20-j}=0.6415\]
на основе функции распределения биномиальной случайной величины (для $N=100$ эта величина будет равна 0.994). Наиболее простым способом контроля FWER (чтобы эта величина была $\leq0.05$) является принцип Бонферрони, который, однако, приводит к слишком низкой мощности. \citet{Hanck2009} предлагает адаптацию работы \citet{RomanoWolf2005}. На первом шаге тестовые статистики для каждого отдельного ряда упорядочиваются по возрастанию (чем меньше статистика, тем более она говорит в пользу стационарности ряда). Далее строится совместная доверительная область для оценок коэффициентов в ADF-регрессии, а (общие для всех рядов) квантили выбираются таким образом, чтобы общий уровень накрытия был равен $1-\xi$. Квантили выбираются на основе бутстрапа. Гипотеза единичного корня для конкретного ряда отвергается, если 0 не попадает в доверительную область. Данный подход (асимптотически) контролирует FWER. На этом этапе процедура не останавливается, и можно построить следующую доверительную область для оставшихся не отвергнутыми $N-N_1$ временных рядов. Новая квантиль для доверительного интервала также вычисляется на основе бутстрапа. Процедура продолжается до тех пор, пока остается невозможным отвергнуть  гипотезу ни для одного ряда.

Подходы \citet{DeckerHanck2014} и \citet{MoonPerron2012} основаны на ложном коэффициенте обнаружения (False Discovery Rate, FDR), который является более слабым условием, чем FWER (и лучше применяется в случае большого количества тестов) (см.  \citet{BenjaminiHochberg1995}). FDR определяется как ожидаемое значение числа ложно отвергнутых гипотез, деленное на общее число отвержений, что является обобщением ошибки I рода в ситуации множественного тестирования. В \citet{DeckerHanck2014} предлагается использовать две  процедуры, предложные в \citet{BenjaminiHochberg1995} и \citet{RSW2008}. Метод \citet{BenjaminiHochberg1995} основан на упорядочивании $p$-значений по возрастанию, а уровни значимости для каждого упорядоченного ряда выбираются равными $\xi_j=(j/N)\xi$. Статистики таким образом упорядочиваются от наиболее значимой к наименее значимой, и процедура начинается с тестирования наименее значимой статистики (см. также \citet{Hanck2013}. \citet{STS2004} предлагают менее консервативную процедуру, в которой в знаменателе $\xi_j$ участвует не $N$, а истинное число $N_0$ верных нулевых гипотез (см. \citep{MoonPerron2012}, где число $N_0$ оценивается согласно методу \citep{Ng2008}). Метод \citep{RSW2008} наоборот тестирует статистики с наиболее значимой к наименее значимой, а критические значения вычисляются на основе бутстрапа последовательно для каждого шага после отвержения соответствующей гипотезы. 

В работах \citet{Kapetanios2003} и \citet{ChortareasKapetanios2009} предлагается метод разделения временных рядов в панели на стационарные и нестационарные (sequential panel selection method, SPSM). На первом шаге проверяется гипотеза о панельном единичном корне, и если данная гипотеза не отвергается, процедура останавливается. В противном случае, если гипотеза отвергается, строятся тестовые статистики для каждого временного ряда (например, ADF), выбирается минимальная из них (минимальная для левосторонних тестов), и ряд, соответствующий этой минимальной статистике, удаляется из панели и снова тестируется гипотеза о панельном единичном корне. Процедура не останавливается до тех пор, пока гипотеза о панельном единичном корне будет отвергаться. Доказывается состоятельность полученной доли стационарных (или нестационарных) временных рядов в панели при $N/T\rightarrow0$ и некоторых дополнительных условиях. В \citep{Kapetanios2003} также предлагается тест для проверки гипотезы о том, что существует кластер временных рядов с одинаковыми коэффициентами при экзогенных переменных, а также аналогичная последовательная процедура для определения временных рядов, входящих в такой кластер.

Похожая идея была предложена в \citet{Smeekes2014}, где разрабатывается бутстраповский последовательный квантильный тест (bootstrap sequential quantile test, BSQT). Пусть $k_0=0,1,\dots,N$ -- число стационарных временных рядов, а  $q_0=k_0/N$ -- соответствующая пропорция этих стационарных рядов. Более формально, пусть $\mathcal{S}=\{i\in \mathcal{N}_N:|\rho_i|<1\}$ -- множество $k_0$ стационарных рядов, так что $k_0=|\mathcal{S}|$, где $|\cdot|$ обозначает количество элементов множества. Аналогично множество $N-k_0$ нестационарных рядов обозначается как $\mathcal{U}=\{i\in \mathcal{N}_N:\rho_i=1\}$.

Пусть $0<q_1<\dots<q_r<1$ -- множество из специфицированных долей стационарных рядов, которые нужно последовательно тестировать. \citet{Smeekes2014} тестируетт нулевую гипотезу $H_0(q_j):|\mathcal{S}|=k_j=[q_jN]$ (что пропорция стационарных рядов равна $q_j$) против альтернативы $H_1(q_{j+1}):|\mathcal{S}|\geq k_{j+1}$ (что по крайней мере пропорция $q_{j+1}$ рядов стационарна). Эта нулевая гипотеза проверяется последовательно для $j=1,\dots,r$, останавливаясь только тогда, когда $H_0(q_j)$ не будет отвергаться. Пусть $\tau(q_j,q-{j+1})$ -- тестовая статистика для проверки нулевой гипотезы $H_0(q_j)$ против альтернативы $H_1(q_{j+1})$. Также пусть $\theta_i$ -- некоторый состоятельный тест на единичный корень для конкретного ряда $i$. В общем случае, асимптотически тесты для разных $i$-ых рядов могут быть коррелированными из-за пространственной зависимости в данных. Пусть $\theta_{(1)},\dots,\theta_{(N)}$ -- порядковые статистики $\theta_1,\dots,\theta_N$, упорядоченные по возрастанию. Тогда тестовой статистикой $\tau(q_j,q-{j+1})$ будет порядковая статистика, соответствующая альтернативе:
\[\tau(q_j,q_{j+1})=\theta_{(k_{j+1})}=\theta_{([q_{j+1}N])}.\]
Из-за зависимости между статистиками для разных объектов предлагается использовать процедуру блочного бутстрапа для получения критических значений. Процедура блочного бутстрапа аналогична \citep{PSU2011} и обоснована для очень широкого класса DGP, включая факторную динамику ошибок и пространственную коинтеграцию.

Важно отметить, что если мы отвергаем гипотезу $H_0(q_j)$ против $H_0(q_{j+1})$, то мы только можем сказать, что $q_j$ рядов стационарны, но ничего не можем сказать о $q_{j+1}$. То есть, если $\hat{q}=q_j$, то это можно интерпретировать только как то, что доля стационарных рядов -- $q_{j-1}$, а нестационарных -- $q_{j+1}$. Другими словами, действительная пропорция стационарных рядов должна лежать в интервале $(q_{j-1},q_{j+1})$. Поэтому данные квантили, $q_1,\dots,q_r$, должны лежать близко друг к другу для более точного результата. С другой стороны, если они лежат далеко друг от друга, это увеличивает мощность процедуры из-за объединения групп рядов, то есть на каждом шаге метод использует кросс-секционную информацию тем же самым способом, что и стандартные тесты на панельные единичные корни. Однако \citet{Smeekes2014} не дает практической рекомендации, как выбрать разбиение $q_j$, указывая, что это зависит от многих факторов, таких как величина $N$ и $T$, конкретная эмпирическая задача и так далее. Можно было бы рассмотреть частный случай, полагая $q_j=(j-1)/N$, $j=1,\dots,N$, что позволяет тестировать каждый временной ряд в панели последовательно, однако такая постановка будет иметь фундаментально иную интерпретацию и будет обоснована для малых $N$. Также отметим, что подход BSQT асимптотически обоснован только относительно $T\rightarrow\infty$, но $N$ предполагается фиксированным.

По сравнению подхода BSQT с подходом SPSM, последний основан на тесте на панельный единичный корень и может иметь низкую мощность, если малое количество рядов в панели являются стационарными.

Как отмечает \citet{Smeekes2014}, подход \citet{MoonPerron2012} слишком затратный с точки зрения вычислительного времени по сравнению с подходами BSQT и SPSM. \citet{Smeekes2014} сравнивает различные методы на конечных выборках. Резюмируя, BSQT работает лучше в панелях с большим $N$, даже если $T$ мало. В панелях с большими $T$ и $N$ метод BSQT работает почти также, как метод MP. В панелях с малым $T$, но большим $N$, метод BSQT работает относительно хорошо, в то время как метод MP значительно ухудшается. Также BSQT работает лучше, чем SPSM. Метод \citet{Ng2008}, который будет описан далее, доминируется всеми остальными методами. Если отношение $T/N$ мало, лучше всего использовать  метод BSQT, в противном случае лучше использовать метод MP, хотя BSQT также работает хорошо.

Иной подход тестирования доли стационарных временных рядов в панели был предложен в \citet{Ng2008}. Пусть доля \textit{не}стационарных рядов обозначается как $\theta=\in(0,1]$. Подход \citet{Ng2008} отличается от других тестов на панельный единичный корень, поскольку агрегирует данные, а не сводит модель к пулу. \citet{Ng2008}  показывает, что для процесса $y_{it}$ с фиксированными эффектами $\mu_i$ популяционная (для $N\rightarrow\infty$) дисперсия $V_{t,\infty}=Var_i(y_{it})$ будет примерно равной $E(\mu_i)+\theta\cdot t+c$, где константа $c>0$. То есть в генеральной совокупности, состоящей из смеси стационарных и нестационарных рядов, кросс-секционная дисперсия будет иметь компоненту тренда, которая растет с темпом $\theta$, долей нестационарных рядов. Из этого следует, что
\[\Delta V_{t,\infty}=V_{t,\infty}-V_{t-1,\infty}=\theta.\]
Взятие разности предотвращает возможность наличия ложной регрессии для дисперсии в предстоящем оценивании. Для оценивания $\theta$ сначала вычисляется дисперсия
\[V_{t,N}=\frac{1}{N}\sum_{i=1}^N{(y_{it}-Y_{t,N})^2},\]
где 
\[Y_{t,N}=\frac{1}{N}\sum_{i=1}^N{y_{it}}.\]
Тогда оценка для $\theta$ будет иметь вид
\[\hat{\theta}=\frac{1}{T}\sum_{t=1}^T{\Delta V_{tN}}.\]
При $N\rightarrow\infty$, следующей за $T\rightarrow\infty$
\[\sqrt{N}(\hat{\theta}-\theta)\Rightarrow N(0,2\theta),\]
так что предельное распределение разрывно в $\theta=0$, поскольку при стационарности всех рядов дисперсия не будет расти. Отметим, что можно тестировать гипотезу $\theta=1$, что эквивалентно тестированию гипотезы о панельном единичном корне. \citet{Ng2008} также обобщает модель на случай серийной корреляции.

При наличии индивидуальных трендов с коэффициентами $\beta_i$  \citet{Ng2008} предлагает добавлять в регрессию разность квадратичного тренда, то есть
\[\Delta \hat{V}_{t,N}=\theta+\beta\Delta t^2+eta_{t,N}.\]
Однако в этом случае даже для больших выборок асимптотическая аппроксимация получается не очень хорошей. В \citet{Westerlund2014c} производится корректировка смещения оценки $\hat{\theta}$, а также анализируется локальная мощность и одновременная (а не последовательная, как в \citet{Ng2008}) асимптотика по $N$ и $T$. Вследствие рассмотрения такой одновременной асимптотики некоторые компоненты не затухают, и оценка будет отличаться от полученной в \citet{Ng2008}. 

\citet{Westerlund2014d} также рассматривает теоретическое обоснование плохого поведения теста \citep{Ng2008} на конечных выборках на основе совместной асимптотики и предлагаются скорректированные на смещение тесты для двух случаев: фиксированный $T\geq2$ и $T\rightarrow\infty$, оба при $N\rightarrow\infty$.

\citet{PVWW2015} разрабатывает подход, больше похожий на подходы для многомерных временных рядов, нежели на подходы по панельным единичным корням, поскольку $N$ предполагается фиксированным. Теория близко связана с теорией на определение ранга коинтеграции в многомерных временных рядах (общие стохастические тренды -- общие I(1) факторы). Авторы допускают широкий спектр процессов порождения данных, включая перекрестную коинтегрированность посредством модели общих (нестационарных) факторов. Предложенный тест на ранг является тестом отношения дисперсий, не зависящий (асимптотически) от мешающих параметров. Нулевой гипотезой является $H_0: 0\leq c\leq N$, где $c$ -- число общих стохастических трендов, а альтернативной -- $H_1: c_1\leq c$. Последовательная процедура заключается в следующем: на первом шаге тестируется гипотеза $c=N$. Если эта гипотеза не отвергается, то все временные ряды в панели I(1), и нет перекрестной коинтеграции, процедура останавливается. В противном случае тестируется гипотеза $c=N-1$, основанная на всех, не не самом большом собственном значении некоторой матрицы. Процедура продолжается до тех пор, последовательно удаляя самые большие собственные значения, пока гипотеза не будет не отвергаться, или пока не будет определяться равный нулю ранг.

\subsection{Другие методы}

В данном разделе мы опишем те подходы, которые по тем или иным причинам нельзя отнести к предыдущим разделам.

\citet{Lopez2009} предлагает тест на единичный корень в панели, основанный на GLS-детрендировании, аналогичный \citep{ERS1996} для временных рядов. На первом шаге каждый временной ряд детрендируется согласно следующей схеме. Пусть $\bar{\rho}=1+\bar{c}/(N^\kappa T)$, где $\kappa=1/2$ и $\bar{c}=-7$ для модели с фиксированными эффектами, и $\kappa=1/4$  и $\bar{c}=-13.5$ для модели с индивидуально специфическими трендами (см. \citet{MPP2007}). Тогда данные GLS-детрендируются как $y^{\bar{c}}_{i1}=y_{i1}$, $y^{\bar{c}}_{it}=y_{it}-\bar{\rho}y_{i,t-1}$ и $z^{\bar{c}}_{1}=z_{1}$, $z^{\bar{c}}_{t}=z_{t}-\bar{\rho}z_{t-1}$, где $z=1$ или $z=(1,t)'$. Отметим, что использование первого наблюдения при (квази) GLS-детрендировании является ключевым фактором для получения невырожденного предельного распределения статистик (см. \citet{Westerlund2015a}). После этого оценивается регрессия $y^{\bar{c}}_{it}$ на $z^{\bar{c}}_{t}$ для каждого $i=1,\dots,N$, и на основе оцененного коэффициента $\hat{\beta}$ строятся GLS-детрендированные ряды $y_{it}^d=y_{it}-\hat{\beta}z_{t}$. После этого система уравнений оценивается как SUR с заранее специфицированным лагом для каждого уравнения (выбранным, например, через MAIC) с ограничением, что все тестируемые коэффициенты одинаковы в каждом уравнении (то есть тестируется гипотеза единичного корня против однородной альтернативы). Автор предлагает вычислять критические значения при помощи решетчатого бутстрапа на основе остатков при нулевой гипотезе. 

\citet{SJO2008} разрабатывают тесты на наличие двух единичных корней с различными методами дефакторизации. 

В \citet{BMW2002} оценивается системы из авторегрессионных уравнений, используя SUR, и для каждого временного ряда строится тестовая статистика и тестируется гипотеза о наличии единичного корня. Полученный тест имеет более высокую мощность, чем обычный тест Дики-Фуллера, но предельное распределение отличается от распределения Дики-Фуллера. Однако предельное распределение зависит от степени корреляции между ошибками в уравнениях. 

В \citet{CostantiniLupi2013} и \citet{Westerlund2015b} обобщается подход \citet{Hansen1995} на панельный случай. Напомним, что подход \citet{Hansen1995} заключается в том, что если временной ряд коррелирован с какой-то внешней переменной, ковариатой (covariate), то можно включить эту переменную в регрессию и получить более высокую мощность теста при ненулевой корреляции. \citet{CostantiniLupi2013} рассматривают неоднородную альтернативу, и предложенный тест основан на комбинации $p$-значений, как было предложено в \citet{Choi2001} и \citet{DHT2006} для учета кросс-секционной зависимости. В \citet{Westerlund2015b} рассматривалась более общая ситуация, в которой каждый из временных рядов в панели может зависеть от своего множества внешних ковариат, однако рассматривается только тест против однородной альтернативы. Также \citet{Westerlund2015b}  анализирует асимптотическую локальную мощность теста и обнаруживает, что ковариаты могу компенсировать проблему потери мощности в случае наличия индивидуальных (случайных, incidental) трендов. Полученный тест, как оказалось, является единственным, кто имеет мощность в окрестности $1/\sqrt{N}T$ нулевой гипотезы, в отличие от всех других тестов, которые имеют локальную мощность только в окрестности $1/N^{1/4}T$. В случае пространственной корреляции в ошибках автор предлагает дефакторизовать временные ряды любым из способов, описанных ранее. В \citet{HKY2015} предлагается процедура, в которой на первом шаге тестируется гипотеза о панельном единичном корне, и в случае, если она отвергается (что говорит о некоторой доле стационарных временных рядов в панели), конкретный временной ряд тестируется на наличие единичного корня, используя остальные ряды в качестве ковариат (в уровнях или в первых разностях в зависимости от наличия или отсутствия единичного корня на основе одномерных тестов). Для выбора количества ковариат авторы предлагают три метода: на основе асимптотической локальной мощности, на основе скорректированного $R^2$ и на основе общего фактора как ковариаты. \citet{juodis2019optimal} получают результаты по асимптотической оптимальности тестов с ковариатами для различных типов детерминированной компоненты. 

\citet{ReeseWesterlund2015} объединяют подходы, основанные на добавлении кросс-секционных средних (CA) для аппроксимации факторов (\citet{Pesaran2007} и \citet{PSY2013}) с подходами, основанными на удалении факторов (PANIC) (\citet{BaiNg2004,BaiNg2010}), поскольку оба этих подхода являются наиболее популярными в тестировании гипотезы о наличии панельного единичного корня. Комбинацию этих подходов авторы называют PANICCA (PANIC + CA). В \citep{ReeseWesterlund2015} предлагается использовать подход PANIC, но не к оцененным компонентам по методу главных компонент, а к оцененным компонентам по методу кросс-секционного усреднения. Более точно, путь данные порождаются как
\begin{eqnarray}
y_{it}&=&\mathbf{\alpha}_i^{\prime}\mathbf{D}_{t}+\mathbf{\lambda}_i\mathbf{F}_t+e_{it}\\
\mathbf{x}_{it}&=&\mathbf{\beta}_i^{\prime}\mathbf{D}_{t}+\mathbf{\Lambda}_i^{\prime}\mathbf{F}_t+\mathbf{u}_{it},
\end{eqnarray}
где $\mathbf{D}_{t}$ является детерминированной компонентой, $\mathbf{x}_{it}$ является вектором $(m\times1)$ дополнительных переменных, а остальные переменные и параметры определяются как и ранее. Определяя вектор $\mathbf{z}_{it}=(y_{it},\mathbf{x}_{it}^{\prime})^{\prime}$, запишем эти уравнения в векторном виде как
\begin{equation}
\mathbf{z}_{it}=\mathbf{B}^{\prime}_i\mathbf{D}_{t}+\mathbf{C}_{i}^{\prime}\mathbf{F}_t+\mathbf{v}_{it},
\end{equation}
где мы используем соответствующие переобозначения. Запишем теперь данное уравнение в первых разностях, меняя размерность $\mathbf{D}_{t}$:
\begin{equation}
\Delta\mathbf{z}_{it}=\mathbf{b}^{\prime}_i\Delta\mathbf{D}_{t}+\mathbf{C}_{i}^{\prime}\Delta\mathbf{F}_t+\Delta\mathbf{v}_{it},
\end{equation}
или, детрендировав каждую из компонент (очистив от $\Delta\mathbf{D}_{t}$),
\begin{equation}
(\Delta\mathbf{z}_{it})^d=\mathbf{C}_{i}^{\prime}(\Delta\mathbf{F}_t)^d+(\Delta\mathbf{v}_{it})^d.
\end{equation}
Поскольку $\mathbf{C}_{i}$ и $(\Delta\mathbf{F}_t)^d$ не идентифицируются по отдельности, вместо оценки методом главных компонент для $(\Delta\mathbf{F}_t)^d$ предлагается CA-оценка $\overline{(\Delta\mathbf{z}_{it})}$, являющаяся средним $(\Delta\mathbf{z}_{it})$. Затем оценка для $\mathbf{C}_{i}$ находится методом наименьших квадратов. Для оценки $\mathbf{F}_t$ предлагается брать накопленную сумму соответствующей разности, как в \citet{BaiNg2004}. В отличие от работ Песарана, здесь средние интерпретируются как не прокси для факторов, а как сами оцененные факторы (см. Remark 2 \citet{ReeseWesterlund2015}). Далее можно тестировать на наличие единичного корня отдельно факторную компоненту и идиосинкразическую компоненту, как в \citet{BaiNg2004}.

\citet{HMS2014} тестируют наличие единичного корня в панели на основе многомерной регрессии и многомерного линейного теста при наличии кросс-секционной коррелированности. 

\citet{Solberger2014} предлагает тест отношения правдоподобий в случае наличия (возможно) интегрированных общих факторов.

Поскольку экономические и финансовые данные часто наблюдаются с некоторой периодичностью, возникает необходимость тестировать наличие сезонных единичных корней, то есть единичных корней на различных частотах. Периодичность может быть как квартальная и месячная, так и дневная и внутридневная. Соответственно, наличие этой периодичности необходимо учитывать при построении модели, а также в предварительном анализе, связанным с тестированием наличия единичных корней в панельных данных. Тестирование наличия сезонных единчных корней в панельных данных было рассмотрено в \citet{LeeShin2006} с  использованием подхода \citet{ShinKang2006}, основанного на инструментальных переменных. Были предложены тесты как на основе модели пула, так и на основе комбинации $p$-значений. Авторы получают классические асимптотические распределения (нормальное и хи-квадрат) для рассматриваемых тестов при фиксированном $N$ и $T\rightarrow\infty$.

Однако данный подход имеет существенный недостаток, который заключается в том, что тестируется наличие единичного корня на всех частотах, а не на какой-то одной конкретной частоте. Для этого \citet{OSG2005} обобщают подход HEGY \citet{HEGY1990} на случай панельных данных. При наличии кросс-секционной корреляции в \citet{OSG2007} предлагается дополнять панельную регрессию HEGY кросс-секционными средними лагированных уровней и первых разностей индивидуальных временных рядов в панели. В качестве альтернативы можно использовать бутстрап. \citep{KunstFranses2011} рассматрвиали проблема тестирования для месячных данных и предлагаются непараметрические тесты.

\subsection{Сравнение тестов}

Исходя из большого количества разработанных тестов на панельный единичный корень, которые были описаны выше, было проведено большое количество исследований на основе симуляций Монте-Карло, чтобы сравнить, какие из этих тестов работают лучше или хуже в различных процессах порождения данных. 

В \citet{JangShin2005} сравнивались тесты \citet{PhillipsSul2003} (далее PS), \citet{BaiNg2004} (далее BN) и \citet{MoonPerron2004} (далее MP). Вспомним, что PS и MP корректируют кросс-секционную зависимость на основе проектирующей матрицы, которая сокращает факторы, в то время как BN удаляет факторы непосредственно путем их вычитания. Также PS и BN комбинируют тесты о $\rho_i$ путем их усреднения, в то время как MP основан на регрессии пула. В \citet{JangShin2005} данные подходы комбинируются с различными методами детрендирования, такими как OLS, GLS, рекурсивное детрендирование, WSLS и безусловный метод максимального правдоподобия. 
Таким образом, \citet{JangShin2005} рассматривают все возможные комбинации методов детрендирования, дефакторизации и комбинирования результатов о коэффициентах $\rho_i$. На основе результатов сравнения размера и мощности авторы рекомендуют использовать OLS-детрендирование (из-за простоты реализации) и метод BN для очистки от факторов. Метод BN дает более стабильный размер, особенно при малых $T$. Проекция для очистки от факторов и оценивание пула приводит к очень плохому размеру, а усреднение индивидуальных тестов дает лучший размер, чем оценивание регрессии пула. Из двух тестов, основанных на усреднении или на регрессии пула, первый имеет лучший размер, одновременно не теряя сильно в мощности по сравнению с последним тестом. 

Независимо \citet{Gutierrez2006} сравнивают те же самые тесты, PS, BN и MN, в дополнение к тесту \citet{Choi2006}. \citet{Gutierrez2006} заключает, что тест MP имеет хорошие размер и мощность при различных спецификациях и различных $T$ и $N$. Тест BN, основанный на регрессии пула, имеет хорошие свойства при GLS-детрендировании, а при OLS-детрендировании имеет низкую мощность. Тест \citet{Choi2006} имеет сильные либеральные искажения размера. Тест PS показывает искажения размера, если имеется более одного общего фактора. Искажения размера часто происходят из-за переоценки количества факторов при малых $N\leq15$ для тестов BN и MP. При наличии трендов все тесты теряют в мощности.

Снова независимо от рассматриваемых работ в работе \citet{GPU2010}сравниваются тесты \citep{Pesaran2007}, MP, BN, а также тесты \citet{BreitungDas2008} и \citet{Sul2009}, которые не учитывают факторную структуру. В \citet{GPU2010} обсуждается различие в процессах порождения данных, лежащих в основе каждого из тестов, а также различия в нулевых гипотезах. Симуляции проводятся также для различных видов факторных процессов порождения данных (с одним или двумя факторами), а факторы и ошибки порождаются как MA(1) процессы. Также рассматривается несколько вариантов того, какая из компонент, факторы или идиосинкразическая ошибка (или оба) имеют единичный корень. Для устранения влияния фиксированных эффектов на тестовые статистики из данных вычитается начальное наблюдение, как в \citet{BreitungDas2008}. \citet{GPU2010} получают следующие результаты. Во-первых, если нестационарность приходится только на общие факторы, но не на идиосинкразическую ошибку, то такую структуру данных могут обнаружить только тесты \citet{BaiNg2004} и \citet{Sul2009}. 
Во-вторых, тест на наличие единичного корня в единственном общем факторе имеет низкую мощность. Данная картина наблюдается и в случае нескольких факторов. В-третьих, при тестировании на наличие единичного корня в идиосинкразической компоненте тест $CIPS$ на основе модели пула имеет лучшую мощность, чем индивидуальные $CADF$ тесты (хотя, поскольку тесты были основаны на работе \citet{Pesaran2007}, они не учитывали возможность наличия более одного фактора). Аналогичный эффект наблюдается и для тестов BN, хотя тест BN, основанный на модели пула, имеет существенные искажения размера при малой временной размерности панели. Тесты MP показывают аналогичные свойства, также как и тесты \citet{BreitungDas2008} и \citet{Sul2009}, хотя последние только при $N<T$. Также \citet{GPU2010} указывают на проблему выбора количества факторов для тестов BN и MP, и данные тесты показывают искажения размера, если число факторов неправильно специфицировано.

В работе \citet{SHT2009} рассматривается аналогичная проблема сравнения различных тестов, но для более широкого диапазона возможных процессов порождения данных (в том числе наличие нескольких факторов, ненормальный процесс ошибок, стохастический единичный корень), однако, авторы сравнивали только тесты BN, MP и \citet{Pesaran2007}. Ключевое отличие от предыдущих работ заключается в анализе большого количества методов выбора количества факторов (предложенных в \citet{BaiNg2002}). Кроме того, \citet{SHT2009} предлагают ряд новых методов, основанных на критерии Хеннана-Куинна.

На основе симуляций \citet{SHT2009} делают вывод, что критерии Хеннана-Куинна 
улучшают свойства на конечных выборках, особенно при малых $N$ (и $T>50$), хотя если и $T$ достаточно мало, ни один из критериев не работает достаточно хорошо. 

Сравнивая тесты на панельный единичный корень, \citet{SHT2009} заключают, что тест Фишера $P_{\hat{e}}$ имеет хорошие свойства размера и мощности при широком классе процессов порождения данных. Тест $Z_{\hat{e}}$ в \citet{BaiNg2004} также имеет хороший размер, но его скорректированный аналог $Z_{\hat{e}}^{+}$ в \citet{WesterlundLarsson2009} редко отвергает нулевую гипотезу.

\section{Тестирование на стационарность для пространственно-коррелированных панелей}

Аналогично случаю с пространственно некоррелированными ошибками, можно рассмотреть аналоги тестов на стационарность для пространственно коррелированных панелей. \citet{BaiNg2005} используют метод дефакторизации, разработанный в \citet{BaiNg2004}. \citet{BaiNg2005} очищают от факторов временной ряд, а затем применяют тест Фишера, основанный на $p$-значениях KPSS-статистик к идиосинкразической ошибке. KPSS-тест можно также применить к оцененным факторам, а не только к идиосинкразической ошибке (аналогично \citet{Hadri2000}). В \citet{ShinSnell2006} утверждается, что вычитание кросс-секционных средних до вычисления KPSS-статистик асимптотически приводит к инвариантности относительно пространственной корреляции. Однако \citet{Jonsson2011} показывает, что при малых $N$ такой подход приводит к существенным искажениям размера, что можно решить путем использования критических значений для конечных выборок (автор получает их путем анализа поверхности отклика). В \citet{HadriKurozumi2012} учет общих факторов производится аналогично \citet{Pesaran2007}. \citet{HadriKurozumi2012} также рассматривают различные методы учета слабой зависимости ошибок, такие как \citet{SPC2005} и \citet{TodaYamamoto1995}. 

\citet{DHT2010} разрабатывают тест на стационарность, учитывающий пространственную корреляцию в ошибках. Норма долгосрочной корреляционной матрицы предполагается неограниченной при $N\rightarrow\infty$. Запишем процесс порождения данных как в \citet{Hadri2000}, 
\begin{eqnarray}
y_{it}&=&u_{it}+\varepsilon_{it},\label{CSCStat1}\\
u_{it}&=&u_{i,t-1}+v_{it},\label{CSCStat2}
\end{eqnarray}
где $\varepsilon_{it}$ являются коррелированными с долгосрочной ковариационной матрицей $\mathbf{\Omega}$ и соответствующей корреляционной матрицей $\mathbf{\Xi}$. Определим
\[S_{it}=\hat{\omega}_{ii}^{-1/2}\sum_{j=1}^t{y_{ij}}=\sum_{j=1}^t{\tilde{y}_{ij}},\]
где $\hat{\omega}_{ii}$ -- состоятельная оценка ${\omega}_{ii}$, $(i,i)$-й элемент матрицы $\mathbf{\Omega}$. Пусть $\mathbf{S}_t=(S-{1t},\dots,S_{Nt})'$. Тогда, используя тот факт, что
\[\frac{1}{\sqrt{T}}\mathbf{S}_{\lfloor rT\rfloor}\Rightarrow\mathbf{\Xi}^{1/2}\mathbf{W}(r),\]
статистика следа, предельное распределение которой не будет зависеть от мешающих параметров, будет иметь вид
\begin{equation}\label{CSCStat3}
\kappa_{\bot}=\frac{1}{T^2}\sum_{t=1}^T\mathbf{S}^{\prime}_t\hat{\mathbf{\Xi}}^{-1}\mathbf{S}_t\Rightarrow\int_0^1{\mathbf{W}(r)^{\prime}\mathbf{W}(r)dr}.
\end{equation}
\citet{DHT2010} рассматривают случай, когда недиагональные элементы корреляционной матрицы $\mathbf{\Xi}$ одинаковые (хотя процедура также хорошо работает при различной корреляции, что было показано на симуляциях). \citet{DHT2010} строят ненормализованную статистику $T^{-2}\sum_{t=1}^T{\mathbf{S}^{\prime}_t}\mathbf{S}_t$ и 
при заданном предположении приводят модифицированную статистику, основанную на коэффициенте корреляции. Полученная статистика будет иметь распределение Крамера-фон Мизеса, как и в одномерном случае для статистики KPSS. Также предлагается процедура состоятельного оценивания величины корреляции. При наличии детерминированной компоненты предлагается использовать рекурсивное детрендирование с изменением тестовых статистик.

\citet{HLM2005} предлагают тестовую статистику при $T\rightarrow\infty$ и фиксированном $N$ в модели
\begin{equation}\label{StatCSD1}
y_{it}=\rho_i y_{i,t-1}+\varepsilon_{it},
\end{equation}
где нулевая гипотеза имеет вид
\[H_0:|\rho_i|<1 \ \text{для всех $i$},\]
а альтернативная записывается как
\[H_0:\rho_i=1 \ \text{для по крайней мере одного $i$}.\]
Тестовая статистика, предложенная \citet{HLM2005}, имеет вид
\begin{equation}\label{StatCSD2}
S_k=\frac{C_k}{\hat{\omega}(a_{k,t})},
\end{equation}
где $C_k=T^{-1/2}\sum_{t=k+1}^Ta_{k,t}$, $a_{k,t}=\sum_{i=1}^N{y_{it}y_{i,t-k}}$, а $\hat{\omega}(a_{k,t})$ -- оценка долгосрочной дисперсии процесса $a_{k,t}$. О статистике $S_k$ можно думать как о стандартизованном среднем ряда $a_{k,t}$, деленном на его долгосрочное стандартное отклонение. При нулевой гипотезе статистика $S_k$ имеет стандартное нормальное распределение и расходится при альтернативе. Основное требование, чтобы $k=\lfloor (\delta T)^{1/2}\rfloor$ для некоторой константы $\delta>0$. В противном случае, если мы запишем $C_k$ как $C_k=\sum_{i=1}^NC_{i,k}$, где
\[C_{i,k}=\frac{1}{(T-k)^{1/2}}\sum_{t=k+1}^T{y_{it}y_{i,t-k}},\] 
и положим, например, $k=1$, то математическое ожидание $C_{i,k}$ будет зависеть от $\rho_i$ (и не будет сходиться к нулю), так что асимптотическая нормальность не будет выполняться. Для практического применения тестовой статистики \citet{HLM2005} предлагают использовать $k=\lfloor \sqrt{3T}\rfloor$. Тест \eqref{StatCSD2} допускает также пространственную корреляцию в ошибках.

При наличии детерминированной компоненты обоснованным асимптотически тестом будет тот же тест, построенный на основе детрендированных и стандартизировнаных остатках. Однако при усреднении по $N$ на конечных выборках может возникнуть существенное смещение, и  \citet{HLM2005} предлагают использовать статистику, скорректированную на смещение для конечных выборок:
\begin{equation}\label{StatCSD3}
\tilde{S}_k=\frac{\tilde{C}_k+\tilde{c}}{\hat{\omega}(\tilde{a}_{k,t})},
\end{equation}
где $\tilde{c}=(T-k)^{-1/2}\sum_{i=1}^N\tilde{c}_i$,
\[\tilde{c}_i=tr\left[\left(T^{-1}\sum_{t=1}^T\mathbf{x}_{it}\mathbf{x}_{it}^{\prime}\right)\hat{\mathbf{\Omega}}\{\mathbf{x}_{it}\tilde{y}_{it}\}\right],\]
$\hat{\mathbf{\Omega}}$ -- оценка ковариационной матрицы $\mathbf{x}_{it}\tilde{y}_{it}$, а остальные элементы статистики с тильдами обозначают соответствующие детрендированные аналоги. 

Если ошибки имеют $\varepsilon_{it}$ факторную структуру, то тест $\tilde{S}_k$ все еще сохраняет свои асимптотические свойства. Однако можно использовать подход \citet{BaiNg2004} для дефакторизации, а затем построить статистику $\tilde{S}_k$ (обозначаемую как $\tilde{S}_k^F$), которая будет иметь лучшие свойства на конечных выборках, даже учитывая тот факт, что оцененные факторы не будут состоятельными при предположении о фиксированном $N$. Оба теста работают хорошо при $N\leq40$. Тест $\tilde{S}_k^F$ имеет существенно более высокую мощность по сравнению с $\tilde{S}_k$, но немного худший размер при наличии кросс-корреляции. Тесты \citet{BaiNg2005} имеют плохой размер при наличии сильно автокоррелированной идиосинкразической  компоненты и сильно чувствительны к неправильной факторной спецификации. Симуляции \citet{HadriKurozumi2012}  показывают, что в случае фиксированных эффектов тест \citet{HLM2005} более или менее контролирует размер (хотя ни один из рассмотренных тестов не превосходит другой в терминах размера и мощности), а предложенный в \citet{HadriKurozumi2012}  тест с коррекцией долгосрочной дисперсии согласно \citet{SPC2005} лучше всего работает в случае трендов. Отметим, что при малых $N$ и пространственной коррелированности ошибок можно также использовать многомерный тест на  стационарность, предложенный в \citet{ChoiAhn1999}.

Использование бутстарпа или сабсемплинга также оправдано для учета кросс-секционной корреляции, см., например, \citet{GOS2009}, где был применен подход \citet{MaddalaWu1999}.

В \citet{LeeShin2009} пространственная корреляция учитывается через временные эффекты, и авторы разрабатывают (локлально наилучший инвариантный) LBI-тест (locally best invariant), предполагая индивидуальные и временные эффекты случайными.

\section{Тестирование единичных корней в панелях при наличии нелинейности}

\subsection{Тестирование при наличии сдвига в детерминированной функции}

Наличие структурных сдвигов может исказить статистические выводы на основе тестов, построенных без учета этих сдвигов. Как было впервые показано в работе \citep{Perron1989} (в контексте одного временного ряда), при наличии структурного сдвига гипотеза о наличии единичного корня часто не отвергается, если процесс порождается как стационарный относительно структурного сдвига.


Для решения это проблемы было предложено добавлять соответствующие дамми переменные в детерминированную компоненту в расширенную регрессию Дики-Фуллера. В \citet{ILT2005} разрабатывается подход для аналогичного тестирования наличия единичных корней в панельных данных. Проблема заключается в том, что один из наиболее популярных тестов IPS требует коррекцию на математическое ожидание и стандартное отклонение усредненной статистики (средней всех индивидуальных статистик), которые, в свою очередь, зависят от местоположения сдвига, поскольку распределение индивидуальной тестовой статистики зависит от этого местоположения. В \citet{ILT2005} предлагается использовать LM-тест, предложенный \citet{SP1992} и \citep{AL1995}, однако, авторы не рассматривают возможность изменения наклона тренда, а только изменение в уровне. Согласно \citet{AL1995}, LM-тест остается инвариантным относительно включенных в модель дамми-переменных, и это свойство сохраняется для панельного теста на единичный корень. В \citet{ILT2005} авторы рассматривают модель следующего вида:
\begin{eqnarray}
y_{it}&=&\mu_i+\beta_it+\delta_iDU_{it}(T_{b,i})+u_{it},\\
u_{it}&=&\rho_i u_{i,t-1}+\varepsilon_{it},
\end{eqnarray}
где $DU(T_{b,i})=\mathbb I(t>T_{b,i})$, $\mathbb I(\cdot)$ -- индикатор-функция, а процесс $\varepsilon_{it}$ является слабо зависимым стационарным процессом, удовлетворяющим стандартным условиям. 

Тестовая статистика для коэффициента $\alpha_i$ строится на основе следующей регрессии:
\begin{equation}\label{eq.break1}
\Delta y_{it}=\gamma_i+\delta_i\Delta DU_{it}(T_{b,i})+\alpha_i\tilde{S}_{i,t-1}+e_{it},
\end{equation}
где 
\[\tilde{S}_{i,t-1}=y_{i,t-1}-\tilde{\gamma}_i(t-1)-\tilde{\delta}_iDU_{i,t-1},\]
оценки $\tilde{\gamma}_i$ и $\tilde{\delta}_i$ получены из OLS-регрессии
\[\Delta y-{it}=\gamma_i+\delta_i\Delta DU_{it}+\varepsilon_{it}.\]
Асимптотическое распределение LM-статистики для $i$-го временного ряда не зависит от местоположения сдвига, $\lambda_i=T_{b,i}/T$, так как зависимость от $\lambda_i$ имеет порядок $T^{-1/2}$. Для того чтобы зависимость была пренебрежимой для суммы статистик, требуется дополнительное условие на величину $N/T\rightarrow\kappa$ (посколкьу сумма разностей в кросс-секционных объектах накапливается с ростом $N$). Для соответствующей статистики IPS-типа тогда нет необходимости вычислять моменты, зависящие от местоположения сдвига - можно использовать моменты при условии отсутствия сдвига. Тогда также выполняется стандартная нормальная асимптотика для теста типа IPS. При наличии серийной корреляции регрессия \eqref{eq.break1} дополняется запаздывающими разностями ряда $\tilde{S}_{it}$.

\citet{CostantiniGutierrez2012} рассматривают модель с инновационным сдвигом:
\begin{eqnarray}
y_{it}&=&d_{it}+u_{it}\\
u_{it}&=&\rho_iu_{i,t-1}+\varepsilon_{it},\\
\varepsilon_{it}&=&\Psi_i(L)e_{it},
\end{eqnarray}
а детерминированная компонента специфицируется как
\begin{equation}
d_{it}=\mu_i+\Psi_i(L)\theta_iDU_{it}.
\end{equation}
Протестировать гипотезу единичного корня можно на основе следующей регрессии:
\begin{equation}
y_{it}=\rho_iy_{i,t-1}+\mu_i^*+\delta_iDU_{i,t-1}+\theta_iD_t(T_{b,i})+\sum_{j=1}^{k_i}{\phi_{ij}\Delta y_{i,t-j}}+e_{it},
\end{equation}
которая является просто переформулированным DGP. Дата сдвига оценивается на основе максимизации абсолютного значения $t$-статистики для $\hat{\theta}_i$. \citet{CostantiniGutierrez2012}  используют тесты, основанные на $p$-значениях \citet{MaddalaWu1999} и \citet{Choi2001} и процедуру решетчатого бутстрапа.

В \citet{LWY2016} сдвиг предполагается гладким, и авторы моделируют его функцией Фурье. Для проверки гипотезы единичного корня используется подход \citet{PSY2013}.

\subsection{Тестирование гипотезы о наличии единичного корня против нелинейной альтернативы}

\citet{UcarOmay2009} рассматривают нелинейный процесс порождения данных в виде экспоненциальной модели авторегерссионного процесса с гладким переходом (exponential smooth transition, ESTAR):
\begin{equation}
\Delta y_{it}=\mu_i+\phi_iy_{i,t-1}+\psi_iy_{i,t-1}\left[1-\exp(-\gamma_iy^2_{i,t-d})\right]+\varepsilon_{it},
\end{equation}
где $d\geq1$ является параметром задержки (delay) и $\gamma_i$ является параметром, обеспечивающим возвращение к среднему для всех $i$. Следуя общеприняой практике, положим $\phi_i=0$, так что средний (внутренний) режим является нестационарным, и $d=1$:
\begin{equation}
\Delta y_{it}=\mu_i+\psi_iy_{i,t-1}\left[1-\exp(-\gamma_iy^2_{i,t-1})\right]+\varepsilon_{it}.
\end{equation}
Тест на наличиие единичного корня основан на гипотезе $\gamma_i=0$ для всех $i$ против альтернативы, что $\theta_i>0$ для некоторых $i$. Однако непосредственное тестирование    $\gamma_i=0$ является затруднительным, поскольку параметр $\psi_i$ неидентифицируем при нулевой гипотезе. Следуя \citet{KSS2003}, применяется разложение Тейлора первого порядка около $\gamma_i=0$ для всех $i$, получая следующую регрессию:
\begin{equation}
\Delta y_{it}=\mu_i+\delta_iy_{i,t-1}^3+\varepsilon_{it},
\end{equation}
где $\delta_i=\psi_i\gamma_i$. После этого можно проверить гипотезу $\delta_i=0$ для всех $i$ (линейная нестационарность) против альтернативы $\delta_i<0$ для некоторых $i$ (нелинейная стационарность). Для тестирования используется стандартная техника панельных единичных корней, в \citet{UcarOmay2009} предлагается использовать усечение статистики аналогично \citet{Pesaran2007} (гарантирующее существование моментов) и решетчатый бутстрап для учета кросс-секционной корреляции.

\subsection{Тестирование изменения инерционности}

Большинство исследований основываются на предположении о постоянном порядке интегрированности временных рядов на всем рассматриваемом периоде. Однако существуют свидетельства того, что в некоторых экономических и финансовых временных рядах инерционность изменяется с I(0) к I(1) или наоборот в определенный момент времени. Это говорит о том, что авторегерссионный параметр не является стабильным с течением времени, что может изменить выводы о тестировании наличия единичного корня. \citet{CostantiniGutierrez2007} предлагают метод проверки гипотезы единичного корня против альтернативы об изменении инерционности для панельных данных. Авторы рассматривают обобщение рекурсивных тестов, предложенных в \citet{BLS1992} и \citet{taylor2005fluctuation}. Пусть процесс ошибок (абстрагируясь от детерминированной компоненты) порождается как
\begin{eqnarray}
y_{it}=\rho_{it}y_{i,t-1}+\varepsilon_{it}.
\end{eqnarray}
Тестируется гипотеза о наличии единичного корня (постоянной инерционности I(1)) для всех $i$, $H_0:\rho_{it}=1$, против альтернативы об изменении инерционности, что $\rho_{it}<1$ для $t=1,\dots,T_b$ и для $i=1,\dots,N_1$ и $\rho_{it}=1$ для $t=T_b+1,\dots,T$ и для $i=N_1,\dots,N$. Аналогично \citet{BLS1992} строятся ADF-регрессии по выборкам $t=1,\dots,\lfloor T\delta\rfloor$, $\delta\in[\delta_0,1]$. Последовательность таких статистик называется форвардными рекурсивными (forward-recursive) статистиками. Обратными рекурсивными (reverse-recursive) статистиками будут статистики, построенные по выборкам $t=T-\lfloor T\delta\rfloor +1,\dots,T$. Также рассматриваются статистики, предложенные в \citet{leybourne2003tests}, полученные при оценивании регрессии для обращенных во времени данных (первое наблюдение становится последним). Затем берутся минимумы для каждой из статистик по $\delta\in[\delta_0,1]$, а также минимумы от некоторых комбинаций различных статистик. Для тестирования гипотезы в панели применяется подход, основанный на $p$-значениях \citet{MaddalaWu1999} и \citet{Choi2001}. Для учета кросс-секционной коррелированности используется бутстрап.

\subsection{Тестирование на стационарность}

\citet{CBL2005}, а далее \citet{HadriRao2008} рассматривают тестирование гипотезы о том, что все временные ряды в панели являются стационарными относительно структурного сдвига. Аналогично тестированию гипотезы единичного корня в одномерных временных рядах, наличие сдвига при тестировании гипотезы о стационарности приводит к слишком частому отвержению этой гипотезы в пользу неверной альтернативы о наличии единичного корня. Тесты, предложенные в \citet{CBL2005} и \citet{HadriRao2008} являются обобщением панельного теста \citet{Hadri2000}. Рассматриваются следующие модели:
\begin{equation}
\text{Модель 0: }y_{it}=\mu_i+u_{it}+\delta_iDU_{it}+\epsilon_{it},
\end{equation}
\begin{equation}
\text{Модель 1: }y_{it}=\mu_i+\beta_it+u_{it}+\delta_iDU_{it}+\epsilon_{it},
\end{equation}
\begin{equation}
\text{Модель 2: }y_{it}=\mu_i+\beta_it+u_{it}+\gamma_iDT_{it}+\epsilon_{it},
\end{equation}
\begin{equation}
\text{Модель 3: }y_{it}=\mu_i+\beta_it+u_{it}+\delta_iDU_{it}+\gamma_iDT_{it}+\epsilon_{it},
\end{equation}
с 
\[u_{it}=u_{i,t-1}+v_{it},\]
где $v_{it}$ является слабо зависимым стационарным процессом. Переменные, отвечающие за сдвиг, определятся как $DU_{it}=\mathbb I(t>T_{b,i})$ и $DT_{it}=(t-T_{b,i})\mathbb I(t>T_{b,i})$. В модели 0 происходит только сдвиг в уровнях, в модели 1 предполагается наличие временного тренда, в Модели 2 предполагается изменение наклона тренда без сдвига в уровнях, а в Модели 3 происходит совместный эффект (сдвиг и в уровня, и в наклоне тренда). Отметим, что в \citet{HadriRao2008} рассматриваются все четыре модели, в то время как в \citet{CBL2005} рассматриваются только модели 0 и 3.

Аналогично случаю с отсутствием сдвигов, тестируется гипотеза
\[H_0:\sigma^2_{v,1}=\sigma^2_{v,2}=\dots=\sigma^2_{v,N}=0\]
против альтернативы
\[H_1:\sigma^2_{u,i}>0,\ i=1,2,\dots,N_1;\quad \sigma^2_{u,i}=0,\ i=N_1+1,\dots,N,\]
то есть, что некоторые временные ряды являются стационарными, а некоторые нестационарными. Тестовые статистики строятся аналогично \citet{Hadri2000}:
\begin{equation}
LM=\frac{1}{N}\sum_{i=1}^N\frac{\frac{1}{T^2}\sum_{t=1}^TS^2_{it}}{\hat{\sigma}^2_{\varepsilon,i}},
\end{equation}
где $S_{it}$ является частичной суммой остатков,
\[S_{it}=\sum_{j=1}^t\hat{\varepsilon}_{jt},\]
а $\hat{\sigma}^2_{\varepsilon,i}$ является оценкой долгосрочной дисперсии для $i$-го процесса, причем \citet{HadriRao2008} предлагают использовать подход \citet{SPC2005} с использованием граничного условия.  Тогда для построения итоговой панельнйо статистики требуется скорректировать значение LM-статистики на среднее и стандартное отклонение, чтобы получить асимптотическое стандартное нормальное распределение. Эти среднее и дисперсия в \citet{HadriRao2008} предлагается вычислять на основе характеристических функций.

\citet{HadriRao2008} дополнительно предлагают модифицированный тест для моделей 0 и 3, который позволяет получить предельное распределение, не зависящее от местоположения сдвига:
\[LM^*=\frac{1}{N}\sum_{i=1}^N\left[\frac{\frac{1}{T_{b,i}^2}\sum_{t=1}^{T_{b,i}}(\sum_{j=1}^t\hat{\varepsilon}_{jt})^2+\frac{1}{(T-T_{b,i})^2}\sum_{t=T_{b,i}+1}^T(\sum_{j=T_{b,i}+1}^t\hat{\varepsilon}_{jt})^2}{\hat{\sigma}^2_{\varepsilon,i}}\right].\]
Асимптотическое распределение выражения в квадратных скобках тогда является суммой двух независимых Винеровских процессов, которые не зависят от сдвига. Поэтому для построения итоговой статистики, скорректированной на среднее и стандартное отклонение, можно использовать фиксированные значения моментов, без учета сдвига.  Для Модели 0 математическое ожидание и дисперсия равны соответственно 1/3 и 2/45, а для Модели 3 они равны соответственно 2/15 и 11/3150. Отметим, что такого результата нельзя получить для Моделей 1 и 2.

Если дата сдвига неизвестна, то можно ее оценить и использовать как истинную, минимизируя сумму квадратов остатков по всем возможным датам сдвига для каждого $i$-го субъекта, причем даты сдвигов могут быть различными для разных $i$. При наличии пространственной корреляции в \citet{HadriRao2008} предлагается использовать решетчатый бутстрап.

\section{Тестирование на единичный корень для коротких панелей}

\subsection{Тестирование на наличие единичного корня при отсутствии структурных сдвигов}

Хотя, как утверждают IPS и LLC, их тесты работают и при малых значениях $T$, рекомендуется использовать симулированные, а не аналитические значения моментов распределений. Кроме того, для того, чтобы работала совместная асимптотика по $T$ и $N$, необходимо условие, чтобы $N/T\rightarrow0$, то есть чтобы $N$ было намного меньше, чем $T$.  \citet{HarrisTzavalis1999} разрабатывают тест на единичный корень в панельной авторегрессии первого порядка с серийно  некоррелированными нормальными ошибками при предположении, что $N\rightarrow\infty$, а $T$ фиксировано, что является естественным для многих эмпирических исследований. Авторы акцентируют внимание на однородной альтернативе.

Снова рассмотрим регрессию
\begin{equation}\label{ShortT0}
\Delta y_{it}=\phi y_{i,t-1}+\varepsilon_{it}.
\end{equation}
При отсутствии детерминированной компоненты
\begin{equation}\label{ShortT1}
\sqrt{N}\hat{\phi}\Rightarrow N(0,C_1),
\end{equation}
где $C_1=2[T(T-1)]^{-1}$, а $\hat{\phi}$ -- оценка параметра $\phi$ в регрессии \eqref{ShortT0}. Доказательство этого факта следует из того, что для фиксированного $T$ числитель в выражении для $\hat{\phi}$ является случайной величиной, независимо и одинаково распределенной среди $i$ с нулевым средним и конечной дисперсией, равной $[T(T-1)/2]\sigma^4_\varepsilon$. При $N\rightarrow\infty$ и фиксированном $T$ числитель сходится со скоростью $\sqrt{N}$ к нормальный случайной величине по центральной предельной теореме, знаменатель сходится по вероятности к неслучайной константе со скоростью $N$, из чего следует утверждение. Деля выражение $\sqrt{N}\hat{\phi}$ на $C_1^{1/2}$ (асимптотическое стандартное отклонение), можно получить стандартную $t$-статистику, имеющую асимптотически стандартное нормальное распределение и зависящую от оцененного параметра и известных значений $T$ и $N$. Можно показать, что при больших $T$ результат в \eqref{ShortT1} совпадает с полученным в LLC, то есть что $T\sqrt{N}\hat{\phi}\Rightarrow N(0,2)$. Отметим, что дисперсия тестовой статистики, когда $T$ фиксировано, больше, чем при предположении об асимптотике по $T$, в $T/(T-1)$ раз. Поэтому при предположении о большом $T$, когда на самом деле $T$ мало, гипотеза единичного корня будет отвергаться чаще, чем номинальный уровень значимости.

При наличии фиксированных эффектов
\begin{equation}\label{ShortT2}
\sqrt{N}(\hat{\phi}-B_2)\Rightarrow N(0,C_2),
\end{equation}
где $B_2=\mathrm{plim}_{N\rightarrow\infty}\hat{\phi}=-3(T+1)^{-1}$ и $C_2=[3(17T^2-20T+17)][5(T-1)(T+1)^3]^{-1}$. Аналогично случаю с отсутствием детерминированной компоненты при больших $T$ результаты совпадают с полученными в LLC. При бесконечном $T$ компонента, корректирующая смещение, больше, дисперсия $C_2$ также больше, так что дисперсия стандартизованной статистики будет меньше. Это приводит к сдвигу распределения стандартизованной тестовой статистики влево (относительно того, когда $T$ предполагается фиксированным) на
\[3\sqrt{N}\left[\left(\frac{5(T-1)(T+1)^3}{3(T+1)^2(17T^2-20T+7)}\right)^{1/2}-\left(\frac{5}{51}\right)^{1/2}\right],\]
и сдвиг увеличивает размер теста относительно номинального уровня. Однако, поскольку дисперсия $t$-статистики меньше при асимптотическом $T$, это приводит к уменьшению эмпирического размера. Если первый эффект доминирует второй, то искажения размера будут либеральными. В противном случае размер будет консервативным, и мощность, чтобы отвергнуть нулевую гипотезу, будет меньше.

При добавлении индивидуально специфических трендов
\begin{equation}\label{ShortT3}
\sqrt{N}(\hat{\phi}-B_3)\Rightarrow N(0,C_3),
\end{equation}
где $B_3=\mathrm{plim}_{N\rightarrow\infty}\hat{\phi}=-15[2(T+2)]^{-1}$ и $C_3=[15(193T^2-728T+1147)][112(T+2)^3(T-2)]^{-1}$. Использование моментов на основе большого значения $T$, а не фиксированного, имеет два противоположных эффекта на размер теста. При асимптотике величина корректирующего фактора больше, дисперсия $C_3$ также больше, и дисперсия стандартизованной статистики меньше. Все это сдвигает распределение стандартизованной статистики влево на величину
\[7.5\sqrt{N}\left[\left(\frac{112(T+2)^3(T-2)}{15(T+2)^2(193T^2-728T+1147)}\right)^{1/2}-\left(\frac{112}{2895}\right)^{1/2}\right],\]
и сдвиг увеличивает размер теста относительно номинального уровня. Поскольку дисперсия $t$-статистики меньше при большом $T$, это приводит к уменьшению эмпирического размера, поэтому  тест может быть как либеральным, так и консервативным. 

\citet{HarrisTzavalis2004} предложили более мощный тест LM типа. Имея ML-оценки параметров детерминированной компоненты при ограничении $\phi=0$, можно получить детрендированный ряд и оценить по нему регрессию Дики-Фуллера, получая оценку $\hat{\phi}$. При нулевой гипотезе асимптотическое смещение оценки $\hat{\phi}$ равно $-3/T$, и скорректированная на смещение оценка $\tilde{\phi}\equiv\hat{\rho}+3/T$ при $N\rightarrow\infty$
\begin{equation}\label{ShortT4}
\sqrt{N}\tilde{\phi}\Rightarrow N(0,C(k_\varepsilon,\sigma^2_\varepsilon,T)),
\end{equation}
где $k_\varepsilon$ -- четвертый момент $\varepsilon_{it}$, $E(\varepsilon_{it}^4)$, и
\[C(k_\varepsilon,\sigma^2_\varepsilon,T)=\frac{1}{20}\left[\frac{(T-2)(T+1)(T-3)}{T(T-1)}k_\varepsilon+\frac{(T-2)(4T-1)(T-3)^2}{T(T-1)}\sigma^4_\varepsilon\right]/\left[\frac{(T-2)T}{6}\sigma^2_\varepsilon\right]^2.\]
Для получения $t$-статистики нам нужны состоятельные оценки параметров $k_\varepsilon$ и $\sigma^2_\varepsilon$. В \citet{HarrisTzavalis2004} предлагаются следующие оценки:
\begin{eqnarray*}
\hat{\sigma}^2_\varepsilon &=&\frac{1}{N(T-1)}\sum_{i=1}^N\sum_{t=1}^T(\Delta y_{it}-\overline{\Delta y_i})^2,\\
\hat{k}_\varepsilon &=& \frac{\frac{T^2}{N}\sum_{i=1}^N\sum_{t=1}^T(\Delta y_{it}-\overline{\Delta y_i})^4-3(T-1)(2T-3)\hat{\sigma}^4_\varepsilon}{(T-1)(T^2-3T+3)\hat{\sigma}^2_\varepsilon}
\end{eqnarray*}

Если ошибка $\varepsilon_{it}$ распределена нормально, то
\begin{equation}\label{ShortT5}
\sqrt{N}\tilde{\phi}\Rightarrow N(0,C(T)),
\end{equation}
где $C(T)=[18(2T-3)(T-3)][5T^3(T-2)]^{-1}$, так что тест не будет зависеть от второго и четвертого момента ошибок, и дисперсия будет зависеть только от $T$. Заметим, что при больших $T$ нет необходимости требовать предположение о нормальности ошибок, поскольку компонента, включающая $k_\varepsilon$ в функции $C(k_\varepsilon,\sigma^2_\varepsilon,T)$, является $O(T^{-3})$ и сокращается при устремлении $T$ к бесконечности. На основе симуляций \citet{HarrisTzavalis2004} заключают, что тестовая статистика при предположении о нормальности ошибок робастна  к отклонениям от нормальности.

\citet{BlanderDhaene2012} обобщают работу \citet{HarrisTzavalis1999}  на случай, когда ошибки представляют собой AR(1) процесс. Это соответствует обобщению теста Дики-Фуллера на ADF(1) в контексте единственного временного ряда. Тестовую регрессию можно записать как (опуская для простоты детерминированную компоненту) 
\begin{equation}\label{ShortT6}
\Delta y_{it}=\phi y_{i,t-1}+\phi_1\Delta y_{i,t-1}+\varepsilon_{it},
\end{equation}
где параметр $\phi_1$ отвечает за дополнительную AR(1) динамику, предполагая ее однородность по всем $i$. \citet{BlanderDhaene2012}  получают величину асимптотического смещения для $\phi$ и $\phi_1$ при $N\rightarrow\infty$ и $T$. Итоговый тест основан на скорректированной на это смещение статистике, и это смещение, в свою очередь, зависит от параметра $\phi_1$. Поэтому для вычисления смещения нужна состоятельная оценка параметра $\phi_1$, которую предлагается получать, численно обращая выражение для вероятностного предела. Коррекция смещения совместно с соответствующим образом модифицированной стандартной ошибкой приводит к асимптотически нормальной $t$-статистике при нулевой гипотезе.

\citet{Madsen2010} выводит асимптотическую локальную мощность тестов \citet{BreitungMeyer1994} и \citet{HarrisTzavalis1999} и показывает, что первый тест имеет большую асимптотическую мощность, чем последний. В \citet{KaraviasTzavalis2014} тест \citet{Breitung2000} обобщается на случай фиксированного $T$, и для него находится асимптотическая локальная мощность.

\citet{WHT2007} предлагают два теста, тест основанный на  GMM и тест, основанный на оценке инструментальных переменных (IV тест). GMM тест допускает серийную корреляцию $p\leq T-2$ и основан на отклонении индивидуальных временных рядов в панели от своих начальных значений (как в \citet{BreitungMeyer1994}). Можно состоятельно оценить параметр $\phi$ на основе моментных условий вида
\[E(z_{is}u_{it}(\phi))=E(z_{is}(u_{it}-\phi u_{i,t-1}))=0,\ t=p+2,\dots,T, \ s=1,\dots,t-p-1,\]
где $z_{it}=y_{it}-y_{i0}$.
Тогда GMM-оценка, основанная на этих моментных условиях, принимает вид
\begin{equation}\label{ShortT7}
\hat{\phi}_{GMM}=\left[\left(\sum_{i=1}^Nz'_{i,-1}W_i\right)\hat{\Omega}^{-1}\left(\sum_{i=1}^NW'_iz_{i,-1}\right)\right]^{-1}\left[\left(\sum_{i=1}^Nz'_{i,-1}W_i\right)\hat{\Omega}^{-1}\left(\sum_{i=1}^NW'_iz_{i}\right),\right]
\end{equation}
где
\[W_i=
\begin{bmatrix}
z_{i1} & 0 & 0 & \dots & 0 & \dots & 0\\
0 & z_{i1} & z_{i2} &  & 0 &  & 0\\
\vdots &&&\ddots&\\
0 & 0 & 0 & \dots & z_{i1} & \dots & z_{i,T-p-1}
\end{bmatrix}
\]
является $(T-p-1)\times ((T-p-1)(T-p)/2)$ матрицей инструментов, $z_i=(z_{i,p+2},\dots,y_{i,T})'$ -- $(T-p-1)$-вектор наблюдений, $z_{i,-1}$ -- его однопериодное лагированное значение, $\hat{\Omega}=N^{-1}\sum_{i=1}^NW'_i\Delta y_i\Delta y'_iW_i$ -- состоятельная (при $N\rightarrow\infty$) оценка оптимальной взвешивающей матрицы $\Omega=N^{-1}\sum_{i=1}^NW'_iu_iu'_iW_i$ при нулевой гипотезе $\phi=1$ (поскольку $u_{it}=\Delta y_{it}$).\footnote{В \citet{Madsen2003} в качестве инструментов используются не уровни, а разности, как в \citet{ArellanoBover1995}, однако, при предположении об $i.i.d.$ ошибках. См. также аналогичный подход в \citet{BNW2005}.} На основе этой оценки можно построить $t$-статистику, имеющую стандартное нормальное распределение. При альтернативной гипотезе оценка $\hat{\phi}_{GMM}$ не является состоятельной (за исключением некоторых частных случаев), что может влиять на мощность теста.

Выбирая матрицу инструментов в виде диагональной, $diag(z_{i,1},\dots,z_{i,T-p-1})$ и суммируя итоговые моментные условия по $t$ в единственное условие вида 
\begin{equation}\label{ShortT9}
E\left[\sum_{t=1}^{T-p-1}z_{it}u_{i,t+p+1}(\phi)\right]=0,\ i=1,\dots,N,
\end{equation}
то GMM-оценка сводится к IV-оценке, определяемой как
\begin{equation}\label{ShortT9}
\hat{\phi}_{IV}=\left(\sum_{i=1}^N\sum_{t=1}^{T-p-1}z_{it}z_{i,t+p}\right)^{-1}\left(\sum_{i=1}^N\sum_{t=1}^{T-p-1}z_{it}z_{i,t+p+1}\right).
\end{equation}
Отметим, что тестовая статистика \citet{BreitungMeyer1994} является частным случаем IV-статистики при $p=0$.

Оба теста, GMM и IV, имеют корректный размер при малой размерности по времени относительно размерности кросс-секций, но при увеличении $T$ GMM-тест имеет сильные искажения размера. Для IV-теста сильно отрицательная автокорреляция приводит к существенной потере мощности, особенно  при альтернативе, далекой от нулевой гипотезы о наличии единичного корня (что является контринтуитивным). При $T\rightarrow\infty$ 
\begin{equation}\label{ShortT9.1}
\frac{T\sqrt{N}}{\sqrt{2}}(\hat{\phi}_{IV}-1)\Rightarrow N(0,1).
\end{equation}

При наличии индивидуальных трендов вместо вычитания из исходных временных рядов начального значения производится вычитание из рядов в первых разностях первой разности начальных значений (подробнее см. в \citet{KaraviasTzavalis2015}).

Локальная мощность для IV-теста была получена в \citet{KaraviasTzavalis2015}, где авторегрессионный коэффициент определяется как $\phi_N=1-c/\sqrt{N}$. При этом в модели с фиксированными эффектами
\begin{equation}\label{ShortT10}
\sqrt{N}V_{IV}^{-1/2}(\hat{\phi}_{IV}-1)\Rightarrow N(-ck_{IV},1)
\end{equation}
при $N\rightarrow\infty$, где $k_{IV}=1/\sqrt{V_{IV}}$, $V_{IV}$ -- дисперсия предельного распределения $\hat{\phi}_{IV}$. Параметр $k_{IV}$ зависит от числа наблюдений по времени $T$, порядка серийной корреляции $p$ и формы серийной корреляции, задаваемой некоторой ковариационной матрицей. При $p=0$ мощность будет максимальной, поскольку используется максимальное число моментных условий в \eqref{ShortT9}. \citet{KaraviasTzavalis2015} также исследовали локальную мощность при MA(1) процессе ошибок. Мощность IV-теста выше при положительной корреляции ошибок, чем при отрицательной.

\citet{KruinigerTzavalis2002} предлагают инвариантную к начальному значению тестовую статистику, проводя преобразование ``within group'':
\begin{equation}\label{ShortT11}
\hat{\phi}_{WG}=\left(\sum_{i=1}^Ny'_{i-1}Q\mathbf{y}_{i-1}\mathbf{M}_1\mathbf{y}_{i-1}\right)^{-1}\left(\sum_{i=1}^Ny'_{i-1}Q\mathbf{y}_{i-1}\mathbf{M}_1\mathbf{y}_{i}\right),
\end{equation}
где $\mathbf{M}_1$ -- матрица ортогонального проектирования $\mathbf{M}_1=\mathbf{I}_T-\mathbf{1}(\mathbf{1}'\mathbf{1})^{-1}\mathbf{1}'$. В \citet{KruinigerTzavalis2002} используется скорректированная на смещение тестовая статистика при $N\rightarrow\infty$
\begin{equation}\label{ShortT12}
\sqrt{N}V^{-1/2}_{WG}\hat{\delta}_{WG}\left(\hat{\phi}_{WG}-1-\frac{\hat{b}_{WG}}{\hat{\delta}_{WG}}\right)\Rightarrow N(0,1).
\end{equation}

\citet{Hayakawa2010} исследует GMM и IV тесты при ARMA$(p,q)$ ошибках, в отличие от \citet{KruinigerTzavalis2002} и \citet{WHT2007}, где предполагались ошибки в виде скользящего среднего. Также в работе \citet{Hayakawa2010} предлагается ряд методов выбора количества параметров $p$ и $q$ в модели.

\citet{KaraviasTzavalis2015} также исследуют локальную мощность WG-теста. При локальной альтернативе $t$-статистика сходится к $N(-ck_{WG},1)$. Поведение локальной мощности данного теста при различных параметрах, связанных с зависимостью ошибок, сильно отличается от поведения мощности IV-теста. IV-тест признается авторами более мощным, чем WG-тест, что связано с дополнительной коррекцией оценки для предотвращения несостоятельности. При $T\rightarrow\infty$ тест IV также является более мощным, WG-тест имеет тривиальную локальную мощность, поскольку корректируется только числитель, в отличие от \citet{HarrisTzavalis1999} (подробности см. в \citet{KaraviasTzavalis2015}).

При наличии трендов WG-тест имеет тривиальную мощность при отсутствии серийной корреляции, в то время как IV-тест имеет нетривиальную мощность в окрестности $N^{-1/2}$ единицы. WG-тест всегда имеет мощность при отрицательной серийной корреляции, и эта нетривиальная мощность является следствием коррекции оценки. На малых выборках IV-тест является смещенным, несмотря на хорошие асимптотические свойства, независимо от знака серийной корреляции ошибок. Это связано с плохой аппроксимацией асимптотической локальной мощности на малых выборках из-за взятия первых разностей и наличии детерминированной компоненты более сложного вида. Поэтому \citet{KaraviasTzavalis2015} предлагают при наличии трендов использовать WG-тест в коротких панелях с серийно коррелированными ошибками.


В \citet{BHP2005} разрабатывался тест на единичный корень c i.i.d. ошибками на основе оценки метода гауссовского квази-максимального правдоподобия (QMLE) для данных в первых разностях, чтобы избавиться от зависимости от фиксированных эффектов. См. также \citet{Kruiniger2008}, где предлагается тест на основе MLE. 

\citet{RSW2015} рассматривают более общую модель, допускающую серийную кросс-секционную корреляцию в ошибках и произвоьный вид детерминированной компоненты. Общие факторы предполагаются фиксированными (в \citet{BaiNg2004} общие факторы являются случайными), так что они могут включать эту детерминировнаную компоненту. \citet{RSW2015} предлагают использовать моментные условия для GMM-оценки авторегрессионного параметра. Рассматриваются оба случая, как с известными, так и с неизвестными (ненаблюдаемыми) факторами. Возможность автокорреляции также рассматрвиается в работе, и учитывается эта автокорреляция через моментные условия. Разработанный авторами тест асимптотически инвариантен к истинной и подобранной детерминированной компоненте, поскольку эта детерминированная компонента рассматривается как дополнительный общий фактор. Следовательно, нет необходимости корректировать тестовую статистику на среднее и дисперсию, и предельное распределение тестовой статистики является асимптотически нормальным независиимо от значения авторегрессионного коэффициента. Другими словами, предельное распределение является нормальным и в случае $|\phi|<1$, и в случае $\phi=1$, и даже $\phi>1$, то есть нет разрывности распределения в окрестности $\phi=1$.

В \citet{Choi2015} предлагается концептуально альтернативный подход для тестирования наличия единичного корня в коротких панелях, допускающий всего два  наблюдения по времени ($T=2$), в то время как все остальные работы требовали наличие минимум трех наблюдений по времени ($T=3$).

\subsection{Тестирование на наличие единичного корня при наличии структурных сдвигов}

Как уже было сказано, аналогично как одномерному, так и многомерному случаям (в частности, при тестировании гипотезы о наличии единичного корня в панельных данных), наличие неучтенного сдвига будет приводить к слишком редкому отвержению нулевой гипотезы о наличии единичного корня во всех временных рядах в панели.

\citet{KaraviasTzavalis2014a} (см. также \citet{Tzavalis2002})  предлагают обобщение работы \citet{HarrisTzavalis1999} на случай наличия общего (происходящего в одну и ту же дату для всех объектов) структурного сдвига при фиксированном $T$ и AR(1) модели, а также рассматривают возможность AR(2) ошибок. Сдвиг может быть как известным, так и неизвестным. Если дата сдвига неизвестна, \citet{KaraviasTzavalis2014a}, следуя подходу \citep{ZA1992}, предлагают вычислять дату сдвига согласно минимизации односторонней стандартизованной тестовой статистики по всем возможным датам сдвига. Предельное распределение получившегося теста является минимумом от фиксированного числа коррелированных стандартных нормальных величин. В работе аналитически вычисляется эта корреляционная матрица, а также приводятся соответствующие критические значения.

В \citet{KaraviasTzavalis2014b} предлагается другой тест для фиксировнаного $T$, допускающий пространственную (''географическую'') зависимость, серийную корреляцию (порядок авторегрессии может быть больше единицы), гетероскедастичность и неоднородность среди объектов. Детерминированная компонента может иметь вид нелинейного тренда, а также может иметь несколько структурных сдвигов (в известные или неизвестные даты). Также структурные сдвиги могут происходить не во всех субъектах. Пространственная (''географическая'') зависимость не обязательно должна задаваться некоторой взвешивающей матрицей для построения тестовой статистики, поскольку используется непараметрическая оценка ковариационной матрицы. 

При неизвестной дате сдвига авторы аналитически выводят предельное распределение как минимальную порядковую статистику между всеми альтернативными последовательными статистиками по каждой возможной дате сдвига. Аналогично \citet{KaraviasTzavalis2014a} предельное распределение этой статистики получается как конечное число коррелированных случайных величин. Форма распределения позволяет авторам аналитически получить критические значения без привлечения симуляций Монте-Карло, что является первым подобным результатом в исследованиях по единичным корням.

\citet{KaraviasTzavalis2016} исследуют асимптотическую локальную мощность двух тестов,  предложенных \citet{KaraviasTzavalis2014a} и \citet{KaraviasTzavalis2014b}.  Предельные распределения этих тестов получены при последовательности локальных альтернатив, и аналитическое выражение показывает, как их средние и дисперсии являются функциями от даты сдвига и временной размерности панели. Рассмотренные тесты имеют нетривиальную мощность в окрестности $N^{-1/2}$ единицы, когда модель включает индивидуальные константы. При наличии случайных трендов мощность становится тривиальной в этой окрестности. Однако эта проблема не всегда имеет место, если тесты допускают серийную корреляцию в ошибках (это увеличение мощности связано с взаимодействием параметров серийной корреляции и непараметрической коррекцией числителя при фиксированном $T$), и полностью исчезает при наличии кросс-секционной корреляции (это связано со способом коррекции числителя в случае фиксированного $T$), что показывает явное различие с локальной асимптотикой при больших $T$. Тест \citet{KaraviasTzavalis2014a} имеет более высокую асимптотическую локальную мощность, поскольку не требует состоятельного оценивания дисперсии ошибок (тест \citet{KaraviasTzavalis2014a} корректируется на несостоятельность и в числителе, и в знаменателе, в то время как тест \citet{KaraviasTzavalis2014b} корректируется только в числителе). В \citet{karavias2019generalized} предлагается общая теория для моделей с трендовой функцией обобщенного вида и обобщенным процессом ошибок, допуская несколько сдвигов, нелинейные тренды и неспецифицированный вид краткосрочной динамик, гетероскедастичности и кросс-секционной неоднородности. \citet{karavias2022missing} рассматривают поведение тестов \citet{HarrisTzavalis2004} и \citet{KaraviasTzavalis2014a} в случае несбалансированных панелей.

\subsection{Тестирование на стационарность}

Аналогично тестам на единичный корень для фиксированного $T$ в \citet{HadriLarsson2005} анализировались тесты на стационарность в панелях. При использовании асимптотического $T$ тест \eqref{Stat5} слишком часто отвергает нулевую гипотезу.  Аналогично \citet{HarrisTzavalis1999} необходимо получить формулы для конечных выборок для двух моментов статистики KPSS, $\eta_{iT}$ в обозначениях в уравнении \eqref{Stat3}. \citet{HadriLarsson2005} получают, что для случая с фиксированными эффектами
\[E(\eta_{iT})=\frac{T+1}{6T}\]
и 
\[E(\eta_{iT}^2)=\frac{T^2+1}{20T^2},\]
так что 
\[Var(\eta_{iT})=\frac{T^2+1}{20T^2}-\left(\frac{T+1}{6T}\right),\]
и можно использовать эти величины для построения скорректированной тестовой статистики аналогично \eqref{Stat5}. При наличии индивидуально специфических трендов выборочные моменты принимают следующие значения:
\[E(\eta_{iT})=\frac{T+2}{15T}\]
и 
\[E(\eta_{iT}^2)=\frac{(T+2)(13T^2+23)}{2100T^3},\]
так что 
\[Var(\eta_{iT})=\frac{(T+2)(13T^2+23)}{2100T^3}-\left(\frac{T+2}{15T}\right).\]


\citet{HLR2012} обобщают работу \citep{HadriRao2008} на случай фиксированного $T$ и $N\rightarrow\infty$, а точные выражения для первых моментов были получены с использованием результата \citet[Corollary 1]{Ghazal1994}.


\section{Практическая реализация}

В настоящее время не так много тестов на панельный единичный корень имплементировано в статистические пакеты анализа данных, особенно учитывающих кросс-секционную зависимость. Ряд команд в пакете Stata можно найти на веб сайте Markus Eberhardt: \url{https://sites.google.com/site/medevecon/code}

Seemon Reese разработал команду \textbf{panicca} для Stata для реализации метода PANICCA \citet{ReeseWesterlund2015},  а также методы \citet{BaiNg2004,BaiNg2010}, см. \url{https://simonreese.weebly.com/code.html}.

Для реализации теста \citet{KaraviasTzavalis2014a} Yiannis Karavias и Elias Tzavalis разработали команду \textbf{xtbunitroot} для Stata, см. также описание в \citet{chen2022panel} и \url{sites.google.com/site/yianniskaravias/files/xtbunitroot}. На веб-сайте  Yiannis Karavias также можно найти код в Matlab для реализации тестов при фиксированном $T$: тест \citet{KaraviasTzavalis2014}. см. \url{https://sites.google.com/site/yianniskaravias/files/fixed-t-breitung-s-panel-data-unit-root-test-code}; тест 
\citet{karavias2019generalized}, см. 
\url{https://sites.google.com/site/yianniskaravias/files/generalized-fixed-t-purts}.

Наконец, Stephan Smeekes и  Ines Wilms разработали пакет \textbf{bootUR} для R, включающий в себя реализацию метода оценки доли стационарных временныз рядов в панели \citet{Smeekes2014} и бутстраповский тест на единичный корень \citet{PSU2011}, см. также описание в \citet{SmeekesWilms2023} со ссылками на другие пакеты в R (\textbf{plm} и \textbf{pdR}), см. \url{https://smeekes.github.io/bootUR/}.

\newpage

\thispagestyle{empty}

\renewcommand{\bibname}{\hfillСПИСОК ЛИТЕРАТУРЫ\hfill\mbox{ }}
\addcontentsline{toc}{chapter}{СПИСОК ЛИТЕРАТУРЫ}

\bibliographystyle{OUPnamed}
\bibliography{svar,bibliography,seasonal,double,panel,nonlinear,explosive}

\end{document}